\documentclass[]{article}

\usepackage[margin=1.0in]{geometry}

\usepackage[utf8]{inputenc}
\usepackage{tabularx}
\usepackage{booktabs}
\usepackage{comment}
\usepackage{listings}
\usepackage{wrapfig}
\usepackage{hyperref}       
\usepackage{url}            
\usepackage{amsfonts}       
\usepackage{nicefrac}       
\usepackage{microtype}      
\usepackage{lipsum}		
\usepackage{graphicx}
\usepackage[numbers]{natbib}
\usepackage{doi}
\usepackage{xcolor}

\definecolor{pgreen}{HTML}{00A99A}
\definecolor{pblue}{HTML}{00A2E3}
\definecolor{pred}{HTML}{ED135A}
\definecolor{pgrey}{HTML}{949698}
\title{Context, Composition, Automation, and Communication - The C$^2$AC Roadmap for Modeling and Simulation}
\date{}

\usepackage{authblk}

\makeatletter
\let\@fnsymbol\@alph
\makeatother

\author{
	Adelinde Uhrmacher\thanks{University of Rostock, Germany, adelinde.uhrmacher@uni-rostock.de},
	Peter Frazier\thanks{Cornell University, USA, pf98@cornell.edu},
	Reiner H\"ahnle\thanks{TU Darmstadt, Germany, haehnle@informatik.tu-darmstadt.de},
	Franziska Klügl\thanks{University of \"Orebro, Sweden, franziska.klugl@oru.se},
	Fabian Lorig\thanks{Malmö University, Sweden, fabian.lorig@mau.se},
	Bertram Ludäscher\thanks{University of Illinois Urbana-Champaign, USA, ludaesch@illinois.edu},
	Laura Nenzi\thanks{University of Trieste, Ital, lnenzi@units.it},
	Cristina Ruiz-Martin\thanks{Carleton University, Canada, cristinaruizmartin@sce.carleton.ca},
	Bernhard Rumpe\thanks{RWTH Aachen, Germany, rumpe@se-rwth.de},
	Claudia Szabo\thanks{University of Adelaide, Australia, claudia.szabo@adelaide.edu.au},
	Gabriel A. Wainer\thanks{Carleton University, Canada, gwainer@sce.carleton.ca},
	Pia Wilsdorf\thanks{University of Rostock, Germany, pia.wilsdorf@uni-rostock.de}
}

\begin{document}


\maketitle
  
\begin{abstract}
Simulation has become, in many application areas, a sine-qua-non. Most recently, COVID-19 has underlined the importance of simulation studies and limitations in current practices and methods. We identify four goals of methodological work for addressing these limitations. The first is to provide better support for capturing, representing, and evaluating the context of simulation studies, including research questions, assumptions, requirements, and activities contributing to a simulation study. In addition, the composition of simulation models and other simulation studies' products must be supported beyond syntactical coherence, including aspects of semantics and purpose, enabling their effective reuse. A higher degree of automating simulation studies will contribute to more systematic, standardized simulation studies and their efficiency. Finally, it is essential to invest increased effort into effectively communicating results and the processes involved in simulation studies to enable their use in research and decision-making. These goals are not pursued independently of each other, but they will benefit from and sometimes even rely on advances in other subfields. In the present paper, we explore the basis and interdependencies evident in current research and practice and delineate future research directions based on these considerations.   
\end{abstract}

\clearpage
\section{Introduction}
\label{sec:introduction}

\begin{figure}
	\centering
	\includegraphics[width=0.95\linewidth]{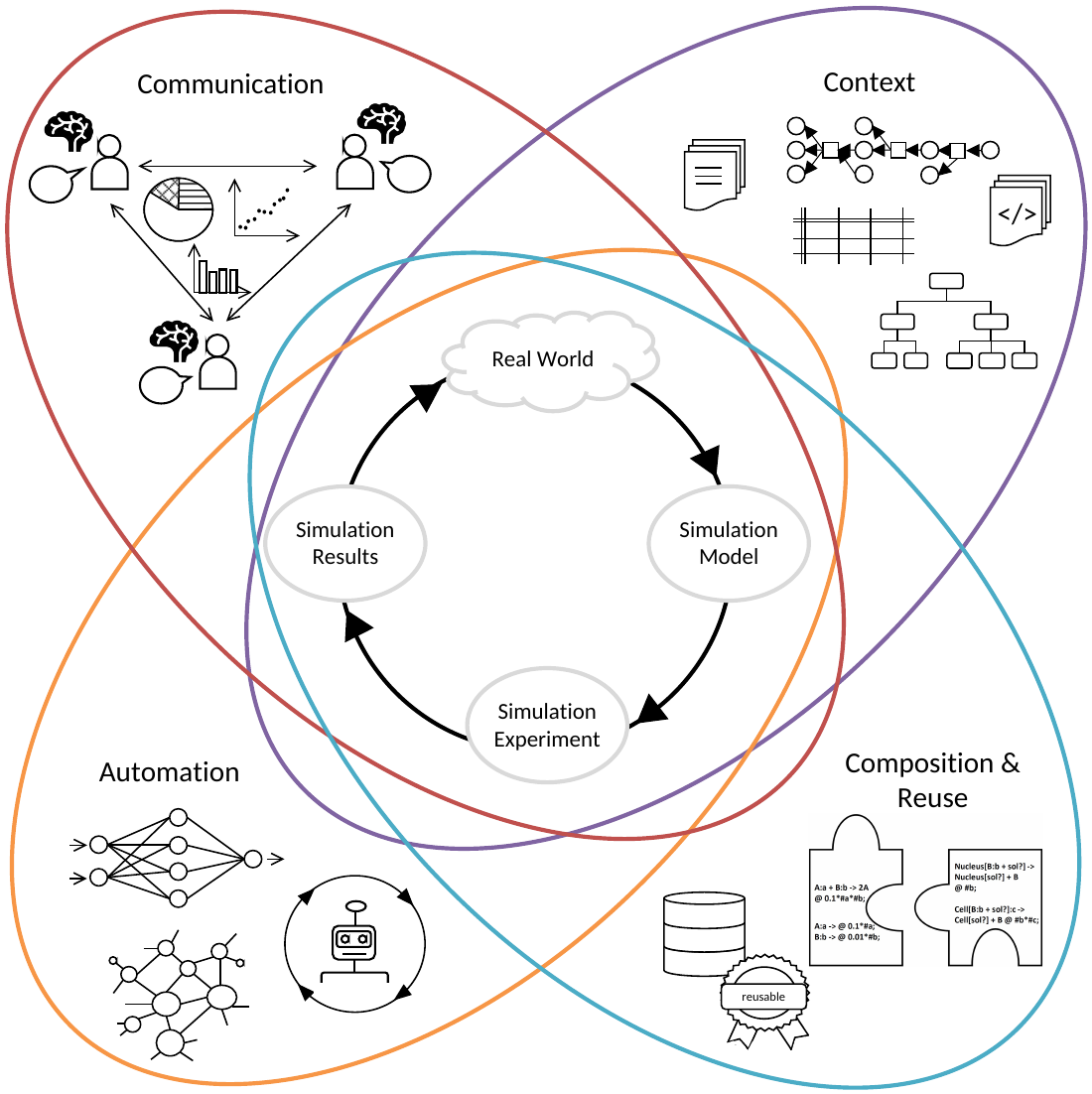}
	\caption{
		The roadmap proposes to support the entire life cycle of modeling and simulation by a) enriching context information beyond the conceptual model that can be deployed for developing the simulation model, executing simulation experiments, and interpreting results \ref{sec:context}, b) providing means for composition and reuse of the different artifacts of the simulation study, such as simulation model, simulation experiments, or behavioral requirements, as well as the needed software and methods \ref{sec:composition}, c) automating large portions of the modeling and simulation cycle, also by exploiting recent developments in artificial intelligence \ref{sec:automation}, and d) fine-tuning the representation of results, models, and activities involved in the simulation study to the mental models and needs of the different users and stakeholders.  
	}
	\label{fig:robinson}
\end{figure}

Simulation has become, in many areas, a sine qua non. Simulation, empirical evaluation, and analytical reasoning are regarded as the three pillars of science \cite{winsberg2010science}. Simulation studies rely on soundly conducting and effectively intertwining steps of analyzing the system of interest, developing and refining the simulation model, executing diverse simulation experiments, and interpreting the (intermediate) results \cite{balci2012life}. In this process, starting from the research question and the system of interest, inputs are selected and modified, assumptions and simplifications are revised in choosing a suitable abstraction for the model, requirements referring to outputs are specified and adapted, and data sources identified that might be used as input or to calibrate or validate the simulation model until a useful approximation has been achieved \cite{law2019build,robinson2014simulation} (Fig. \ref{fig:robinson}). Thus, each step is empowered and constrained by the methods used, as well as the knowledge and experiences of the modeler. In addition to the modeler -  possibly joined by data analysts, programmers, and visualization experts -   
domain experts and decision-makers might become involved at various points. 

Recently, the COVID-19 pandemic has underlined the importance of simulation studies \cite{lorig2021agent}. Simulations were widely used during the pandemic to make forecasts \cite{shinde2020forecasting} and support decisions made by governments \cite{bassoanalytics,ferguson2020impact}, hospitals \cite{garcia2022hospital}, industry \cite{ivanov2020predicting}
and universities \citep{frazier2022modeling}.  Simulations revealed some current limitations in conducting such studies \cite{wolf2020,edeling_impact_2021}, including how quickly useful models can be developed, how the results can be interpreted, and how results and crucial aspects of simulation studies can be communicated to domain experts, 
decision-makers, and the general public \cite{winsberg2020government}.  

To address these limitations in conducting and communicating simulation studies,
further methodological research is needed:
\begin{enumerate}
	\item Ensure that simulation studies come with \textbf{context}. Context is crucial for helping modelers and domain experts to interpret results and reuse simulation products. It is equally important to explain simulation results to decision-makers confidently.
	\item Improve model \textbf{composition} and reuse.  Model composition and reuse avoid building models from scratch. This saves time and, in addition, improves analysis quality because reuse is an important incentive for designing high-quality models. 
	\item Increase simulation \textbf{automation}. Central artifacts of a simulation study, e.g., simulation models and experiments, may be generated automatically. In addition, conducting and documenting the simulation study will benefit from intelligent guidance and support. Automation may save time for the modeler and  
	contribute to the overall quality of simulation studies. 
	\item Facilitate \textbf{communication}. This refers to communications between the modeler and domain expert and between the modeler and decision-makers. Problems in communication appear to be a central limiting factor in effectively using simulation for decisions. 
	Better communication would 
	also reduce the time required to produce impact. 
\end{enumerate}

Any progress towards these goals relies largely on an unambiguous and accessible representation of the simulation study, its activities, sources, and products, and a goal-directed and situation-specific processing of this knowledge. 
Therefore, in the following section, we will first look at the current state of the art in terms of how simulation models (with a focus on discrete and stochastic models), simulation experiments, and behavioral requirements are specified. 
These considerations will be revisited when scrutinizing promising research avenues toward achieving the identified four goals. The present paper builds upon discussions during the Dagstuhl seminar ``Computer Science Methods for Effective and Sustainable Simulation Studies (Dagstuhl Seminar 22401)''   \cite{dagstuhlReport2023}.

\section{State of the Art in Formal Approaches to Modeling and Simulation}
\label{sec:background}

Modeling means structuring and capturing knowledge about a given system in a suitably abstract manner. With the separation between the simulation model and executing the simulation model, the simulation model becomes explicit, accessible, and interpretable (possibly by different simulators) \cite{Zeigler2000}. The value of this separation of concerns and an explicitly and formally specified simulation model is undisputed and reflected in the development of diverse formalisms \cite{Zeigler2000,balbo2000introduction,hillston2005process},
by pragmatically augmented and extended general modeling languages, such as UML and SysML
\cite{UMLuser98,Rum17,Fow97,Wei06,FMS11,JPR+22},
and by the development of application-specific
modeling languages \cite{blinov2004bionetgen,boutillier2018kappa,Helms2017semantics,Rum17,Reinhardt2022language} (see Table \ref{tab:mobd}).


The credibility crisis in simulation \cite{pawlikowski2002credibility} and the desire to promote the reproduction of computational results \cite{merali2010computational} strengthened the case for the accessibility of simulation models and code. In addition, they moved the explicit specification of simulation experiments into the focus of interest \cite{ewald2014sessl,waltemath2011reproducible,salecker2019nlrx,gorlach2011conventional}, and motivated reporting and documentation guidelines for simulation experiments \cite{waltemath2011minimum} as well as for entire simulation studies \cite{rahmandad2012reporting,grimm2014towards,monks2019strengthening}. 
Increasingly, behavioral requirements, i.e., expectations referring to simulation model outputs, are expressed formally in temporal logic, or a domain-specific language to be checkable by (statistical) model checking \cite{agha2018survey, steffen_statistical_2019} or customized algorithms \cite{lorig2017formal}. Many specification languages come with a formal semantics. In Table \ref{tab:mobd}, we exemplarily summarize approaches used for simulation models, simulation experiments, and behavioral requirements, which will be discussed in greater detail in the following. 
It should be noted that the distinction between formalism and domain-specific language (DSL) is not that crisp. However, with DSLs, we refer to approaches and developments that consider the concrete syntax and questions referring to the realization as a usable and comfortable language right from the beginning, while formalisms concentrate more on puristic core concepts.

\begin{table}[]
	\caption{Exemplary approaches for specifying simulation models, simulation experiments, and behavioral requirements.}
	\label{tab:mobd}
	\begin{tabularx}{\textwidth}{XXXX}
		\toprule
		&  Simulation Models & Simulation Experiments &  Requirements \\
		\midrule
		Formalisms    &  DEVS \cite{zeigler1976, Zeigler1977,zeigler2018theory}, Stochastic Petri Nets \cite{balbo2000introduction}, Stochastic process algebras \cite{hillston2005process} & (DEVS) \cite{denil2017experiment}, DAGs \cite{DEELMAN201517} & (Spatio-)Temporal logic  \cite{bartocci_specification-based_2018survey, NenziBBL22} \\
		\midrule
		External DSLs & SysML \cite{JPR+22,Wei06}, BioNetGen \cite{blinov2004bionetgen}
		& SED-ML \cite{waltemath2011reproducible}, MDE-based approach \cite{wilsdorf2022model}, BPMN \cite{BPMN}
		& FITS \cite{lorig2017formal}, BPSL \cite{mitra2019pybionetfit} \\
		\midrule
		Internal DSLs, APIs & Repast \cite{North2013complex}, ABS \cite{abs,kmh21} & SESSL \cite{warnke2018complex}, NLRX \cite{salecker2019nlrx} & -- \\
		\bottomrule
	\end{tabularx}
\end{table}

\subsection{Formalisms and Theoretical Approaches}\label{Formalisms}
\label{sec:formalism}

Formalisms allow an implementation-independent specification of simulation models and other entities of relevance in the simulation life cycle, such as simulation experiments or (behavioral) requirements (Table \ref{tab:mobd}). Formalisms cater to broad applicability. Their classification happens typically at the level of an entire system class such as discrete stepwise, discrete event-based, continuous, or hybrid systems modeling \cite{Zeigler2000}.  We focus on modeling and simulating discrete-event (stochastic) systems in the following.

\subsubsection{Formalisms for Simulation Models}
\label{sec:formalismsimmodel}

Several formalisms were developed to describe simulation models of discrete event systems, including
DEVS (Discrete EVent Systems Specification) \cite{Zeigler2000}, stochastic Petri nets \cite{balbo2000introduction}, and processes \cite{hillston2005process}.

DEVS is a formalism for discrete event modeling that allows the modeler to define hierarchical modular models.
The formalism has been inspired by systems theory \cite{bertalanffy1968general}, which emphasizes a clear boundary between a system and its environment via inputs and outputs.  A system decides how to react to inputs in terms of its state changes and which events to produce, and a system may be composed of interacting sub-systems.
This perception results in DEVS's modular modeling approach of loosely coupled components that can be composed to form model hierarchies. Such a design facilitates model reuse and reduces the effort required for development and testing.

In contrast to DEVS, Petri nets interpret a system as a network of dependent entities and causally interrelated concurrent processes.
Petri nets form directed, bipartite graphs, with places and transitions (forming the nodes) connected by directed edges. In contrast to DEVS, whose time model is continuous so that events can occur at arbitrary times, in the original formulation of Petri Nets, the update of a state (due to the firing of a transition) occurs without an explicit notion of time.
However, extensions exist, such as stochastic Petri nets based on continuous time \cite{balbo2000introduction}.

Likewise, process algebras, such as the $\pi$-calculus, targeting the modeling and analysis of concurrent processes \cite{milner1992calculus}, originally lacked a notion of time and later were extended to describe dynamic systems as communicating, stochastic concurrent processes in continuous time \cite{priami1995stochastic}. In contrast to DEVS or Petri nets, the interaction structure of the $\pi$-calculus is not fixed but dynamic: new processes and new channels for letting processes interact are frequently generated.

Whereas the syntax in which a model is written is quite different in stochastic Petri Nets and stochastic $\pi$-calculus, the processes (i.e., the semantics) in either case are Continuous Time Markov Chains (CTMCs). Thus, in contrast to DEVS, they are based on stochastic semantics. All three formalisms clearly separate model syntax, semantics, and implementation; the same model can be implemented on different platforms supporting reliability and correctness (see Section \ref{sec:ao}). An explicitly defined semantics enables verifying simulation algorithms, e.g., 
\cite{Cardelli2007}.

Specifications in the above formalisms may not be succinct or too limited for specific model classes. This has resulted in further extensions of the formalisms, e.g., the introduction of colored tokens in the case of (stochastic) Petri nets \cite{jensen1996coloured} and attributes in the (stochastic) $\pi$-calculus \cite{John2010}.
Similarly, in DEVS, we find various extensions, e.g., to capture variable model structures \cite{Barros1997}.

\subsubsection{Formalisms for Simulation Experiments}
\label{sec:formalismexperiment}

A simulation can be interpreted as an experiment performed with a model and an experiment as "the process of extracting data from a system by exerting it through its inputs" \cite[p.4]{cellier2013continuous}.
Therefore, in principle, the above formalisms can also be used to model this process and, thus, specify experiments.
Already in the 1970s, Zeigler \cite{Zeigler1977} emphasized the role of explicitly defining experiments conducted with a model by introducing the concept of experimental frame.
An experimental frame is intended to specify the conditions under which a system is observed or experimented with \cite{Zeigler2000}.
Experimental frames consist of three model components: a generator that is responsible for generating input (traces),
a transducer that analyzes the simulation outputs (e.g., conducting summary statistics), and the acceptor that decides whether the experimental conditions are met.
DEVS has not been designed (nor have the other formalisms above) for specifying simulation experiments, such as sensitivity analysis or simulation-based optimization. The use of formalizing a simulation experiment as a dynamic system might be limited, but with experimental frames, crucial ingredients of simulation experiments, such as scanning the parameter space, monitoring the output, and properties that need to be checked, have been identified, and later work could build upon it 
\cite{hillston1991case,traore2006capturing,van2020exploring}.

In the quest for formalisms used for specifying simulation experiments, applying workflows for specifying and conducting simulation experiments requires further consideration \cite{ribault2012using,gorlach2011conventional,page2012goal}. Scientific workflows aim to accelerate scientific discovery in various ways, e.g., by providing workflow automation, scaling, abstraction, and provenance support \cite{cuevas2012scientific} (see also Section \ref{sec:context}). The most basic model treats a scientific workflow as a directed, acyclic graph (DAG) of computational tasks and their dependencies, i.e., a subsequent task can only be executed once all upstream tasks it depends on have been completed.
More advanced computational models view workflows as \emph{process networks} \cite{kahn1976coroutines}, synchronized by the data flow. Implementations of such process networks employ FIFO queues on input ports, resulting in a stream-based execution model \cite{ludascher2006scientific, oinn2006taverna}.

\subsubsection{Formalisms for Specifying Requirements}
\label{sec:formalismrequirements}

The calibration and validation of simulation models imply the execution of various (types) of simulation experiments \cite{leye2009discussion}. Thereby, behavioral requirements play a central role. Often, they are defined in terms of data that are supposed to be replicated by the simulation or in terms of formally specified properties that the simulation output is expected to fulfill. The latter has received increasing attention during the last two decades \cite{batt2006working}. 
A relevant and frequently used formalism to describe requirements relating to expected simulation results is the class of \emph{temporal logics} \cite{Clarke1999}, e.g.,  \emph{Linear-Time Temporal Logic} (LTL) or \emph{Signal-Temporal Logic} (STL)~\cite{donze2013efficient}.
Temporal logics are modal logics with specific \emph{temporal} operators that permit the specification of properties over time; for example, the {\it always} operator is a universal quantifier used to describe that a specification holds at all time instances, the \emph{eventually} operator is used to describe that a specification holds at some point in the future.  Describing subsequent events using the \emph{until} operator is also possible. There are many extensions to specify more refined behavior. \emph{Signal Spatio-Temporal Logic} (SSTL)~\cite{nenzi2018qualitative} and the \emph{Spatio-Temporal Reach and Escape Logic} (STREL) are extensions of STL with certain spatial operators and permit to describe complex emergent spatio-temporal behavior as the formation of patterns.
STL-\textasteriskcentered~\cite{BrimDSV14} and \emph{Time-Frequency Logic} (TFL)~\cite{DonzeMBNGS12} extend STL with means to express oscillatory behavior.
The \emph{Three-valued Spatio-Temporal Logic} (TSTL) enriches SSTL with a three-valued semantics. In this logic, statements about spatiotemporal trajectories can be true, false, and unknown, accounting for the simulations' intrinsic uncertainty and statistical analysis~\cite{Vissat2019analysis}. 
It should be noted that these behavioral requirements are only one type of requirement that simulation studies face \cite{balci2012life,robinson2014simulation}.  

\subsection{Domain-specific Languages}
\label{sec:dsl}

Domain-specific Languages (DSLs), in contrast to General-purpose Programming Languages (GPL),
are designed for a specific application domain \cite{Fowler2010domain}.
Similar to formalisms, they might be applied to simulation models and other artifacts of the modeling and simulation life cycle (Table \ref{tab:mobd}). External and internal DSLs are distinguished. An external DSL is parsed independently of the host general-purpose language, so the model is explicitly accessible as a syntax tree.  External DSLs have their own custom syntax and parser to process them.  
In contrast, internal DSLs are embedded within a GPL, i.e., they are a kind of API designed to exhibit a natural reading flow: the host language is used in a way that gives the feel of a specific language 
\cite{Fowler2010domain}, are Turing-complete and, typically,  
more difficult to analyze than an external DSL.

\subsubsection{Domain-specific Languages for Modeling}
\label{sec:ao}

Modeling languages aim for reuse, better understanding, and communicative abilities of the underlying model and, thus, better sustainability of the model and simulation results. Furthermore, modeling languages can be equipped with sophisticated static analysis techniques for specific properties, giving developers quick feedback and thus considerably increasing efficiency during development and simulation.
DSLs for modeling map the formalisms
discussed in Section \ref{sec:formalism}
into specific languages, including a concrete syntax, to be executed by simulation algorithms according to the formalism's semantics.
They feature various convenient domain- and problem-specific modeling constructs to simplify modeling tasks.

DSL-based simulations have the advantage that the design of DSLs promotes a clear separation of concern between the model and execution engine.
This allows the modeler to focus on the model and, since the syntax and semantics of the DSL are given explicitly, to analyze the model for certain properties, e.g., by model checking techniques \cite{Clarke1999} or a structural comparison of data structures \cite{MRR11b}.
This is particularly the case for external domain-specific modeling languages.

The design of models based on DSLs stands in contrast to highly optimized general-purpose simulation programs, in which a model and its execution are encoded together.
Due to a lack of separation of concern, the model is not easily accessible and reusable either by a human modeler or by another inference program, e.g., to analyze the model statically. In addition, the approach results in other problems, as the simulation engine has not been verified independently of the model. Both can threaten the credibility and validity of computational results \cite{merali2010computational}.

The \emph{Unified Modeling Language} (UML)
\cite{UMLuser98,Fow97,Rum16,Rum17} has been designed
as a standardized modeling language consisting of 14 different explicit modeling
sublanguages for different aspects of software systems.
UML's class diagrams and object diagrams
focus on structure, while Statecharts and activity
diagrams focus on behavior. Semantics has been defined, e.g., in \cite{EFLR99}
to make UML precise \cite{EBF+98}, but due to the general use of UML,
no generally accepted semantics exists. 
Typically, specific profiles of UML are used if a model is to be developed to improve the system understanding via simulation. 
These 
tools synthesize executable simulation code from UML models, typically connected with a core simulation framework. \cite{Bocetal19} discusses UML and, in a broader sense, MBSE approaches to simulation and their merits and drawbacks.

The \emph{Systems Modeling Language} (SysML) is based on 
UML for Systems Engineering, 
and thus, both languages share many common modeling concepts. SysML has received widespread use, e.g., in mechanical engineering \cite{CASSE20171,FMS11,Wei06}. SysML provides additional diagrams to model distributed processes and components with a static structure. SysML also supports discrete event simulation as well as continuous systems simulation.
Today, SysML is mainly used for higher-level systems definitions, including 
evaluating design alternatives, calculating what-if scenarios, and conducting requirements compliance analysis,
including necessary quality assurances and similar engineering tasks. Many of these are based on \cite{NKT+15} or refer to \cite{tolk2013reference} simulation. Because of the increasing necessity to simulate engineered systems virtually before the first physical prototypes emerge, it can be assumed that simulation using DSLs, e.g., based on SysML \cite{CBCR15}, will become a major technique in engineering.


The syntax and semantics of a DSL reflect the primary modeling metaphor(s) and needs of the application domain or community.
This becomes particularly evident if a DSL for modeling focuses on a particular application domain with a well-established modeling metaphor, such as studying gene-regulatory or biochemical systems. Consequently, the syntax of various DSLs for biochemical systems that have been developed over the last two decades builds on the reaction (or rule-) metaphor \cite{faeder2009rule}. The semantics of these DSLs vary between continuous system semantics, transforming a set of reactions to a set of ODEs to be solved by numerical integration, or taking the stochasticity of the system into account by executing the model by stochastic simulation algorithms (SSA) \cite{hoops2006copasi} (and thus interpreting the model as a CTMC \cite{Helms2017semantics}), or, even considering the spatial heterogeneity, interpreting reactions as collisions between particles in space \cite{blinov2017compartmental}.
Whereas switching between spatial and non-spatial semantics requires additional information, e.g., about the diffusion constant or size of reactants, the switching between ODE-based and SSA execution (or a combination thereof) is supported to occur transparently to the simulation model by various simulation tools in the field. This includes established modeling and simulation tools such as COPASI \cite{hoops2006copasi} and research tools such as BioPepa \cite{ciocchetta2009bio}. BioPepa is a DSL derived from the process algebra Pepa \cite{gilmore1994pepa} and has been equipped with different types of semantics to be executed as a set of ODEs or by SSAs, depending on the biochemical system being investigated.



The role of internal DSLs for modeling grows, particularly if the subject of modeling is a hardly constrained class of models that can easily be mapped, e.g., to an object-oriented GPL. This is the case in agent-based \cite{ABAR2017} and DEVS-based modeling for simulation \cite{devs1,wainer2001defining}. Internal or embedded DSLs allow us to use all the features of the host language, including inheritance and type systems, and to program, e.g., agents, as the modeler likes \cite{north2006experiences,luke2005mason}.

Selecting a suitable host language is a crucial first step in designing an internal DSL. The ease of realizing a modeling language as an internal DSL depends on how the programming paradigm of the host language and the offered features fit the requirements of the envisioned DSL and how widely used the host language is; one advantage of an internal DSL is not to learn a new language.
For example the spread of the Java language in the late 90s, with its object-oriented programming paradigm and its convenient features such as simplicity, platform independence, type system, reflection, and support for distributed execution, has led to the development of various \cite{howell1998simjava,mcnab1996using}, in particular agent-based \cite{north2006experiences,luke2005mason}, modeling and simulation tools. 
Based on these tools, specialized internal DSLs can be created tailored to specific models' sub-classes, e.g., for modeling continuous-time agent-based models with CTMC semantics (Fig. \ref{lst:repast-rulebased}). In these cases where an object-oriented programming paradigm is adopted for modeling and simulation, the use of UML diagrams, in particular the class, sequence, state, and activity diagrams, is advocated, e.g., during development and for the documentation of agent-based simulation models \cite{bersini2012uml}, or for conceptual modeling of discrete event systems \cite{Wagner2012UML}.

\begin{figure}[]
	\begin{lstlisting}[language=Java]
		public class SIRAgent extends Agent {
			
			/* ... */
			
			addRule(	() -> this.isInfectious(),
			() -> exp(recoverRate),
			() -> this.infectionState = InfectionState.RECOVERED);
			
			addRule(	() -> this.isSusceptible(),
			() -> exp(infectionRate * neighbours(SIRAgent.class).
			filter((SIRAgent agent) -> agent.isInfectious()).size()),
			() -> this.infectionState = InfectionState.INFECTIOUS);
		}
	\end{lstlisting}
	\caption{
		Model code snippet from a rule-based Repast implementation of an agent-based simple epidemic (susceptible, infected, recovered) model. An adaptation layer enables a compact description of agent-based CTMC models in a style that resembles rule-based languages and allows the execution of agents in Repast Simphony \cite{warnke2016population}.
		The \texttt{addRule} method is provided by an abstract \texttt{Agent} class. It is called in the constructor of a concrete agent class.		
		For the definition of the condition, waiting-time distribution, and effect, the anonymous functions of Java 8 are exploited.		
	}
	\label{lst:repast-rulebased}
\end{figure}

New programming languages or paradigms with compelling features (either for modeling or simulation) always spur interest in the modeling and simulation community.
Active objects (AO) \cite{ActiveObjects17,ABS-SoTA23} are such a 
programming paradigm. Its characteristic feature is that tasks are executed on objects, each with exclusive resource access to the object's memory and processor. Consequently, no interleaving occurs at the statement level but only at the task level. Suspension and resumption of tasks are governed by guards that watch for time- or data-driven events. AO languages are designed to scale to thousands of objects \cite{SJKT22}. They constitute a language paradigm that permits event-driven simulation at scale while abstracting away from low-level concurrency. Language features such as strong data encapsulation, modules, and type safety support modular rule design similar to Figure~\ref{lst:repast-rulebased}. 
Active objects have been used to simulate complex systems, for example, the safety mechanisms of railway operations \cite{KHS18}, high-performance computing interconnection networks \cite{EMHP21}, or container frameworks \cite{TBDJTD20}. The active object paradigm can be extended to the simulation of real-time \cite{JST12,KKS18,Sirjani19} and hybrid \cite{kmh21} systems.
An interesting aspect of AO languages is that their explicit synchronization enables advanced static analysis techniques, including deductive verification \cite{DBH15,KDHJ20}, deadlock detection \cite{GLL16} or worst-case resource analysis \cite{AFPR15}. Specifically, the AO language ABS \cite{abs} was designed with the capability of analysis in mind \cite{BMH14}. 
%
Some AO languages, for example, ABS \cite{abs}, are 
independently executable from any host language, while others are conceived as libraries \cite{HenrioRochas17}. Nevertheless, AO languages can be classified as \emph{internal} DSLs in the present setting because they build upon an object-oriented or object-based core language.
%

The appeal of a GPL as a host language for designing an internal DSL for modeling (and simulation) depends not only on features related to the ease of modeling and the execution efficiency but also on features that are related to the execution and analysis of simulation experiments, i.e., how rich the ecosystem is that a potential host language offers for conducting and analyzing a wide variety of simulation experiments. Thus, also due to Python's widespread use, low threshold, and, in particular, libraries offered for data sciences, several Python-based modeling and simulation tools have been developed in the last decade \cite{matloff2008introduction}, including implementations of Petri Nets \cite{pourbafrani2021python} and DEVS-based simulation tools \cite{van2014modular}.

\subsubsection{DSLs for Simulation Experiments}
\label{sec:dsl-sim}

The increasing awareness about the role of simulation experiments in developing simulation models and conducting simulation studies, on the one hand, and about the credibility crisis of simulation, on the other hand, pushed the development of internal and external DSLs for specifying simulation experiments.
An example of an internal DSL is SESSL: the Simulation Experiment Specification via a Scala Layer \cite{ewald2014sessl}. It relies on bindings to simulation tools and experiment libraries to offer a wide range of simulation experiments \cite{warnke2018complex}, including parameter scans, sensitivity analysis, simulation-based optimization, bifurcation analysis, and statistical model checking. 
Another example is NLRX, a package embedded in R that supports the specification and execution of various experiments with NetLogo models \cite{salecker2019nlrx}.

Also, GPLs aimed at data sciences, such as Python, or computational science, such as Julia, increasingly support the specification of simulation experiments via libraries that offer experiment design, sensitivity analysis, and optimization methods \cite{douglas2020certain,Pythonwsc2021}. These libraries also show the fluent transition between internal DSLs and APIs \cite{Fowler2010domain}.
The advantage of internal DSLs is their flexibility and the range of tools that ship with the GPL. 
If combined with a thorough design of the DSL, internal DSLs enable an executable and, at the same time, highly succinct and readable specification of simulation experiments and help to establish those as first-class objects of simulation studies.

One of the drawbacks of using an internal DSL for specifying simulation experiments is that automatically interpreting, adapting, and reusing these scripts requires significant effort \cite{PengWHU16}.
This is the virtue of external DSLs. During parsing, the crucial parts of the simulation experiment specifications can be easily identified and accessed.
SED-ML is an external DSL, an XML-based format in which simulation experiments can be encoded  \cite{waltemath2011reproducible}.  
SED-ML is a community standard introduced to facilitate the reuse of simulation experiments across simulation tools (Sec. \ref{sec:composition}).  
In RASE (Reuse and Adapt framework for Simulation Experiments~\cite{wilsdorf2022model}), the simulation experiments are also encoded in a tool-independent format, namely, JSON \cite{wilsdorf2022automatic}. 
In SED-ML and RASE, using ontologies referring to the methods used, e.g., specific simulation algorithms or optimization methods, is crucial (see Sec. \ref{sec:composition}).

Whereas the above DSLs have been explicitly designed for specifying simulation experiments, workflow languages are also applicable in principle. The business process modeling community has embraced BPMN \cite{BPMN} as a control-flow-oriented language suitable for business workflow applications. 
The scientific workflow community, on the other hand, 
has different requirements due to the data-intensive and compute-intensive nature of computational science applications and thus has not embraced business workflow models and standards such as BPMN or BPEL
\cite{LudascherWMB09}. 
Instead, specific dataflow-oriented languages and models have been developed, sometimes with specialized features to aid workflow design and comprehension (e.g., COMAD \cite{McPhillips2009workflows} for collection-oriented modeling and design, see also \cite{ZinnBML09}). Subsequently, the Common Workflow Language (CWL) for computational data-analysis workflows was defined \cite{crusoe_methods_2022}. However, control structures play a role in simulation, and consequently, adaptations of BPEL have been successfully applied for specifying and conducting simulation experiments \cite{gorlach2011conventional}.

\subsubsection{DSLs for Requirements}

Another area where DSLs are of importance is in specifying requirements. 
Formally specified behavioral requirements, e.g., in a logic-based language, can be tested automatically.
This applies to deterministic models \cite{dlr99941} as well as to stochastic models \cite{agha2018survey}.
To specify behavioral requirements, such as desirable properties of simulation output, variants of temporal logics are used as a basis (see Section \ref{sec:formalismrequirements}), as are custom-built languages for specifying properties or hypotheses of the simulation model \cite{lorig2017formal,mitra2019pybionetfit,Wilsdorf2023}.
The latter efforts aim at providing languages tailored to expressing requirements or hypotheses by a modeler in "a natural manner." Thus, the usability of the languages by users without a background in computer science propels research on these languages, even though they may forestall analysis capabilities.

Using a logic-based language (with formal syntax and model-theoretic semantics) can decrease the possibility of creating incompatible requirements and help to standardize their definition. 
In addition, established model-checking techniques permit one to automatically verify the satisfaction of expressions in a logic language, avoiding \emph{ad hoc} creation of property test code or even manual inspection of simulations. 

In a stochastic setting, probabilistic model checking is a well-established verification technique that can compute the probability that a property expressed in temporal logic may be satisfied by a given stochastic process.
However, standard model checking techniques~\cite{KwiatkowskaNP05} are not feasible for large-scale stochastic systems. In this case, the standard procedure is to use \emph{Statistical Model Checking} (SMC)~\cite{agha2018survey, steffen_statistical_2019}.
The underlying idea is to approximate the probability of satisfaction of a given formula statistically utilizing simulation, checking only a subset of the whole trajectory space, with usually a guarantee of asymptotic correctness. There are several approaches: qualitative SMC (based on hypothesis testing), quantitative SMC (based on confidence intervals), Bayesian SMC, and SMC for rare events.
See~\cite{agha2018survey, steffen_statistical_2019} for surveys on the topic.
SMC is an efficient technique when the model is fully specified. Still, it is computationally too expensive to analyze a model with uncertain parameters if we want to study some parameters or input space of the model. A 
method to overcome this situation is smoothed model checking (smMC)~\cite{BortolussiMS16}. 
Another consideration when using a logic-based approach in modeling and simulation is that formal methods can be fashioned to infer the requirements directly and automatically for trajectories. Mining logic specifications from data is a promising and challenging new line of research \cite{BartocciMNN22}, which also circumvents the need for the modeler to specify the logic formulas.

Behavioral requirements are only one type of requirement, although likely the most obvious one relating to modeling and simulation. Several UML sub-languages support abstractly capturing structural, behavioral, and interaction requirements. With its Object Constraint Language (OCL), UML provides this textual logic language OCL, which is roughly an executable subset of first-order logic with operations for container structures and associations that can be used to define properties such as requirements, invariants, pre- or postconditions. It is useful to provide mechanisms for underspecification \cite{PR94} in the language in various forms to be able to capture known behavior but abstract away from unknown or irrelevant details. This allows iterative refinement during a development process \cite{PR97} and non-deterministic and probabilistic simulations. 
In the context of software product lines, where variability management is considered a key aspect, requirements modeling is of central importance. It has motivated the development of a larger number of requirement modeling languages \cite{SEPULVEDA201616}. 
The potential of this perception and
the developed languages still wait to be exploited for simulation studies, particularly in the context of model composition and reuse (see Section \ref{sec:composition}).

\subsubsection{The Role of Metamodeling in DSLs}

When developing an external DSL for simulation purposes,
defining the language in its constituents is necessary  \cite{CFJ+16}.
There are three main approaches to developing a DSL: from scratch,
through reuse and variant building of a previously given DSL \cite{CGR09},
and via customization and adaptation of a more general modeling language,
such as SysML. 
For textual languages, one often uses
grammar to describe concrete and abstract syntax \cite{HKR21}.
For diagrammatic languages, metamodeling
\cite{CESW04,AK03,Kle08,ZKD+09} is the best option
to define abstract syntax as an essential core of a modeling language.
Metamodeling became prominent with the Meta Object Facility (MOF)
\cite{OMG08b} and was first used as the syntactic foundation
for the UML.

Metamodeling can be used to define a language's structure and various
additional ingredients, such as internal data types, default values,
predefined functions, simulation schedulers, etc.
These forms of language definitions
support a compact and human-readable specification of simulation
models, simulation experiments, or requirements, while if an appropriate code generator is available, the specifications can be directly mapped to executable simulation models, simulation experiments, 
or algorithms that analyze the results.
For instance, a metamodel specification may be used to map models
in SysML to models for a specific DEVS
simulator~\cite{kapos2019declarative}, or models may be mapped
from BPNM to DEVS, and then from DEVS to executable Java
code~\cite{Cetinkaya2012model}. For that purpose, it helps if the
source language, here SysML, was extended by DEVS-specific constructs
to simplify the mapping  \cite{CBCR15}.

In multi-formalism or multi-paradigm modeling, a complex system's components are expressed through different formalisms, for example, Petri nets, Statecharts, and ordinary differential equations~\cite{Sanders2003,DeLara2004meta}. The various formalisms are represented using an abstract syntax graph, i.e., a ``model of formalism''. Via graph grammars, their metamodels can be transformed into a common formalism, and code can be generated for simulation execution and further analysis. 
With respect to simulation experiments, the model-driven architecture (incl.\ metamodels) \cite{mellor2002model} has been applied to the generation of experiment designs~\cite{Teran-Somohano2015model} and, more generally, for specifying and generating different types of simulation experiments in a back end-independent format, such as JSON~\cite{wilsdorf2022model}.

More widespread use of model-driven engineering of simulation studies as a whole would increase the reusability and self-explainability of models, their requirements and assumptions, simulation experiments, and input and output data and allow further analysis techniques. 
For example, Zschaler and Polack propose a model-driven approach based on a family of DSLs~\cite{zschaler2020family}. As a central feature, they include a language for fitness-for-purpose argumentation, adapted from the Goal Structuring Notation (GSN)~\cite{kelly2004goal}. They argue that this combination of languages allows for building trustworthy and scientifically robust simulation models. 

\section{Supporting context within simulation studies}
\label{sec:context}

Context relates to any important information to conduct and interpret a simulation study. 
Collecting, revising, and representing suitable context information is also at the heart of conceptual modeling as defined in \cite{robinson2008conceptual}.
It should be noted that in contrast to other areas of computer science, such as software engineering, no agreed-upon definition exists for the conceptual model. However, its importance is undisputed in the modeling and simulation field  \cite{robinson2015,fujimoto2017research}.
Definitions range from defining the conceptual model as an abstract description of the simulation model, e.g., exploiting qualitative modeling methods such as UML (see Section \ref{sec:background}), to forming a conglomerate of all information possibly helpful in conducting and consequently, interpreting the results of a simulation study \cite{wilsdorf2020conceptual}.  The conceptual model in \cite{robinson2008conceptual} subsumes research questions;  requirements and general project objectives regarding, e.g., visualization or simulation speed; model inputs, outputs, as well as the data used; scope, level of detail, assumptions, and simplifications; entities, equations (or rules) referring to the system and its dynamics to be modeled, and which modeling approach to use; and, finally, justifications for each design choice. 
In addition to the conceptual model, simulation experiments that have been performed with the model for calibration, validation, or analysis hold important information to interpret and reuse the results of a simulation study and consequently belong to its context as well. Similarly, previous versions of a simulation model and how they have been refined form valuable information about a simulation study \cite{Haack2020}. Both emphasize a more process-oriented view of the context, i.e., how the different sources and (sub-)products are interrelated and have been used to generate the (final)  results.     

\subsection{State of the Art}

To identify what is typically considered information important to interpret a simulation study, we will take a closer look at documentation standards to move from there to a process-oriented view of context and supporting methods that can be exploited for representing and processing this information, such as provenance standards and workflows.  

\subsubsection{Reporting Guidelines for Simulation Studies.}
\label{sec:reporting_guidelines}

The wish to document information about a simulation study that helps to reproduce, interpret, and reuse its results has led to various reporting guidelines. These take the particular demands of the type of simulation model and experiments being executed into account, e.g., systems dynamics \cite{rahmandad2012reporting}, agent-based models \cite{grimm2017documenting,grimm2020odd}, or finite element methods \cite{erdemir2012considerations}. In addition, they may build on sources and conventions of the application field, such as ontologies \cite{krause2010annotation,waltemath2011minimum}. 
Independently of the type of simulation model, specific context information about a simulation study, such as research questions, assumptions, data used, and simulation experiments, is an intrinsic part of its documentation. Documentation guidelines --- examples are TRACE (TRAnsparent and Comprehensive model Evaluation) \cite{grimm2014towards} and STRESS (Strengthening the Reporting of Empirical Simulation Studies) \cite{monks2019strengthening}---aim at documenting all of the essential steps, sources, and products of a modeling and simulation life cycle \cite{balci2012life}. The checklist of TRACE, e.g., comprises problem formulation, model description, data evaluation, conceptual model evaluation, implementation, verification, model output verification, model analysis, and model corroboration. 

\subsubsection{Provenance and Provenance Standards.}

Reporting guidelines concern provenance, i.e., providing ``information about entities, activities, and people involved in producing a piece of data or thing'' \cite{moreau2013provenance}. However, reporting guidelines typically refer to the final results of simulation studies, not the processes by which those have been generated, including variations of the different artifacts or unsuccessful attempts. Provenance opens up a specific view on context, i.e., focusing on the production process of entities in which activities put sources and (intermediate) products of the simulation study into relation to each other by being used and generated by activities.  
Adopting provenance standards \cite{moreau2013provenance} allows modeling these processes qualitatively \cite{ruscheinski2017provenance}. 
Thereby, entities, such as research question, simulation model, simulation experiment (specification), simulation data, parameter, requirement (referring to output behavior), qualitative model, and assumption, are related by activities such as creating a simulation model, refining a simulation model, re-implementing a simulation model, calibrating, analyzing, and validating a simulation model \cite{wilsdorf2022automatic}.

Variations of simulation models are of interest not only to document how a valid simulation model is finally developed based on a successive refinement \cite{Haack2020} but also in relating simulation models across simulation studies and thereby forming families of simulation models \cite{budde2021relating}. 
Thus, the provenance of simulation models might be partly cast as a \emph{variability management} problem (see Section~\ref{sec:variability}), and thus open challenges approached by adopting methods from this area of software engineering.

\subsubsection{Workflows for Knowledge-intensive Processes.}

Workflows are an approach closely related to capturing provenance. Workflows and workflow systems have a long history, e.g., in databases and business process modeling. Already in the 1980s and increasingly in the 1990s and 2000s, the specific requirements of scientific data management led to the development of scientific workflow systems \cite{ludascher2009scientifica,liew2016scientific}.
Since workflow systems provide a controlled execution environment, they often support capturing provenance information at various levels of granularity \cite{herschel2017survey}. Often, the very fine-grained capturing of provenance requires post-processing of the collected provenance information, e.g., user-specific filtering, 
to provide abstractions on demand, as shown for simulation studies in \cite{ruscheinski2019capturing}.

A downside of workflow systems is that they often treat tasks as black boxes whose semantics are opaque to users. As a result, user-oriented workflow design and meaningful provenance capture are often challenging, as the required information is unavailable. 
Workflows have been applied to specify (see section \ref{sec:background}) and execute individual simulation experiments since workflows enable flexible reuse of repetitive processes. 
However, applying traditional workflows to entire simulation studies is challenging, as model refinement phases are intertwined with model analysis, calibration, and validation activities. To support these knowledge-intensive processes, which are driven by the user's expertise, declarative workflows may offer the required flexibility \cite{vaculin2011declarative}. In the study \cite{ruscheinski2019artifact}, an artifact-based workflow approach is applied, specifying declaratively the life cycle of central artifacts, such as the conceptual model (with a focus on formally defined requirements), the simulation model, and the simulation experiment and their interdependencies, and exploiting inference mechanisms based on the defined constraints to guide and support the user in conducting the simulation study. 

\subsection{Future Research Directions}

Challenges to be faced when equipping simulation studies with context include agreeing on what is needed as context and settling on accessible representations; providing support for storing and collecting context information; managing and maintaining the evolution of context information; and finally, developing methods for exploiting context for (automatically) conducting and interpreting simulation studies. 

\begin{figure}
	\centering
	\includegraphics[width=\textwidth]{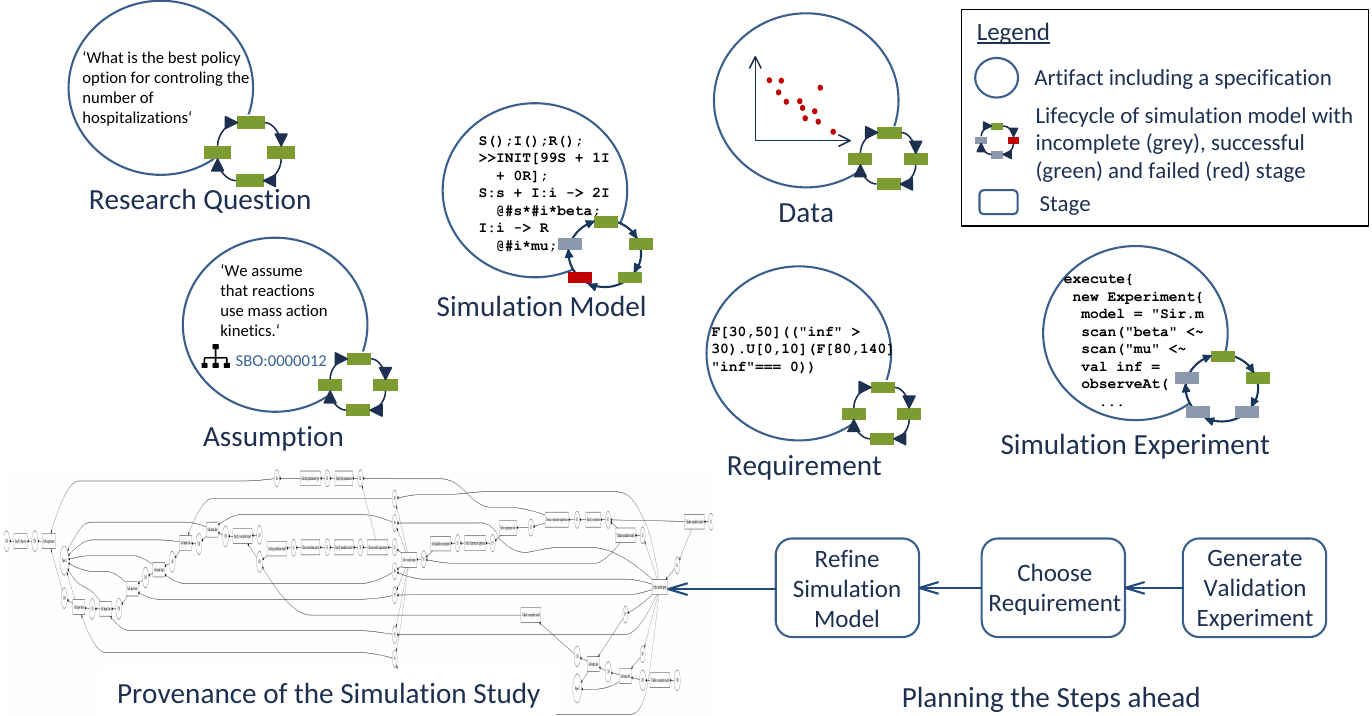}
	\caption{To move forward, the context of simulation studies needs to be explicitly and unambiguously represented. Therefore, the various artifacts play a role (here depicted with their individual life cycle), including conceptual model (research questions, assumptions, requirements, data, etc.), simulation experiments, and simulation models inspired by \cite{ruscheinski2019artifact}. Methods are needed to interrelate those with provenance retrospectively as well as to exploit them prospectively, guiding (based on a workflow-based view) or even automatically generating the next steps (see Section \ref{sec:automation}).}
	\label{fig:context}
\end{figure}

\subsubsection{Standardization of the Nature of Context and How to Represent it.}
\label{sec:standard-context}

As stated, various documentation guidelines exist in different areas of modeling and simulation. Identifying shared and distinctive features would further communication between application domains and insights into the respective simulation practices. 
Whereas research outlets such as journals should encourage policies that each simulation model or study should come with suitable context documentation, these policies must be accompanied by suitable methods to keep the effort manageable.
This question is directly linked to how to represent the context information. Textual documentation of simulation studies is notoriously lengthy. For example, describing an agent-based model following the ODD protocol as encouraged by the JASSS journal results easily in producing 30 pages \cite{klabunde2015agent}. Therefore, 
it is argued \cite{grimm2020odd} that although such documentation is useful as a supplement, means for succinct documentation are needed to be included in the main body of a publication. Adopting provenance standards such as PROV-DM provides a bird's eye view of how sources and (intermediate) products contributed via specific activities to the overall results and might be a first step in this direction \cite{Haack2020}.


In addition, to support exploiting context information, the more unambiguous and computationally accessible the representation is, the better. Therefore, whenever possible, DSLs (Section \ref{sec:background}) should be applied to specify the simulation model, the simulation experiment, and the requirements formally (\cite{Schuetzel2014,zschaler2020family}). However, further ingredients of the documentation might withstand a formal representation. Here, developing domain-specific ontologies will play an important role. For example, to specify assumptions about biochemical models, the systems biology ontology (SBO) could be applied to match a statement about some proteins being ``degraded very slowly, we assume that their concentrations remain constant throughout the time course'' to the concept Concentration Conservation Law (\emph{ID 362}) of the SBO \cite{budde2021relating}. The methods developed in the context of ontology learning offer new opportunities \cite{al2020automatic} that need to be explored to lend ontology developments new momentum (see also section \ref{sec:composition}). 

\subsubsection{Storing and Collecting Context Information}

Various methods are available for storing context information. 
Archives may bundle relevant information about the simulation model, data, and experiments in one place \cite{bergmann2014combine}.   
Also, web interfaces and graph databases as a backend can facilitate the documentation and particularly the retrieval of relevant information about simulation studies \cite{budde2021relating}. For providing a set of suitable and maintained tools, community efforts are required not only to push the standardizations of what does and what does not belong to the context of a simulation study (see above) but also to maintain the required tools and model repositories, and allow them to evolve (see also Section \ref{sec:composition}).

This does not yet address the crucial question: How can we facilitate the collection of context information?
For the textual documentation of context, applying tools, such as Jupyter Notebook, can ease the burden of documenting simulation studies and enrich the documentation with interactive, computational elements \cite{ayllon2021keeping}. However, this documentation still relies on manual efforts. 
Ideally, context information collection should happen automatically and be transparent to the user (see the section on automation \ref{sec:automation}). 
If workflow systems are used, those will automatically generate a corresponding documentation of what has been done, including activities and (sub-)products  \cite{herschel2017survey}. 
Automatically collected information grows typically rapidly, preventing easy access and communication. This requires some means of automatic (user-specific) filtering, interpretation, and abstraction, for which (semi-)formal approaches and encoding heuristics about simulation studies have proved crucial ingredients \cite{ruscheinski2019capturing}. 
Alternatively, instead of workflows, 
approaches that collect provenance information from scripts, such as \cite{pimentel2017noworkflow}, could be combined with means for effectively monitoring the modeler and methods that infer implicit context information, e.g., whether in-between variations of simulation models simulation experiments are being executed. These methods must be made broadly available for the modeling and simulation community, supporting an automatic and systematic collection of crucial context during the simulation study. 

\subsubsection{Context Maintenance and Evolution}
\label{sec:provenance-vm}

With provenance, we focus on how sources and products of the simulation study are related by activities and emphasize a process-oriented view of context. Adopting provenance standards helps to query provenance information beyond individual simulation studies \cite{ruscheinski2017provenance}. However, the question remains about how the different artifacts and their evolution can be described. DSLs are an established approach, particularly in keeping track of the different artifacts. However, they are not tuned to factor out commonality among different versions, which can result in redundancy and pose challenges to maintainability.

Feature description languages \cite{ABKS13,SHT06} were developed to capture commonality and variability in software artifacts. They also have a formal semantics \cite{SHT06}. 
Moreover, model transformation characterized and driven by features is available in the form of the delta-oriented programming paradigm (see \cite{dop} and Section~\ref{sec:variability}).
These deltas can be equipped with formal assertions describing their requirements and effects, so it is possible to specify and verify properties of delta application formally \cite{TSHA12}. This capability could be useful for making implicit assumptions about artifact variations explicit. This would support reasoning about variations of simulation models within an explorative simulation study \cite{davis2000exploratory}. 

Of course, there are substantial differences between 
variability within simulation studies and variability encountered in software products. For example,
the context of simulation studies encompasses a wide variety of artifacts, some of which are informal, such as assumptions or research questions (see section \ref{sec:background}). In addition, the provenance of simulation studies records ``successful'' product variants and, equally, failed attempts \cite{Haack2020}. Perhaps most importantly, provenance encompasses \emph{unplanned} variability as artifacts in simulation studies are open to frequent, often on the fly,  revisions. 
This is in stark contrast to variability as encountered in software design (see Section \ref{sec:variability}), where variability engineering is factored out as much as possible into an early process phase \cite{SWPL05}. These differences create research problems that need to be addressed, but there are potential advantages from a feature-oriented view on provenance management. E.g., features aggregate multiple model changes, which yields a flexible notion of abstraction that is of interest to facilitate communication with different stakeholders (see also Section \ref{sec:communication}). Due to their formal semantics, features open up further possibilities for exploiting provenance information, e.g., for consistency checks within a simulation study or across different ones.

\subsubsection{Exploiting Context Information}

Provenance can be exploited prospectively as well as retrospectively during simulation studies.
Prospectively, context information can guide the modeler to conduct simulation studies that comply with specific standards and thus increase the quality of simulation studies. Traditionally, this interpretation of context is closely related to workflow research. As stated above, workflow support for entire simulation studies is still rare. Developing workflow systems for knowledge-intensive problems, such as simulation studies, requires significant efforts, which exploiting process discovery methods might help reduce \cite{slaats2020declarative}. 

Context information is also invaluable for enlarging the portion of simulation studies that can be automated (a more detailed discussion is in Section \ref{sec:automation}).
In \cite{ruscheinski2019artifact}, context information about requirements is used to generate validation experiments automatically, while in \cite{ruscheinski2022artefact}, convergence tests for a finite element analysis are automatically generated based on a given threshold for the discretization error. In both cases, an artifact-based workflow approach is utilized. In other work \cite{wilsdorf2022automatic}, provenance information enables automatically reusing simulation experiments and adapting them for newly revised and extended models. However, many open challenges remain, particularly when informal context information needs to be mapped into formal, i.e., computationally accessible, specifications, independent of whether those are assumptions, requirements, simulation experiments, or simulation models.

Even the fundamental purpose of context in simulation studies, i.e., facilitating the interpretation of its products, holds major open research questions. To facilitate interpretation, context needs to be presented at the appropriate level of abstraction and fine-tuned to the subject of inquiry (see also section \ref{sec:communication}). The non-formality of many aspects of context and its heterogeneity aggravate the induced challenges. 
In summary, the interplay between the conceptual model employed in the ``third layer'' (scientific assumptions, requirements, etc.), the trace-based view (retrospective provenance), and the workflow-based view (prospective provenance) all need to be considered simultaneously to communicate (see section \ref{sec:communication}) and exploit context effectively (Fig.~\ref{fig:context}).

\section{Composition and Reuse}
\label{sec:composition}

Simulation models are often composed of separate sub-models to cope with the complexity of a system to be modeled. Consequently, a simulation model consists of a set of interacting components, each of which ideally has been designed for reuse. Similarly, other artifacts, such as simulation experiments, can be composed and reused. Generally, a composition-based design facilitates and, in some cases, enables reuse. 
Reuse in the context of modeling and simulation can include everything from ``code scavenging'' to the reuse of model components, up to the reuse of an entire model \cite{pidd2002simulation,robinson2004simulation,petty2019model}. 
Reusing existing simulation artifacts promises to reduce development time and cost \cite{paul2002use,szabo2007syntactic,robinson2004simulation} and helps proliferate knowledge across a wider user community. Simulation model reuse requires methodologies for abstraction, retrieval, selection, integration, and execution \cite{janssen2020code,robinson2004simulation}. 


\subsection{State of the Art}


During the COVID-19 crisis, when simulation models needed to be developed quickly to inform policymaking, the lack of respectively the benefits of reusing model components and composing models became apparent. Popper et al. \cite{popper2020synthetic} describe how they reused a generic agent-based model of the Austrian population developed before the pandemic \cite{bicher2018gepoc} and that large parts of their COVID-19 model could build on independently validated components which the final model benefitted from. As also proposed in \cite{zhu2019reusability}, they exploited the layered architecture of their agent-based modeling and simulation framework. The challenge remains in understanding what is needed to achieve this on larger scales, including methodological approaches and initiatives that can be advanced. 


\subsubsection{Model Composition and Reuse at Different Levels}

The Covid-19 example mentioned above highlights the need for fast model development times, which can be best achieved through the composition and reuse of existing models or components appropriately within the given context. The benefits of composition (and reuse) of software are well known and have also been exploited for the design of modeling and simulation frameworks \cite{phil,osa,devs1,Himmelspach2007} and the reuse, automatic generation, and execution of various simulation experiments \cite{wilsdorf2022automatic}. 
The composition and reuse of simulation models have a different quality, as a simulation model encapsulates a specific relation to the system to be studied, partly reflected in its context that needs consideration if the composed model shall work as intended, i.e., reliably answering the current questions about the system of interest. This is the reason why model composition and reuse have been identified as a central challenge of modeling and simulation \cite{pidd2002simulation,fujimoto2017research}.  

The Level of Conceptual Interoperability Model (LCIM) \cite{tolk2003levels} describes seven levels of model interoperability to characterize which level of interoperability has been ensured in the current reuse (or composition) of models. Interoperability can refer to letting simulators interact (thus, the simulation algorithm and model are treated as a non-separable unit) or composing simulation models. In the latter case, the composed model is executed by a simulation engine. Level 1 refers to technical interoperability. Most formalisms or languages for modeling support some form of composition \cite{Zeigler2000,priami1995stochastic,hillston2005process}. The first interesting level is the level of syntactic interoperability \cite{szabo2007syntactic}. At this level, a more elaborate definition of interfaces and the integration of type systems becomes crucial. Interfaces kept separately from concrete model implementations guarantee that the coupling of components is syntactically correct and supports successive refinement and compatibility analysis \cite{Roehl2008,osa}.

If the ontologies of the application domain are accessed to specify the components' interfaces, we move to the next, the semantic level. According to \cite{wang2009interoperability}, exchanging content is what the semantic level is about. Further up, the pragmatic, dynamic, and conceptual levels are distinguished, requiring increasing levels of information that the components have about each other's context to correctly interpret the meaning (Section \ref{sec:context}). However, most efforts that support composition do so at lower levels. This usually implies ensuring that input and output ports are correctly connected and that data flows in the correct format without an in-depth understanding of the assumptions and constraints of each of the connected components. In addition, as the composition may involve different modeling paradigms (multi-formalism modeling) and even integrate discrete or continuous components, suitable means for model transformations \cite{lara2002atom} or synchronization schemes for simulation are required \cite{Liu2002component}.
Semantic and higher-level composability is significantly more difficult to achieve, demanding knowledge and alignment of model assumptions, constraints, and a common understanding of the simulation context (see section \ref{context}).

\subsubsection{Reporting Guidelines and Formats for Reuse and Composition}


Besides access to the source code of a model, which often requires the use of a specific simulation framework, the successful reuse of a model or some of its components within a given context also requires a comprehensive and explicit description of the model's structure, underlying assumptions, configuration, and other information that might be relevant for potential future users. Likewise, reproducing experiments necessitates information on the data and model that have been used, how the experiments were conducted, and how the outputs were conducted. Common guidelines and formats facilitate the sharing of this information.

Examples of guidelines for describing models and simulation experiments include \emph{MIRIAM} (Minimal Information Requested In the Annotation of biochemical Models) \cite{novere2005minimum} and \emph{MIASE} (Minimum Information About a Simulation Experiment) \cite{waltemath2011minimum}. What characterizes many of these guidelines (also those that aim to capture entire simulation studies, see Section \ref{sec:reporting_guidelines}) is that they were proposed by a community, meaning that a larger group of researchers developed them and signaled their commitment.
There is no direct link between verbal description and the model's components and code.
Still, when applied consistently, documentation guidelines enable a high level of interoperability (and reuse) as they make assumptions, constraints, and simplifications of the model explicit. However, this heavily relies on the author's rigor when describing the model.

The above reporting guidelines result in documents in the form of structured text with little to no formalization. The following formats are aimed at automatic reuse by different simulation tools. 
\emph{CellML} is an XML-based model description language that originated in biology but can be used for different types of mathematical models \cite{lloyd2004cellml}. Its goal is to facilitate the storing and exchange of models and the reuse of model components independent of the software that has been used for model building. The language can be used for describing both the structure of the model, i.e., its components and how they are connected, as well as metadata for the annotation of the model, i.e., purpose, authorship, and references, but also for the reproduction of simulations and the visualization of outputs. A closely related language with a similar purpose is \emph{SMBL}, the Systems Biology Markup Language,  \cite{hucka2003systems}. It is possible to translate SMBL to CellML, and vice versa \cite{smith2014sbml}.
Both formats enforce or at least encourage using annotations and ontologies, e.g., to uniquely identify variables and parameters, which supports semantically meaningful reuse and composition of simulation models \cite{krause2010annotation}. Thereby, they also address computational challenges of model composition   \cite{page1999observations}.  
Standardization efforts, such as SBML and CellML, have been facilitated by the momentum in systems biology in the early 2000s, existing ontologies in the application domain, and, last but not least, the structural similarity of simulation models being developed, i.e., species reaction systems often expressed as ODEs. To support other types of models, the standard core of SBML is extended by specialized packages, e.g., to support spatial models \cite{schaff2023sbml}.      

Likewise, facilitating the exchangeability and reproducibility of simulation experiments requires the specification and description of the experiments using a common interchange format or language. This might include the data and the model that has been used, potential modifications that need to be applied to the model before experimentation, and how the output data should be analyzed. The Simulation Experiment Description Markup Language (SED-ML, see also Section~\ref{sec:dsl-sim}) is an example of a format that can be used to specify simulation setups \cite{waltemath2011reproducible,smith2021simulation}. In practice, it is typically not used by modelers but to export and import simulation experiments between different tools or frameworks. SED-ML offers a semantic annotation of elements using ontologies of the application domain and simulation methods.

Even though markup and exchange formats and documentation guidelines pursue different approaches, they can be connected. SED-ML, for instance, enables encoding information required by the MIASE guidelines. Yet, little work exists on the automatic conversion between exchange formats and documentation guidelines, nor assessing the consistency between both.

\subsubsection{Model Repositories}

Janssen et al. \cite{janssen2020code} discuss different types of model sharing that exist for making model code available, i.e., archives (e.g., open science model libraries), web-based version control repositories (e.g., GitHub and SourceForge), journals, personal or organizational storages (e.g., Dropbox or institutional websites), and distinct framework repositories. 
General code repositories such as GitHub\footnote{\url{https://github.com/}} or SourceForge\footnote{\url{https://sourceforge.net/}} host a large number of models. A simple search of the term `simulation model' on GitHub returned nearly 2,000 hits.
Several challenges exist with general-purpose code repositories. The discovery and selection of existing models are nearly impossible without prior knowledge of a specific model. The discovery relies heavily on appropriate keywords being used within the model metadata, which benefits from ontologies of the application domain and ontologies specific to the methods being used \cite{silver2011ontology}.  
For the selection, additional information becomes crucial.

To address this, purpose-built model repositories exist. 
The \emph{NetLogo Model Library}\footnote{\url{https://ccl.northwestern.edu/netlogo/models/}} contains hundreds of NetLogo models that contain code, instructions on how to run and modify them. 
Inclusion within the library requires the submission to a central database where the model is checked and released for wider community use. \emph{ComSES.Net}, Network for Computational Modeling in Social and Ecological Sciences, is an open community of researchers, educators, and professionals focused on improving the development and reuse of agent-based and computational models to study social and ecological systems. The community develops and maintains the CoMSES Model Library. 
Upon submission, authors can request their model to be peer-reviewed for structural completeness and to fulfill the ComSES community standards. The library contains over 7,500 publications of models, their metadata, and citations. Around 2,500 models are currently stored in the BioModels repository \cite{li2010biomodels}, most of which are specified in SBML with metadata that include references to established ontologies. More than 1000 of these models have been curated. This implies in Biomodels that a model has been independently tested and checked, whether the simulation results stated in the publication could be reproduced based on simulation experiments. Publishing artifacts, such as simulation models, in repositories is increasingly accompanied by some assessment of quality, which requires a separate review. 
This development aligns with an ACM initiative that introduces separate reviewing processes for the artifacts associated with ACM publications \cite{acm}. Five badges can be assigned to the publication after the artifact is reviewed, including \emph{Results Reproduced}, \emph{Artifacts Available}, and \emph{Artifacts Reusable}. 

\begin{figure}[htbp]
	\begin{center}
		\includegraphics[width=0.8\textwidth]{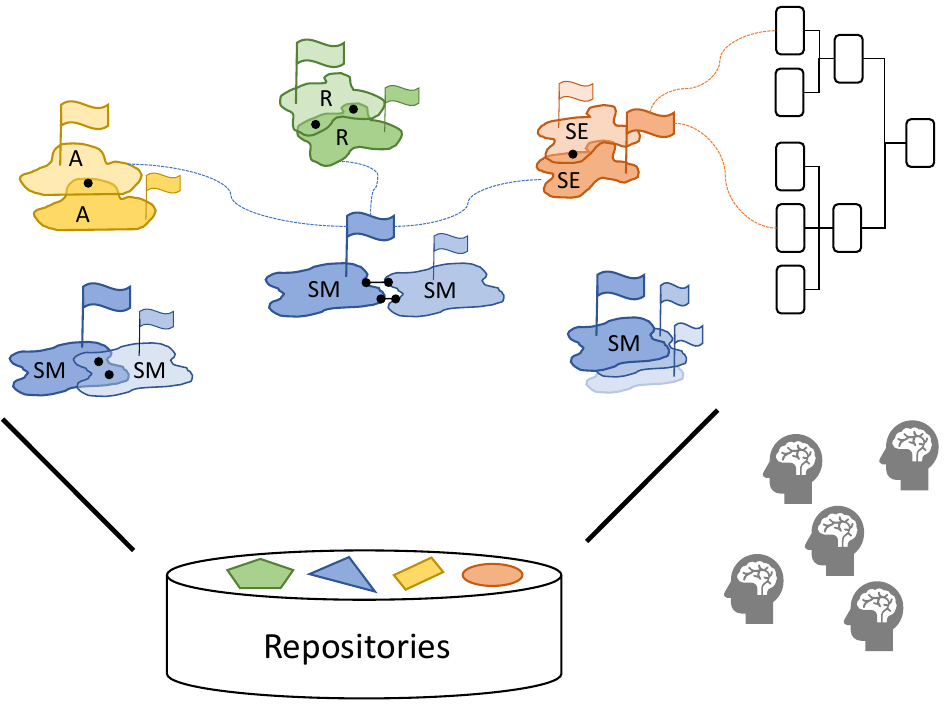}
	\end{center}
	\caption{Community efforts are required to maintain repositories, to develop standards and ontologies - not only for simulation models (SM) but also for other artifacts of simulation studies (Assumptions (A), Simulation experiments (SE), and Requirements (R)), - to enhance composition and reuse. These efforts need to be accompanied by methodological advances that support unambiguous and succinct annotations with suitable meta-information (e.g., a simulation model's context, see section \ref{sec:context}), flexible (e.g., white-box) and powerful (e.g., pragmatic level) composition and analysis methods (e.g., comparing and interpreting variations and automatically testing assumptions and requirements, see section \ref{sec:automation}).   
	}
	\label{fig:reuse}
\end{figure}

\subsection{Future Research Directions}


Model composition and reuse have the potential to enable efficient model development. However, thorough documentation, proper revalidation, sharing platforms, and incentive structures are only some community efforts required to enable reuse. In this chapter, we discuss different existing approaches and methods that facilitate specific aspects of composition and reuse. To systematically address existing shortcomings in model reuse, further advances in community engagement and documentation formats are required, as well as new mechanisms for composing models.   


\subsubsection{Community and User Engagement.}

To be able to reuse simulation models or their components and simulation experiments, they must be made available in the first place. A study on the availability of agent-based models indicates an upward trend regarding the share of publications that make their models available. However, model availability is still generally low (under 20\% in 2018 \cite{janssen2020code}). Nevertheless, the availability of simulation artifacts alone does not promote reuse. 
There is a need for empirical studies focused on practitioners and researchers to understand the process of model reuse better and identify requirements. Numerous empirical studies exist on reusing software components \cite{mili1995reusing, mohagheghi2007quality}. 
Similar empirical studies in the realm of simulation models would allow a more in-depth understanding of the barriers to model reuse. It could guide the design and implementation of solutions. These could include specific DSLs to capture research questions or model assumptions or to motivate the development of application-specific ontologies. 

New methodological developments to further composition or reuse without well-functioning tools broadly supported within the community will not advance model reuse. There is a need for community support in contributing to, testing, and trialing various supporting tools. Drawing again from the software engineering community, where tool demonstrations and competitions are commonplace, there is a need for more tool-focused papers within major conferences and journals.

\subsubsection{Standards and Formalizations.}

As discussed above, there is a need for standardized languages or formats to specify models 
and means to specify their context unambiguously to understand various aspects of a model and how to reuse it. However, for any new language to be successful in its facilitation of reuse, requires 
broad community buy-in. This can be achieved
through joint development of new languages across the community, ensuring that everyone contributes to their continuous development.

For example, in the area of agent-based modeling and simulation, one might build on the existing momentum of ODD, perhaps through extending ODD to various specialized domains, such as building templates for specifying models of agents with decision capabilities based on ODD+D \cite{muller2013describing}. As stated in \cite{grimm2020odd}, there is a need to complement the ODD documentation guidelines with more formal (and succinct) approaches. 
Transformation methods must be designed, allowing for easy translation from existing specifications to new, more formal languages or protocols.

These considerations apply to the documentation or specification of the simulation model and aspects of a simulation model's context (Section \ref{sec:context}). For requirements, different formal approaches mostly based on temporal logic exist (see Section \ref{sec:background}); what is missing are means to facilitate their adoption by a community, be this in terms of languages with a suitable expressiveness, or even a repository of domain-specific requirements, e.g., in the form of stylized facts \cite{RePEc:arx:papers:1812.02726}, that can be reused to check a simulation model's validity automatically within a particular domain.  



\subsubsection{New Mechanisms for Composing Simulation Models}
\label{sec:compositionfwcomposition}

Existing composition mechanisms, such as import/export, inheritance, refinement, delegation, etc., appear insufficient to compose simulation models from given constituents in a semantically valid manner.
One reason is that the composition of simulation models is an intrinsically \textit{parallel} way of composition: two sub-models do not execute independently of each other, but they interfere.
For example, one model might rely on the constant concentration of a certain species (and accordingly, the parameters have been calibrated), and this model shall now be extended by being composed with another sub-model that produces this species. How can these relationships be captured?
One necessary ingredient for the composition of subsystems is a suitable notion of scope that lets one define the \textit{boundary} of a subsystem, including the quantities it may depend on and the ones that it might possibly change.

Exchanging information via output and input events might prove cumbersome or insufficient to model certain situations. For example, to capture upward and downward causation of multi-level systems, the upper levels might directly access the states of models at the lower levels and vice versa, to change their states accordingly
\cite{Steiniger2016}. These interactions form a kind of value coupling, 
i.e., different variables in different sub-components have the same value during simulation 
\cite{elmqvist1997modelica}, so interfaces must be enriched by other interaction means between simulation models.
In many areas, the composition does not happen as a black box composition (via traditional interfaces) but can occur as a fusion \cite{Randhawa2010} or merging \cite{schulz2006sbmlmerge} of simulation models, in which also the internals of simulation models are accessed (Fig. \ref{fig:reuse}). 
\textit{Invariants} describe what does \emph{not} change during the execution of a sub-model or a composition of sub-models. They are a way to define constraints that contribute to a valid composition and fusion of models. If we interpret invariants as properties expressed on simulation results, e.g., in terms of temporal logic, behavioral requirements could be rechecked whether they hold for the composed model as they did for each component, e.g., \cite{PengWHU16}. 

\subsubsection{Representation and Evaluation of Variability} \label{sec:variability}
Hussain et al. \cite{hussain2022approaches} distinguish model reuse, whether individual components are reused, if the model is developed as a composition of existing ones, or if an entire simulation model is reused, and how many adaptations (variations) are introduced. These variations of simulation models can also be observed if various models have been generated over time, reflecting increasing knowledge about a system of interest, its mechanisms and behavior, and different research questions \cite{budde2021relating}. 
In software engineering, systematic management of variability (and commonality) is well-established as a design approach known as \emph{Software Product Lines} (SPL) \cite{SWPL05,ClementsNorthrop01}. Differences between related model variants are represented abstractly as \emph{features}, which may be parameterized. A set of features and parameter values then characterizes one concrete model, called a \emph{product} in SPL terminology.
Variability modeling techniques are fairly agnostic with respect to the underlying implementation paradigm \cite{SetyautamiHaehnle21} and, thus,  might be used in connection with various simulation models. Variability modeling also seems a promising approach to represent and reason about model variants that are stored in model repositories: 
to keep track of different versions that evolve over time, to interpret similarities and differences within models \cite{henkel2018notions}, to record incompatibility between model features or to state specific features as being required, see also Section~\ref{sec:provenance-vm}.
In any case, domain knowledge is needed to define and interpret commonalities and differences.


\section{Automation}
\label{sec:automation}

Automation of the modeling and simulation life cycle promises to increase complex simulation studies' efficiency, quality, and reproducibility.
To this end, 
automation also bears the potential to ``close the loop'' such that model adaptations and new experiments can be iteratively derived from previous results and studies~\cite{waltz2009automating}.
The problem of automating feedback loops is also a central challenge in digital twins. A digital twin needs to mirror the state of its physical counterpart reliably. Hence, based on data obtained by monitoring the physical system, the model must be updated, and new simulation experiments must be executed automatically.

A variety of approaches for automation have been explored in the field of modeling and simulation or may be transferred from related research fields.
However, automation is challenging because most knowledge rests as implicit assumptions in the modeler's or domain expert's mind, and the cognitive processes behind modeling and simulation are poorly understood.

\subsection{State of the Art}
When discussing automation of modeling and simulation studies, the various tasks of a simulation study have to be considered.
These include conceptual modeling, building the simulation model, specifying and executing simulation experiments, data analysis, and visualization.
Alongside these tasks, automation should guarantee the simulation study's reproducibility, reusability, and credibility.

\subsubsection{Conceptual Modeling and Model Building} 
Machine learning approaches can automatically construct conceptual models from verbal descriptions. 
Named entity recognition, association rule learning, link prediction, ontology mapping, and process discovery are just some of the many techniques for rule, text, and graph mining discussed in the context of conceptual modeling~\cite{maass2021pairing}.
So far, for instance, a semi-automatic approach has been developed for generating conceptual model diagrams from verbal narratives about agent-based models based on pattern-based rules and grammar about the concepts and relationships~\cite{Shuttleworth2022narratives}.
The automatic extraction may be supported by knowledge graphs that connect knowledge of an entire domain from diverse sources and allow for semantic querying~\cite{DBpedia}.
The CovidGraph, for instance, interrelates publications, patents, and clinical trials with biomedical ontologies~\cite{Guetebier2022covid}.

For the automatic construction of simulation models, formal transformations between domain-independent, concept-level models (also known as metamodels) and executable models in domain-specific modeling languages were developed~\cite{kapos2019declarative}.
In between the high-level conceptual model and the implementation-level simulation model, various additional layers of abstraction may need to be generated to cater to the needs of the different stakeholders.
Here, techniques from process mining may come into play to produce models with differing complexity~\cite{maneschijn2022methodology}.
In addition to accommodating different views of the simulated problem, there has also been an interest in learning model abstractions to speed up simulations~\cite{CAIROLI2023114169}.

To take a ``shortcut'' from verbal narratives to executable code of simulation models~\cite{jackson2023natural}, there have been first attempts to use Generative Pre-trained Transformer language models for model building (see the GPT family of models~\cite{brown2020language}).
These types of natural language models have the capability to generate and organize semantic concepts~\cite{Hansen2022computation}. 

Another major class of approaches aims to generate simulation models that can accurately capture some (measured) time series data~\cite{DZEROSKI2008360, north2022review}.
This includes discovering the underlying nonlinear differential equations and their parametrizations using symbolic regression, which is based on 
genetic programming principles~\cite{sun2019data}. 
However, symbolic regression is computationally expensive and prone to overfitting. 
To overcome these challenges, sparse regression has been used for identifying nonlinear dynamics (SINDy) \cite{brunton}.
This approach is based on the assumption that only a few terms define the dynamics of a system.
Sparse identification has also been tailored for biochemical reaction networks by introducing a library of candidate components that may be involved in a reaction system \cite{burrage_using_2024}. In contrast to SINDy, this approach considers system components not only individually but also as couplings between them. In addition, the formulated regression problem can be solved by a non-negative least squares algorithm.
Methods for sparse Bayesian inference can additionally provide uncertainty estimates \cite{Jiang}. 
To effectively recommend models that achieve the desired behavior, the automatic retrieval and incorporation of context information from literature was investigated~\cite{Ahmed2022context}.

\subsubsection{Simulation Experiments and Model Execution}
With respect to simulation experiments, various approaches have focused on their unambiguous design, generation, and reuse. 
Consequently, languages for specifying efficient experiment designs based on hypotheses~\cite{lorig2019hypothesis}, logics for checking temporal and spatial properties~\cite{bartocci_specification-based_2018survey, nenzi2018qualitative,NenziBBL22}, and metamodels for making the ingredients of different types of simulation experiments explicit~\cite{wilsdorf2022model} were developed and applied in generating simulation experiments automatically~\cite{wilsdorf2022automatic} (see Section~\ref{sec:background}).
Also, frameworks exist that provide general guidance for the experimentation process, e.g., the SAFE simulation automation framework for experiments guides its users through the initialization of model parameters, the configuration of parallel simulation execution, the processing of output data, and the visualization of the results \cite{perrone2012safe}.

To lend further support, assistance for simulation experiments has been tailored to the specific type of simulation experiment at hand.
In particular, which methods and parametrization to use, e.g., variance-based analysis versus partial rank correlation coefficients~\cite{wilsdorf2021exploiting} in sensitivity analysis or batch mean versus moving window in steady-state estimation~\cite{leye2014composing}, has been addressed.
With such specialized guidance for setting up these analyses and means for executing the experiments automatically, problems regarding the validity and reproducibility of a model can be identified or even avoided.
In the study~\cite{edeling_impact_2021}, for example, the importance of sensitivity and uncertainty
analysis was demonstrated for applying and interpreting a COVID-19 model. 
The model of the pandemic was shown to be highly sensitive with respect to several of the intervention, disease, and geographic parameters: Uncertainty in these input parameters amplified the uncertainty in the model output by 300\ \%.
Providing such information automatically, in addition to the simulation result itself, is crucial for the decision-makers to interpret adequately, e.g., the number of available ICU beds predicted by the model.

Furthermore, generating and executing simulation experiments automatically may support model-building decisions and drive the progress of an entire simulation study.  In the approach of sensitivity-driven simulation
development, e.g., the model is refined or reduced depending on the outcome of sensitivity analysis~\cite{suleimenova2012sensitivity}.
However, clear guidelines for when to conduct which analysis may not exist. 
In addition, the question of which specific method to select cannot easily be answered. 
For example, in the context of optimization, choosing the right method proved difficult as the response surface of the objective function would have to be known a priori~\cite{Villaverde2018benchmarking}.
Gradient-based optimization methods, e.g., assume smoothness of the response surface. However, this is not the case for many simulation optimization problems.
To deal with non-smooth response surfaces, novel approaches for automatic differentiation over discontinuous functions with smooth interpretation can be employed \cite{kreikemeyer2023}.

There is a growing pool of machine learning approaches with the goal of choosing which method to apply to solve a problem in an automated way.
Work in this area includes an automated selection of methods and hyperparameters in integer programming solvers, e.g., via optimized algorithm portfolios~\cite{Koenig2022speeding}, synthetic problem solvers for composing algorithms for various subtasks of simulation experiments~\cite{leye2014composing}, premise and strategy selection in automated theorem proving~\cite{Irving2016deepmath}, and automated selection of the architectures and hyperparameters in deep neural networks~\cite{Zela2018towards}.

Similarly, approaches for adaptively selecting the most efficient simulation algorithm~\cite{Helms2015}, adaptive methods for an evenly distributed Pareto set in multiobjective optimization \cite{Hunter2019introduction}, or adaptive parallelization of simulations in heterogeneous hardware environments~\cite{xiao2020openablext} have been investigated.
When dealing with limited resources, approaches for test prioritization could filter out the experiments or simulation runs that are most critical, e.g., in testing the validity of a simulation model~\cite{Noor2015similarity}.
In addition, a metamodel-based approach has been applied to automate the implementation and deployment of distributed simulations in High-Level Architecture (HLA) compliant cloud services~\cite{Bocciarelli2013saas}.

\subsubsection{Data Analysis and Visualization}

Another research target is the automatic analysis, interpretation, and visualization of simulation data and data used as input for calibration or validation.
Various supervised and unsupervised machine learning methods can be combined in a knowledge discovery process for simulation data~\cite{feldkamp2020knowledge}.
These, in addition to advances in clustering and classification of time series data~\cite{Ali2019clustering}, as well as detection of oscillations~\cite{Dambros2019oscillation} or outliers~\cite{Blazquez-Garcia2022}, will be crucial for providing automatic support in the data analysis, visualization, and interpretation phase, and to go beyond manual validation.
With the increasing push for sustainability and efficiency of computing, specifically in simulation studies (``green simulations''), parts of simulation outputs may be stored and reused for answering new questions~\cite{feng2017green}.

\subsubsection{Reproducibility, Reusability, and Credibility}

Ensuring the reproducibility, reusability, and credibility of models and associated artifacts are ongoing challenges (see Sections \ref{sec:context} and \ref{sec:composition}). Accordingly, approaches for recording provenance traces of entire simulation studies in a non-intrusive manner and presenting aggregated views on provenance are of high relevance~\cite{blount2021observed, ruscheinski2019capturing}.
To capture provenance and, therefore, to document a simulation study automatically, approaches such as system wrappers, application reporting, operating system observation, or log file parsing~\cite{Allen2010provenance} have been explored.
They allow observing modelers in their usual working space, e.g., specific IDEs (integrated development environments), consoles, or libraries.

To correctly capture the provenance traces and to understand their meaning, methods for detecting, interpreting, and visualizing the differences between model versions are required~\cite{gebhardt2022exploring}.
The variability in model versions may, e.g., refer to different parameter settings, level of detail, the goals and hypotheses addressed, and the choice of modeling formalism.
Approaches to managing the variability of models (particularly over time) are discussed in Section~\ref{sec:variability}.

Provenance traces of previous simulation studies may be used to construct workflow models that can be applied in future simulation studies to execute suitable next actions automatically.
Here, process mining techniques may be employed to generate data science workflows from code~\cite{berti2019process} automatically and to provide case-based support~\cite{leake2008}.

\subsection{Future Research Directions}
Open challenges in the automation of simulation studies include the necessity for semantically annotated knowledge as a prerequisite for automation, the effective application of diverse machine learning methods, the imposition of constraints on automation (particularly human involvement), and the demonstration of the benefits of newly developed automation methods for modelers and other stakeholders in simulation studies.

\subsubsection{Semantically Annotated Knowledge}

Automating simulation studies requires explicit and formally specified or annotated knowledge about all modeling and simulation life cycle phases.
This includes knowledge about the goals and intentions behind the activities of the modeler (e.g., calibration, validation, or prediction), specific hypotheses regarding the model behavior under certain circumstances, and knowledge of established methodologies of a domain.

When applying machine learning approaches for automation, knowledge is required, e.g., for annotating the training, test, and validation data, for selecting relevant features, and for interpreting the results.
Also, when applying rule-based inference systems, explicit and semantically unambiguous knowledge is the foundation for automatic reasoning about simulation studies.

The approaches presented in Section~\ref{sec:background}, including model-based approaches, ontologies, DSLs, and temporal logics, will be essential for unambiguously and machine-accessibly representing the various knowledge.
In addition, provenance graphs and other documentation standards discussed in Section~\ref{sec:context} will be crucial, as they contain valuable information about how the different artifacts of a simulation study are related. 
Moreover, open-source software and open repositories need to be facilitated to enable automatic retrieval and exploitation of the explicitly specified knowledge. 

The overall body of knowledge about modeling and simulation studies will be growing by leveraging the various formal and open methods.
This knowledge and the methods for automation may be shared and exploited within but also across application domains and simulation approaches (e.g., finite element analysis), as simulation studies often share important characteristics~\cite{wilsdorf2022model}.

When reusing information within and across domains and approaches, beyond formally specifying the information needed, there is the additional challenge of dealing with the different terminologies used.
Without controlled, structured vocabularies (ontologies) that provide clear semantics to the expressions, more general support will remain elusive.

Efforts towards collecting and consolidating that knowledge need to be truly community-driven.
This includes regular opportunities for consultation and exchange via simulation working groups and the collaborative development of tools via hackathons. 
To facilitate these processes, they may be integrated with existing (domain-specific) organizations, such as COmputational Modeling in BIology NEtwork (COMBINE) or the Open Modeling Foundation (OMF).

\begin{figure}
\centering
\includegraphics[width=\textwidth]{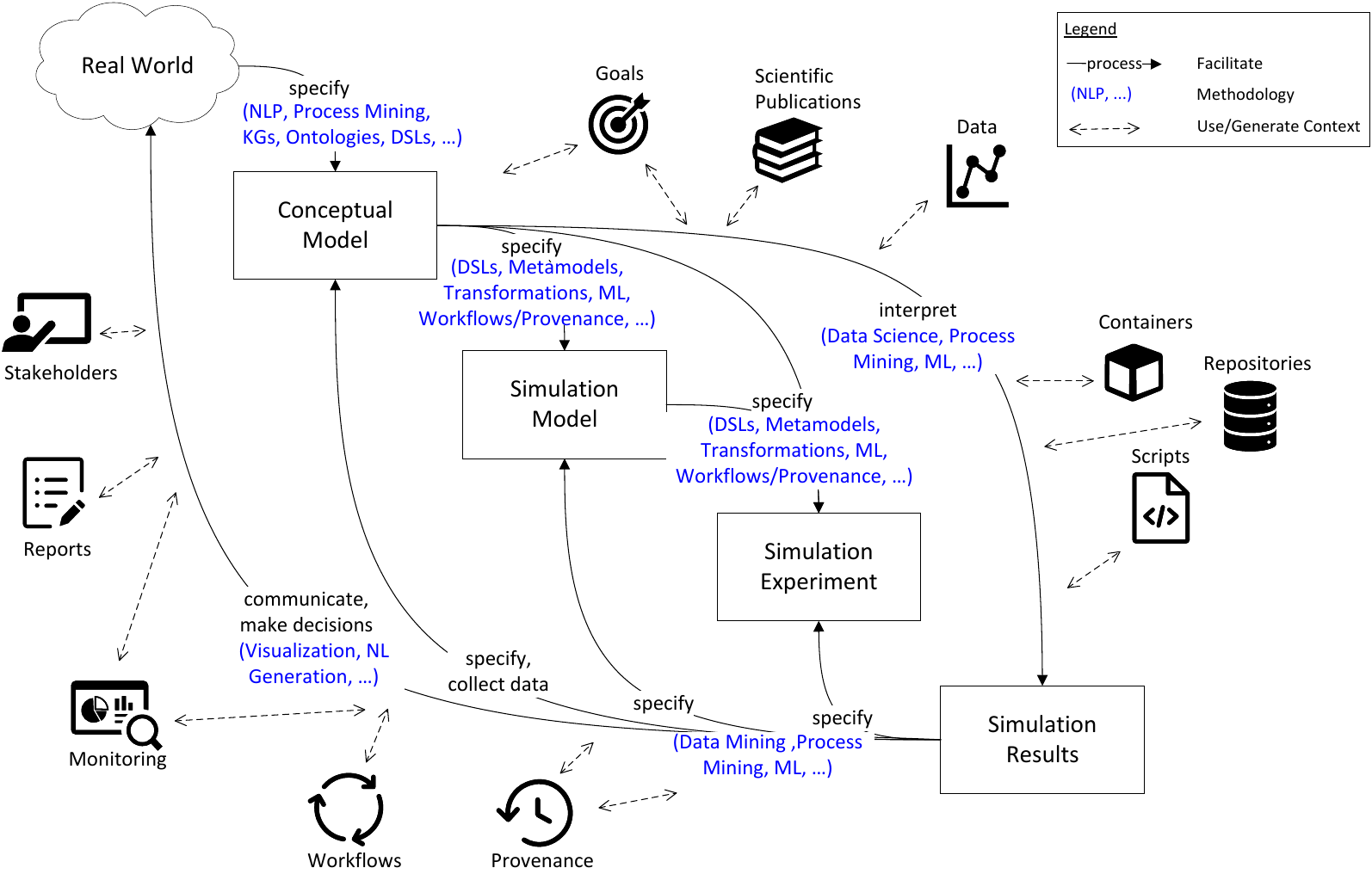}
\caption{Enriching the different tasks of the modeling and simulation life cycle (adapted from \cite{balci2012life}) with intelligent methods to enhance the automation in generating the conceptual model, the simulation model, and simulation experiments, as well as in interpreting and communicating the simulation results.}
\label{fig:automation}
\end{figure}

\subsubsection{Intelligent Modeling and Simulation Life Cycle}

So far, modeling and simulation studies automation has merely targeted individual steps of the modeling and simulation (M\&S) life cycle, e.g., conceptual modeling, model building, experiment specification, experiment execution, output analysis, etc.
Future efforts thus should not only be directed toward improving the automation of these steps but also toward combining and integrating existing and newly developed approaches to support modeling and simulation studies as a whole. 
Fig.~\ref{fig:automation} shows the tasks of the M\&S life cycle and lists methodologies used for enhancing the degree of automation in these tasks. In addition, the various sources that provide context information for automation are depicted.
The developed approaches, frameworks, and tools combined should be able to make (semi-)automatic, intelligent decisions during the modeling and simulation life cycle.

Questions that are still not answered automatically include, for example, when and how to refine a model further or reuse and compose existing model components (see Section~\ref{sec:composition}) or how to generate a model from scratch when no data is available for model fitting. 
Not only simulation models but also simulation experiments, requirements, and even assumptions may be automatically generated depending on the context provided.
In addition, automation may assist in deciding which type of simulation experiment to conduct using what method and parameterization or when to collect more data and what kind.
These decisions might be based on state-of-the-art machine learning methods.
A considerable challenge in this regard is acquiring data for training, testing, and validating the learning models.
To be suitable input data for machine learning, existing simulation studies need to be semantically annotated with context.
Ontologies may, e.g., provide context information about the type and role of simulation experiments, as well as the methods applied.

However, this manual annotation and data collection entails a substantial workload overhead for the modelers, and the potential payoff in terms of development time may not sufficiently justify these efforts.
One solution to circumvent the manual effort could be the generation of synthetic data (i.e., synthetic
simulation studies) to train machine learning models.
Either way, the challenge of dealing with incorrect, inconsistent, or incomplete knowledge when building learning models needs to be addressed.

\subsubsection{Large Language Models for Simulation}

Following the release of ChatGPT on 30 November 2022, new developments in natural language generation have changed how we think and work, including in science and engineering.
ChatGPT and other large language models (LLMs) are based on transformer architectures and pre-trained on massive data sets, which may further be fine-tuned to specific applications~\cite{LIN2022111}.

The use of LLMs to support simulation development and to automate the entire M\&S life cycle is becoming a topic of growing interest. 
For instance,~\cite{giabbanelli2023gptbased} identified four main M\&S tasks where text generation via LLMs may be effectively applied.
The first one is explaining the simulation models' narrative to be understandable for all stakeholders, which will also facilitate participatory modeling (see section \ref{sec:communication}).
The second one is focused on summarizing simulation outputs and conveying the main differences between what-if scenarios.
The third task is generating textual reports to aid the interpretation of simulation visualizations.
Lastly, LLMs may assist in finding and explaining simulation errors and offering guidance to resolve them, thereby assisting in verifying and validating simulation models.

Recently, LLMs have also been trained to generate code in various programming languages~\cite{Vaithilingam2022}.
For simulation models, the automatic generation of fully functional and executable simulation models has been discussed~\cite{jackson2023natural}.
However, generating simulation experiments, requirements, or assumptions in domain-specific languages also needs to be investigated.
In software engineering, there has been a trend towards no-code and low-code development for several years, intending to minimize the amount of manual coding required~\cite{woo2020rise}.
Certainly, LLM-powered tools will soon also drastically change how we develop and analyze simulation models and implement modeling and simulation tools.

\subsubsection{Human in the Loop}

Most simulations are designed to be interpreted by humans.
Keeping humans in the loop during the modeling and simulation life cycle via approaches like visual modeling or interactive, exploratory analysis of results is, therefore, one of the primary goals of providing automatic support for simulation studies.
Different levels of information need to be conveyed to the decision makers, domain experts, and modelers (see also Section~\ref{sec:communication}).
Even if substantial parts of a simulation study are conducted automatically, humans should retain control over the intelligent M\&S process. This includes establishing trust in the automation, e.g., based on methods for explainable AI.
Furthermore, parts or sequences of the M\&S life cycle may require user-specific or project-specific workflows instead of a one-fits-all approach.
Thus, approaches for tailoring the M\&S workflow, e.g., via preference learning, will be more than welcome as long as some ground rules, such as ``do not use the same data for calibration and validation of the model", are obeyed.
These approaches should ideally learn ``as you go'', with as few customizations by the user as possible.
However, a compromise will have to be made that trades off the user overhead of manually entering additional information against increased support 
when conducting a simulation study.

\subsubsection{Evaluation}
\label{sec:automation:evaluation}

In modeling and simulation research, new software is typically evaluated with respect to its runtime performance and the number of steps required using benchmark models~\cite{JESCHKE20112562}.
Moreover, illustrative case studies from diverse application areas can be used for demonstrating new methods and tools, such as in~\cite{Kleijnen1995sensitivity}.
Nevertheless, there is still a need for suitable benchmarks and measures for assessing the gain in productivity and the reduction of error in simulation studies by automatic, intelligent support.
In particular, there is a need for both quantitative and qualitative metrics for how efficient (with respect to time and other resources), effective (in terms of results and information gain), and how accurate (without technical or methodological errors) the automation is.
Realistic case studies with representative user groups are required to evaluate the superiority of automatic support compared to the fully manual or randomized case.
Depending on what task is evaluated (e.g., automatic model generation, algorithm selection, or output interpretation), different measures and different study designs may be required.

For users, including modelers and stakeholders, well-designed studies are crucial in providing the necessary argument for the widespread adoption of the developed automation methods and tools in their daily practice.
For researchers, on the other hand, a thorough---and if possible quantitative---evaluation will improve 
understanding of the developed method and their impact
and guide future research directions.
In the field of visual computing, quantitative methods, and user studies became a separate research field, which is reflected in the numerous projects and research centers, as well as recommendations on the topic (e.g.,~\cite{SFB-TRR161, Bylinskii2022towards}). There, publishers increasingly demanding explicit reporting of user studies (e.g., the journal ``Visual Computing for Industry, Biomedicine, and Art'' expects documentation via the STROBE guidelines for observational studies~\cite{vonElm2007strengthening}).
Similar initiatives will have to be pursued in modeling and simulation.

\section{Communication and Stakeholder Understanding}
\label{sec:communication}

Effective communication with stakeholders (decision-makers and others with a stake in the outcome of the simulation study) is a prerequisite for successfully influencing decisions with simulation \cite{sadowski1999tips}. Effective communication includes working with stakeholders to understand their beliefs and preconceptions, the underlying goals of the simulation study, and the broader organizational context in which decisions based on simulation will be made \cite{sadowski1999tips,sturrock2020tested}. Effective communication also includes explaining simulation study conclusions in clear language tailored toward stakeholders' viewpoints. 
Although the simulation analyst may be confident in the simulation's accuracy because of deep knowledge of the simulation implementation, stakeholders often lack this basis for confidence. The simulation analyst may try to convey this knowledge by explaining how the simulation works, but this faces obstacles.
First, many decision makers (e.g., public officials and leaders in industry) lack training in computer-based simulation and other quantitative prediction methods, which makes it difficult for them to judge the accuracy of a simulation based on detailed technical explanations. Second, decision-makers tend to be busy, which makes even the technically trained among them unwilling to invest the time needed to parse detailed explanations.
Third, simulation models are unable to represent every detail of the real world and often have input parameters that are hard to estimate accurately. Thus, even a complete understanding of how a simulator works may be insufficient to give high confidence in the accuracy of its predictions.

Other forms of communication and analysis, better tailored to the stakeholders and their situation, may be required to build a basis for confidence in a simulation's accuracy and to inform a decision properly. For example, imagine an agent-based simulation model that predicts the effect of public health interventions (e.g., COVID-19 vaccinations) on health outcomes.  The simulation might predict that allocating \emph{fewer} vaccines to older individuals and more toward younger ones \emph{decreases} mortality. Stakeholders may find this counter-intuitive because they think (correctly) that older individuals have a higher mortality risk when infected with SARS-CoV-2. A simple, clear explanation would be, ``Young people are more social, and vaccinating them prevents the fast spread of the virus, which indirectly reduces infections in older people.''.

\subsection{State of the Art}

Successful simulation analysts must communicate directly and extensively with stakeholders to understand their viewpoints and explain simulation outputs. While partial automated support exists for these tasks (described below, including visualization, utility, and prior elicitation methods for learning stakeholder goals and beliefs), many communication tasks are not sufficiently supported by current computer science methods.

\subsubsection{Best Practices from Simulation Practitioners}

Experienced simulation practitioners have written about their experience using simulations to inform stakeholders. 
The studies \cite{sadowski1999tips,sturrock2015tutorial} argue for the importance of understanding stakeholders: How they will potentially use simulation results, how they define success, their background, and the broader power structure and organizational context in which they operate.  Both also point to the danger of being asked to perform a simulation study that ``justifies'' the correctness of a decision in hindsight.  This phenomenon also arises using evidence in public health policy 
\cite{baldwin2004simulation}.
The paper \cite{sadowski1999tips} argues for the importance of delivering results in a timely manner that aligns with deadlines when decisions must be made. 
Sturrock \cite{sturrock2015tutorial} points out that stakeholders, who often have substantial domain expertise, can be valuable partners in validating a simulation model.
He also advises simulation analysts presenting results to avoid excessive detail, to avoid overemphasizing the accuracy of their output data, and to contextualize information from the simulation by explaining how it relates to stakeholders' needs. 
In the context of healthcare, \cite{baldwin2004simulation} argues for the value of an iterative approach where simulation analysts closely collaborate with stakeholders to build a simulation model, reminiscent of co-design approaches to public policy \citep{blomkamp2018promise}. This helps create a shared understanding among stakeholders of how the simulation works and the reasoning behind its design. 
This is highly related to the concept of Participatory Modelling (see Subsection \ref{sec:participatory_modelling} below).

\subsubsection{Communicating about Scientific Evidence with Policymakers}

A line of research reviewed in  \citep{orton2011use,smith2015black, oliver2014systematic} studies how scientific evidence and models influence policy decisions, with much of the literature focusing on public health.
They find a significant gap between policy decisions and scientific evidence that could support these decisions. 
Indeed, \cite{smith2015black} goes as far as to argue that the primary value of evidence tools to public health policymakers is actually not that it often leads to better decisions but that its use can signal to others that the policy maker is making ``good'' decisions.

Factors that support the use of evidence for influencing policy decisions include
decision makers' perceptions of evidence quality \citep{orton2011use},
a culture of using evidence to make decisions \citep{orton2011use},
timeliness and relevance of the evidence \citep{oliver2014systematic},
the need to account for the practical context in which policy decisions are made \citep{orton2011use}, and
the strength of the relationship and level of collaboration between policymakers and researchers \citep{oliver2014systematic}.
While much of this literature focuses on the influence of evidence reviews, some work \citep{TaylorRobinson2008} specifically considers the influence of quantitative models.
This work points again to the importance of decision-makers perceptions of evidence quality,
which can be supported by peer review, transparency, and the value of user support for models.

Science Communication \cite{burns2003science, kappel2019science} is a related endeavor that includes methods for creating scientific awareness, understanding, literacy, and culture among stakeholders, decision-makers, and the general public. For more than 50 years, Science Communication has been seen as a research field with tools and techniques mostly drawn from social and behavioral sciences \cite{Guenther&Jouberg2017}. As many challenges are shared, relevant principles and techniques can be borrowed to communicate simulation studies to decision-makers and stakeholders. A good example is research towards tools that summarize scientific articles in understandable language (e.g., \cite{Guo+2021} for biomedical research articles).

\subsubsection{Learning Stakeholder Preferences and Beliefs}
\label{sec:stakeholder}

As argued above, the successful use of simulation to support decisions requires the simulation analyst to know how the simulation results will be used to support decision making \cite{sadowski1999tips,sturrock2020tested}. 
The most common approach for learning about decision criteria is to talk with a decision-maker. 
However, with increasing size and complexity of models, detailed walk-troughs become infeasible with respect to timeliness with which decisions need to be taken, as for example argued by \cite{Davis2016} for defense applications.
Similar experiences are stated by \cite{GEMS2020} even for cost-effective and innovative ways to test new ideas or prototypes. The report discusses decision-makers need rather information from quick-term analyses while experts are not equipped to provide those.

Different approaches have been suggested to systematically determine the relevant decision criteria to be taken into account when developing simulators to support decision-makers:    
Multi-attribute utility theory models how humans make decisions when multiple outcomes matter.
Utility elicitation and preference learning methods \cite{chen2004survey} estimate such utility functions from stakeholder feedback.
Closely related to utility elicitation methods, prior elicitation methods \cite{best2020prior} estimate these probability distributions from decision-maker feedback.

There is a large and closely related literature on multi-criteria decision-making and multi-attribute utility theory \cite{wallenius2008multiple}.
In this literature, the focus is on supporting one or a group of decision-makers in coming to a decision. In the context of a simulation study, this includes lowering the cognitive effort required to identify the most preferred option among those that have been simulated.
Much of this literature can be understood as helping decision-makers explore a Pareto frontier -- the set of non-dominated outcome vectors -- with high-dimensional outcome vectors.

In conjunction with multi-criteria decision-making, there is literature on multi-objective optimization combining simulation and multi-objective Bayesian optimization. Here, there are two lines of literature: one 
estimates the Pareto frontier without modeling the human decision-maker \cite{knowles2006parego,daulton2021parallel}, and another 
models the decision-makers utility function to help focus simulation effort on the parts of the Pareto frontier that are most important to a final decision \cite{astudillo2020multi,lin2022preference}.
There is also a line of literature on preferential Bayesian optimization \cite{gonzalez2017preferential} 
which works directly with pairwise comparison feedback from a decision maker over potential decisions.
Algorithms developed seek to minimize the number of pairwise comparisons needed to help the decision-maker find a good decision.

Existing automated methods for understanding stakeholder preferences and beliefs have several shortcomings
that could be improved via future work in the context of simulation studies.
First, they assume access to stakeholders is sufficient to support collecting time-intensive pairwise comparisons.
This may be unrealistic for some stakeholders, especially prominent politicians or business leaders.
Second, they assume that decision-makers believe the simulation results; this may not be the case also, due to a lack of trust 
\cite{Harper+2021trust}. 
Third, they assume that the criteria used to make a decision are fully captured by the presented predicted outputs.  
Stakeholder consideration over non-included criteria may be missed.

\subsubsection{Visualization}



Model visualization provides access to the actual model with visual representations and presentations of the (static) model itself, i.e., its structure and logic.  
These are often based on conceptual diagrams \cite{Vernon-Bido+2015}, such as causal loop diagrams and stock/flow diagrams in System Dynamics \cite{Lane2008}, and UML diagrams in Agent-Based Simulation \cite{Siebers+Klugl2017}.
When using formalisms and DSLs for modeling (see Section \ref{sec:dsl}), a higher abstraction level can be automatically derived and, possibly, graphically represented, e.g., focusing on the network structure as in Petri nets or reaction-based models, or on the model hierarchy as in DEVS.
Hierarchical, modular, composite modeling approaches (see Section \ref{sec:composition}) inherently provide abstraction levels that allow one to zoom in and out on demand. For visually exploring these models, various methods of graph \cite{tominski2009cgv} or tree visualizations \cite{schulz2011treevis} are available.

Simulation-run visualization is essential in revealing insights into the model's dynamics. 
There are basically two ways: 
model state and simulation output can be animated during a simulation run, and output data can be aggregated, analyzed, and presented after (first) simulation runs are finished. St-Aubin et al.~\cite{St-Aubin+2023} survey visualization support of 50 simulation platforms -- mostly for discrete event and agent-based simulation. Hereby, they not only distinguish between visualization of graphs, 2D, 3D, but also survey logging methods and integration of analytics.
Motivated by the large amount of data generated by simulation,
increasingly advanced methods for accessing and visualizing these data have been developed and successfully applied, e.g., \cite{Unger&Schumann2009,Eichner+2014,Feldkamp+2020}. This interactive visualization, integrated with data mining algorithms, happens under the umbrella of exploratory data analysis and visual analytics.

Visualization of simulation data has often been restricted to analyzing the generated data. However, this perspective foregoes the particular chances that the combination of visualization and simulation offers, i.e., to integrate visualization more deeply into the data-generating process of simulation experiments.  In the paper \cite{Kresimir2018}, important concepts for effective integration of visual analytics and conducting simulation experiments are presented, e.g., for optimization and simulation experiment automation \cite{Kresimir2022a}.
By more deeply integrating visualization into modeling and simulation studies, visualization becomes an additional tool, not merely in conducting simulation experiments with models, but also in the model development cycle, as in \cite{andrienko2018viewing}.

The conceptualization and implementation of these visual methods imply significant effort and knowledge in the area of visualization. Therefore, although important for the acceptance and use of a simulation tool, more elaborate visualizations and GUIs are mostly found in commercial rather than academic research simulation tools \cite{St-Aubin+2023}. 
To make matters worse, problem- and stakeholder-specific visualization solutions are required to communicate simulation results effectively. 
Even when similar systems are simulated, different visual analytics solutions are needed for different purposes: For example, to inspect the spreading of COVID-19 infections due to contacts between individuals along with associated metadata \cite{sondag2022visual} or to support public health officials in planning and ensuring the availability of resources (such as hospital beds) under different spreading scenarios \cite{afzal2020visual}.


\subsubsection{Participatory Modelling}
\label{sec:participatory_modelling}

Directly involving stakeholders and decision makers not merely in formulating requirements and using the results but in developing and testing the simulation model facilitates communication and increases trust in simulation results. This type of collaboration can already be found in early system dynamics models, also under the term ``group model building'', e.g., \cite{Vennix+1999}. Using principles and techniques from participatory research, stakeholders take over an active role, e.g., by contributing to Joint Application Design Workshops, by exploring prototypes, or through participating in user panels (for a more technical view, see \cite{Ramanath&Gilbert2004}, a survey of methods in \cite{Voinov+2018}). Participatory approaches  (\cite{Barreteau+2017} gives an overview) were successfully applied in decision-making contexts where communication with diverse stakeholders is essential, such as environmental management \cite{Barnaud+2008,Sahraoui+2021,Bruggen+2019}. This was systematized in the so-called companion modeling approach \cite{Barreteau+2003}. Will et al.~\cite{Will+2021} view the exchange frequency between modelers and stakeholders as the most critical aspect for models supporting decision-making in socio-environmental scenarios. 

\begin{figure}
\centering
\includegraphics[width=0.85\textwidth]{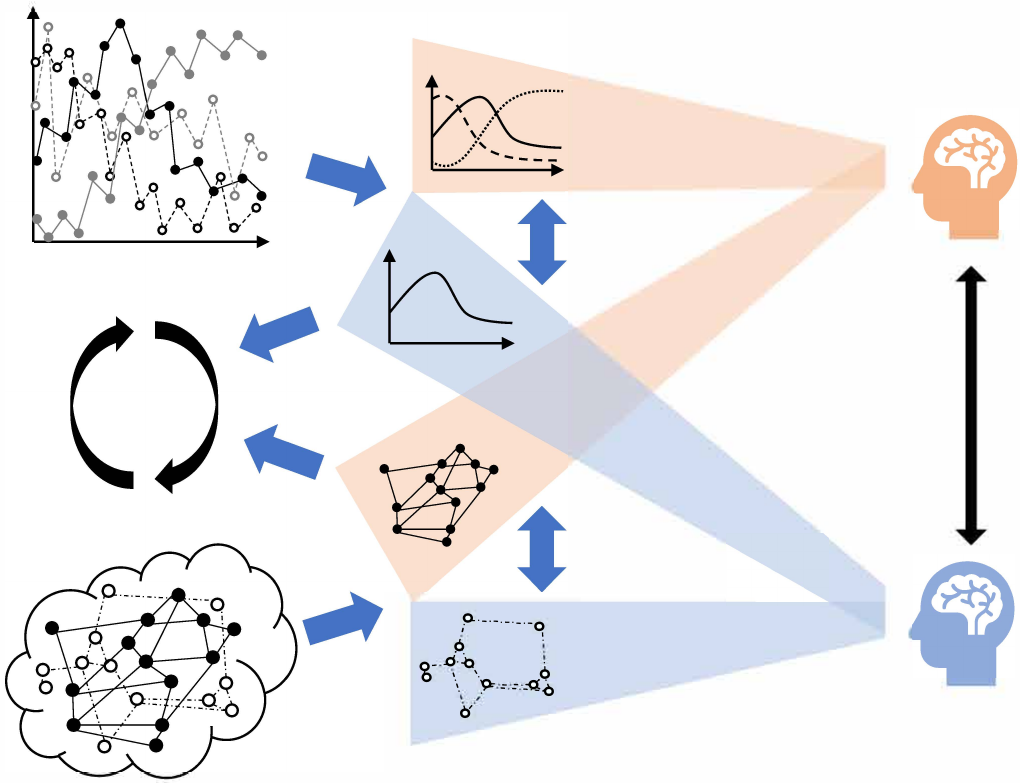}
\caption{Taking the stakeholder into the loop and the stakeholder's specific view into account: generating stakeholder-specific abstractions and explanations about both model and simulation results automatically. Different stakeholders have different needs, so tuning interactions and participatory opportunities is essential.}
\label{fig:communication}
\end{figure}

\subsection{Future Research Directions}

Understanding the audience is critical for simulation success in advice from simulation practitioners \citep{sadowski1999tips,sturrock2015tutorial}. Therefore, more efforts need to be invested into better understanding stakeholders, using the toolbox of social and psychological empirical research, including quantitative models that build on game theory and Bayesian decision theories. 
The insights gained about the background, preferences, and expectations of stakeholders will allow for improving communication with stakeholders (and other users) by revealing and exploiting differences in perspectives, automatically generating stakeholder-specific explanations of the results, and offering interaction possibilities with the simulation study and (ideally) with the modeler in a participatory approach, tuned to the constraints of 
the stakeholder, all of which will increase understanding of simulation studies and results (Figure \ref{fig:communication}).  



\subsubsection{Revealing and Exploiting Differences in Perspectives}

Generally, we can not expect stakeholders' backgrounds, preferences, and expectations to align with the context of the simulation study and its results. In addition, temporal constraints will aggravate the problem. 
Consider, for instance, informing a busy stakeholder about a simulator's prediction $f(x)$ for an outcome across a range of inputs $x$.
The stakeholder has their own estimate of $g(x)$ for this outcome based on their domain expertise. They are also interested in the outcome that varies with $x$ and depends on how they plan to use the simulator's predictions.
The number of inputs $x$ is large, making it prohibitively time-consuming to tell the stakeholder $f(x)$ for every $x$. Instead, we would like to prioritize communicating about those $x$s with a large difference between the simulator's prediction $f(x)$ and the stakeholder's estimate $g(x)$, and where the stakeholder has a high level of interest. A major challenge in doing this is that the simulation analyst may not know the stakeholder's estimate $g(x)$ or their level of interest. Moreover, methods that learn about the stakeholder's $g(x)$ by directly asking them to provide estimates for some collection of $x$ are limited in the number of $x$. 

When human simulation analysts take on this task, they leverage a mechanistic model of how the stakeholder thinks to learn the ``simulation model inside their head''.
They then show simulation results where they differ from the estimated mechanistic model and explain the difference. This is a lot of work for the simulation analyst.
Within this context, we see several opportunities for future work:
\begin{itemize}
\item Automated methods for processing stakeholder speech or conducting and analyzing short interviews to derive the stakeholder's mental model and $g(x)$.
\item Methods for designing simulation experiments to identify $x$ where $f(x)$ and $g(x)$ are very different, given an estimate of $g(x)$.
\item Interactive visualization methods for helping stakeholders explore $f(x)$, and for helping the stakeholder understand how $f(x)$ and $g(x)$ are different.
\end{itemize}

These new approaches must use stakeholder's and decision-maker's time as efficiently as possible. Here, automation has the advantage that elicitation results are formalized and may support automation in later stages. In addition, assumptions directing the elicitation process can be made explicit and used in communication.

\subsubsection{Generating Explanations Automatically}

To help build stakeholder confidence in simulation results, successful simulation analysts must sometimes explain 
why a simulator outputs a particular value. 
A challenge is that many simulators are complex, take high-dimensional inputs, produce high-dimensional outputs, and encode complex processes. 
This can make literal explanations, or in-detail walk-throughs, for why a simulator's prediction is reasonable, too complex to be useful, especially for non-technical stakeholders.
In such situations, it can be useful to provide a largely qualitative explanation that is only approximately correct.
For example:
``The simulated number of infections decreases when masking is mandated because masks reduce the chance that an infected person infects others.'' or
``Simulated hospitalizations decrease by 30\% when masking is mandated, while simulated infections decrease by only 10\%.
Simulated infections are largely driven by younger people who are infected at social gatherings and have low compliance with masking.
Because younger people are less likely to develop severe symptoms, these infections do not contribute significantly to hospitalizations. Simulated hospitalizations are largely driven by older and more vulnerable individuals, who are more likely to comply with mask mandates.''   
Automated generation of explanations makes the stakeholders more independent from the simulation analyst, taking potentially preconceived expectations out of the analysis. We would expect that automatically creating explanations 
increases the trust of stakeholders in the simulation output \cite{Harper+2021trust}; the ability of the simulation analyst to generate explanations is augmented, not replaced. 
In this context, the potential of large language models, such as chatGPT, could also be explored to generate natural language explanations for certain aspects of a simulation study and to improve understandability. 

One productive avenue toward generating explanations 
is first to observe that each explanation above is essentially a causal graph approximating the simulation model. Causal loop diagrams, one form of causal graphs, are an established qualitative abstraction of simulation models used for conceptualization  \cite{bouchet2022using}. 
Alternative causal representations might represent not only variables as vertices but, for example, may correspond to an aggregation of microscopic entities within the simulation \cite{parry2008comparative}. 

Causal abstractions might be derived automatically from formal representations of the simulation model (see Section \ref{sec:background}), e.g., by exploiting model-driven reverse engineering from software engineering, or they might be learned by observing the input-output behavior of the simulation. In the ideal case, they would use information about the simulation model, the context (see Section \ref{sec:context}), and information gained from (automatically) executing various simulation experiments (see Section \ref{sec:automation}). Simulation is a data-generating process. Therefore, in addition to filtering or further processing of generated data 
according to stakeholders' expectations to support explanations, data can be generated for the purpose of explanations. 
Explanations are not only constrained to causality; they might also refer to revealing context information necessary for interpreting the simulation output, e.g., which part of the context should be made explicit and, if so, how to the stakeholder? Are in the given situation of the stakeholder the assumptions more important, or should the stakeholder be made aware based on which data the model has been validated, again constraining possibly its application (see also Section \ref{sec:composition} on model reuse)?

\subsubsection{Interactions for Understanding} 
A more engaging approach 
to building stakeholder understanding is providing methods 
to help the stakeholder explore the model's behavior and, if time allows, to do so jointly with other stakeholders and the modelers, even in a participatory manner during model building.
Being more engaged in a simulation study by interaction will increase trust in the stakeholder-model relation and, if participatory approaches are used, for example, during model development 
strengthen the stakeholder-modeler relation \cite{Harper+2021trust}. However, the temporal constraints of stakeholders will limit the practicability of the approach. Efficient visualization, interaction, and automatic analysis methods fine-tuned to the stakeholder's questions will be essential in realizing this.  

Animation is an established tool for showing stakeholders what entities are modeled in a simulation and how they interact to produce outcomes.
However, a single, one-size-fits-all animation cannot give a stakeholder a comprehensive view of a simulation model's behavior. 
Adopting state-of-the-art visual analytics methods and the corresponding established techniques, such as linked views, and offering individualization of 
pipelines, e.g., "analyze first, show the important, zoom, filter and analyze further, details on demand” \cite{Keim2008}, will help to communicate simulation results more effectively to the individual stakeholder. 
New methodological developments are needed, given that effective visualizations need to be 
tuned to the application or even to a stakeholder's question of interest. Thus, they need to be highly adaptable. Exploiting the full potential of visual analytics for modeling and simulation requires expertise in visualization methods, their development, and application, and thus, demands close cooperation with the visual analytics community. For computationally intensive simulations, providing a responsive real-time experience, efficient simulation algorithms are needed that may trade accuracy for speed. Anticipating the stakeholders' interest -- supported by automated elicitation approaches, as given above -- would also allow for generating simulation runs in advance to have the results when stakeholders want to examine them interactively.

\section{Discussion and Conclusion}
\label{sec:conclusion}
In the following, we will briefly summarize our discussions on how the four goals are interrelated and which foundation, challenges, and strategies they share to move ahead. 

\subsection{Four Interrelated Goals}

\begin{figure}
\begin{center}
	\includegraphics[width=0.75\textwidth]{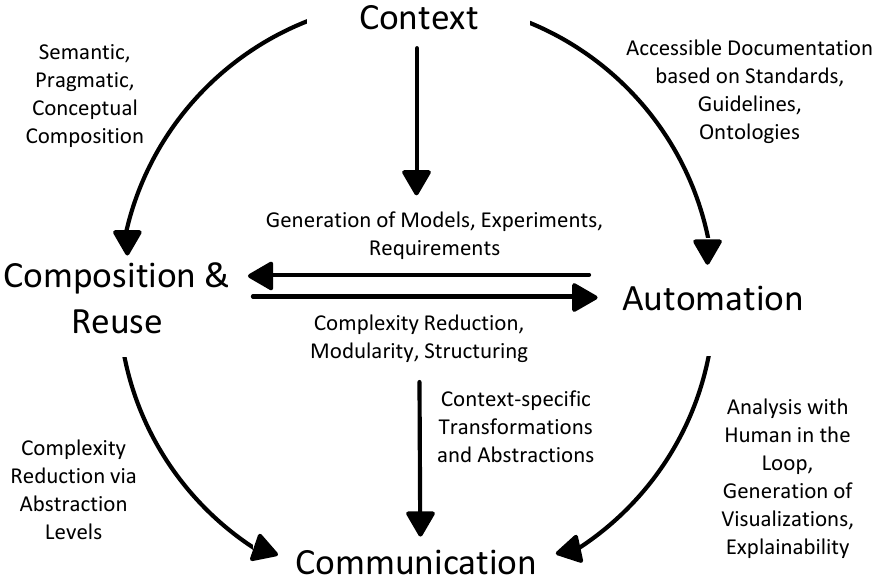}
\end{center}
\caption{Interrelations between the four goals \emph{context}, \emph{composition \& reuse}, \emph{automation}, and \emph{communication} in which the different goals enable or facilitate each other by specific means. }
\label{fig:goals_relations}
\end{figure}

The
\textbf{
C$^2$AC Roadmap for Modeling and Simulation} proposes four goals to move methods and practice of modeling and simulation studies forward within this millennium: to provide computational support for representing and evaluating \emph{context}, to support \emph{composition and reuse} in simulation studies, to extend the degree of \emph{automating} simulation studies, and to enhance the \emph{communication} with different stakeholders. 
We showed that the four goals are tightly coupled and exhibit many interdependencies; see Fig. \ref{fig:goals_relations}.

With \emph{context}, we extend the conceptual model, i.e., everything useful in building a simulation model or conducting the simulation study, with information about every constituent's actual usage. A provenance-based approach makes the different artifacts' role in the knowledge-intensive process of conducting a simulation study explicit. This is a pre-requisite for knowledge about artifacts as well as the involved processes to be exploited retrospectively for more comprehensive documentation, analysis, and assessment of simulation studies, and, in addition, prospectively, to guide the modeler through simulation studies or even to automate these studies. 

\emph{Composition and reuse} have been shown to be central to mastering the complexity of building a simulation model. Still, too little support exists to ensure semantically valid composition or reuse, and the same applies to reasoning about similarities and differences. Other modeling and simulation study artifacts, such as simulation experiments, requirements, or assumptions, hold valuable information that future work needs to build on. At the same time, developing those artifacts (and thus the entire simulation study) will benefit from methods and efforts aimed at facilitating composition and reuse, such as introducing standards and formalization.

Increasing the fraction of simulation studies that can be executed \emph{automatically} boosts the efficiency of conducting simulation studies. It holds the promise of more systematically conducted simulation studies than at present. Automation relies on making knowledge in a modeling and simulation study accessible and interpretable. This refers to knowledge about the application domain of the simulation study as well as modeling and simulation methods and their application. Therefore, in addition to an unambiguous representation of the various artifacts, suitable annotations and ontologies will play a central role, as does the exploitation of machine learning methods.

\emph{Communication} of simulation results relies on understanding the different users, whether these are domain experts or decision-makers. 
Their view on the system of interest, expectations, and questions must be considered when communicating and explaining simulation results to users. This applies to automatically generated (visual) abstractions, simulation experiments, explanations, and interactions that allow users to gain insights into the system and explore possible answers to their questions. Accordances and discrepancies between the perceptions of stakeholders and the "reality" of the simulation need to be built upon respectively revealed and actively exploited to enhance understanding of the simulation. Supporting communication more effectively will increase the impact of simulation studies. It will enrich the canon of simulation methods to generate explanations, provide interactive visualization, and exploit participatory approaches tuned to stakeholders' demands and temporal constraints.  

\subsection{Common Foundations and Challenges}

All goals rely on a clear separation of concerns and the unambiguous representation of central artifacts of the modeling and simulation study. Whereas in the past, attention was focused on simulation models and their formal representation, more recent efforts have looked more closely at how to succinctly and unambiguously specify simulation experiments and behavioral requirements. These efforts to provide a user-friendly representation with clear semantics must be intensified and expanded to other artifacts, such as assumptions and research questions, and the processes of the simulation studies to achieve any of the four goals. In this pursuit, domain-specific language design, accessible annotations, ontologies, and the integration of machine learning will play a central role.   

The methodological developments induced by pursuing the suggested roadmap must be evaluated suitably. Evaluation might refer to efficiency (concerning time and other resources), effectiveness (in terms of the intended purpose), and accuracy (without technical or methodological errors). This will propose new research challenges. For example, in addition to increasing the reliability of simulation results, increasing understanding and trust in modeling and simulation studies is a major concern of the roadmap, but how can the level of understanding or trust be reliably assessed, and how can methods and protocols be designed to evaluate a change in those? Clear evaluation procedures are a pre-requisite to measure the impact of the developed methods on modeling and simulation, and thus crucial for any advance in research. 

Developing suitable evaluation methods will open up an entire area of new modeling and simulation research, which can build on metrics, evaluation methods, and best practices in other areas of computer science, such as software engineering and human-computer interaction. In fact, the proposed roadmap will \emph{require} close interaction with other areas of computer science, e.g., with visual analytics, which plays a crucial role in communication, but also, more generally, in keeping the modeler or user within the loop, particularly, if substantial parts of the modeling and simulation life cycle become automated. Similarly, adopting state-of-the-art methods developed in software engineering is key for making progress in modeling and simulation as envisioned in our roadmap. This includes methods that support the design of DSLs and formats for the diverse artifacts of a simulation study, as well as for reverse engineering to generate abstractions to cater to the needs of different users. Reasoning about variations or components of artifacts may benefit from developments such as feature languages and other annotations. The latter leads us to artificial intelligence. Knowledge-based annotations and suitable inference strategies will be crucial for any meaningful composition and reuse beyond the syntactical level and for realizing a more significant degree of automation. Therefore, combining knowledge-based deductive with machine-learning methods is a promising avenue to automate various tasks within the modeling and simulation life cycle. Our roadmap requires but also propels reaching out to other computer science areas to engulf the state of the art in diverse areas, with an expected win-win for all involved. 

As cooperation with other computer science disciplines is needed to move forward, so are additional efforts of the modeling and simulation communities. For example, the maintenance of tools and repositories requires further attention. Efforts need to be directed towards defining and utilizing languages with clear semantics, introducing standards and best practices, developing ontologies of the application domain and modeling and simulation methods, and making those accessible. These research efforts may not be restricted to simulation models but need to be expanded to other modeling and simulation artifacts, such as research questions, assumptions, requirements, simulation experiments, and diverse processes and activities, to treat those as first-class citizens of the simulation study.

\section*{Acknowledgements}
We would like to thank Schloss Dagstuhl (\url{www.dagstuhl.de}). The paper summarizes and extends the discussions of one of the three working groups of the Dagstuhl Seminar 22401: Computer Science Methods for Effective and Sustainable Simulation Studies (03.\ Oct.--07.\ Oct, 2022).

\subsection*{Funding}
A.M.U. and P.W. are funded by the German Research Foundation (DFG) grant 320435134 ``GrEASE - Towards Generating and Executing Automatically Simulation Experiments''.

C. R-M. and G.W. are funded by NSERC - Canada.

F.L. is funded by the Wallenberg AI, Autonomous Systems and Software Program – Humanities and Society (WASP-HS) funded by the Marianne and Marcus Wallenberg Foundation and the Marcus and Amalia Wallenberg Foundation.

\bibliographystyle{unsrtnat}
\bibliography{refs}

\begin{thebibliography}{326}
\providecommand{\natexlab}[1]{#1}
\providecommand{\url}[1]{\texttt{#1}}
\expandafter\ifx\csname urlstyle\endcsname\relax
  \providecommand{\doi}[1]{doi: #1}\else
  \providecommand{\doi}{doi: \begingroup \urlstyle{rm}\Url}\fi

\bibitem[Winsberg(2010)]{winsberg2010science}
Eric Winsberg.
\newblock \emph{Science in the Age of Computer Simulation}.
\newblock University of Chicago Press, 2010.

\bibitem[Balci(2012)]{balci2012life}
Osman Balci.
\newblock A life cycle for modeling and simulation.
\newblock \emph{Simulation}, 88\penalty0 (7):\penalty0 870--883, 2012.

\bibitem[Law(2019)]{law2019build}
Averill~M Law.
\newblock How to build valid and credible simulation models.
\newblock In \emph{2019 Winter Simulation Conference (WSC)}, pages 1402--1414.
  IEEE, 2019.

\bibitem[Robinson(2014)]{robinson2014simulation}
Stewart Robinson.
\newblock \emph{Simulation: The Practice of Model Development and Use}.
\newblock Bloomsbury Publishing, 2014.

\bibitem[Lorig et~al.(2021)Lorig, Johansson, and Davidsson]{lorig2021agent}
Fabian Lorig, Emil Johansson, and Paul Davidsson.
\newblock Agent-based social simulation of the covid-19 pandemic: A systematic
  review.
\newblock \emph{JASSS: Journal of Artificial Societies and Social Simulation},
  24\penalty0 (3), 2021.

\bibitem[Shinde et~al.(2020)Shinde, Kalamkar, Mahalle, Dey, Chaki, and
  Hassanien]{shinde2020forecasting}
Gitanjali~R Shinde, Asmita~B Kalamkar, Parikshit~N Mahalle, Nilanjan Dey,
  Jyotismita Chaki, and Aboul~Ella Hassanien.
\newblock Forecasting models for coronavirus disease (covid-19): A survey of
  the state-of-the-art.
\newblock \emph{SN Computer Science}, 1\penalty0 (4):\penalty0 1--15, 2020.

\bibitem[Basso et~al.(2022)Basso, Goic, Olivares, Saur{\'e}, Thraves, Carranza,
  Weintraub, Covarrubias, Escobedo, Jara, et~al.]{bassoanalytics}
Leonardo~J Basso, Marcel Goic, Marcelo Olivares, Denis Saur{\'e}, Charles
  Thraves, Aldo Carranza, Gabriel~Y Weintraub, Julio Covarrubias, Cristian
  Escobedo, Natalia Jara, et~al.
\newblock Analytics saves lives during the covid crisis in chile.
\newblock Technical report, 2022.
\newblock Franz Edelman Award.

\bibitem[Ferguson et~al.(2020)Ferguson, Laydon, Nedjati-Gilani, Imai, Ainslie,
  Baguelin, Bhatia, Boonyasiri, Cucunub{\'a}, Cuomo-Dannenburg,
  et~al.]{ferguson2020impact}
Neil~M Ferguson, Daniel Laydon, Gemma Nedjati-Gilani, Natsuko Imai, Kylie
  Ainslie, Marc Baguelin, Sangeeta Bhatia, Adhiratha Boonyasiri, Zulma
  Cucunub{\'a}, Gina Cuomo-Dannenburg, et~al.
\newblock Impact of non-pharmaceutical interventions (npis) to reduce covid-19
  mortality and healthcare demand.
\newblock Technical report, Imperial College COVID-19 Response Team London,
  2020.

\bibitem[Garcia-Vicu{\~n}a et~al.(2022)Garcia-Vicu{\~n}a, Esparza, and
  Mallor]{garcia2022hospital}
Daniel Garcia-Vicu{\~n}a, Laida Esparza, and Fermin Mallor.
\newblock Hospital preparedness during epidemics using simulation: the case of
  covid-19.
\newblock \emph{Central European Journal of Operations Research}, 30\penalty0
  (1):\penalty0 213--249, 2022.

\bibitem[Ivanov(2020)]{ivanov2020predicting}
Dmitry Ivanov.
\newblock Predicting the impacts of epidemic outbreaks on global supply chains:
  A simulation-based analysis on the coronavirus outbreak (covid-19/sars-cov-2)
  case.
\newblock \emph{Transportation Research Part E: Logistics and Transportation
  Review}, 136:\penalty0 101922, 2020.

\bibitem[Frazier et~al.(2022)Frazier, Cashore, Duan, Henderson, Janmohamed,
  Liu, Shmoys, Wan, and Zhang]{frazier2022modeling}
Peter~I Frazier, J~Massey Cashore, Ning Duan, Shane~G Henderson, Alyf
  Janmohamed, Brian Liu, David~B Shmoys, Jiayue Wan, and Yujia Zhang.
\newblock Modeling for covid-19 college reopening decisions: Cornell, a case
  study.
\newblock \emph{Proceedings of the National Academy of Sciences}, 119\penalty0
  (2):\penalty0 e2112532119, 2022.

\bibitem[Gro{\ss}mann et~al.(2020)Gro{\ss}mann, Backenk{\"{o}}hler, and
  Wolf]{wolf2020}
Gerrit Gro{\ss}mann, Michael Backenk{\"{o}}hler, and Verena Wolf.
\newblock Importance of interaction structure and stochasticity for epidemic
  spreading: {A} {COVID-19} case study.
\newblock In Marco Gribaudo, David~N. Jansen, and Anne Remke, editors,
  \emph{Quantitative Evaluation of Systems - 17th International Conference,
  {QEST} 2020, Vienna, Austria, August 31 - September 3, 2020, Proceedings},
  volume 12289 of \emph{Lecture Notes in Computer Science}, pages 211--229.
  Springer, 2020.
\newblock \doi{10.1007/978-3-030-59854-9\_16}.
\newblock URL \url{https://doi.org/10.1007/978-3-030-59854-9\_16}.

\bibitem[Edeling et~al.(2021)Edeling, Arabnejad, Sinclair, Suleimenova,
  Gopalakrishnan, Bosak, Groen, Mahmood, Crommelin, and
  Coveney]{edeling_impact_2021}
Wouter Edeling, Hamid Arabnejad, Robbie Sinclair, Diana Suleimenova,
  Krishnakumar Gopalakrishnan, Bartosz Bosak, Derek Groen, Imran Mahmood, Daan
  Crommelin, and Peter~V. Coveney.
\newblock The impact of uncertainty on predictions of the {CovidSim}
  epidemiological code.
\newblock \emph{Nature Computational Science}, 1\penalty0 (2), February 2021.
\newblock ISSN 2662-8457.
\newblock \doi{10.1038/s43588-021-00028-9}.
\newblock URL \url{https://www.nature.com/articles/s43588-021-00028-9}.

\bibitem[Winsberg et~al.(2020)Winsberg, Brennan, and
  Surprenant]{winsberg2020government}
Eric Winsberg, Jason Brennan, and Chris~W Surprenant.
\newblock How government leaders violated their epistemic duties during the
  sars-cov-2 crisis.
\newblock \emph{Kennedy Institute of Ethics Journal}, 30\penalty0 (3):\penalty0
  215--242, 2020.

\bibitem[Cai et~al.(2023)Cai, Carothers, Nicol, and
  Uhrmacher]{dagstuhlReport2023}
Wentong Cai, Christopher Carothers, David~M. Nicol, and Adelinde~M. Uhrmacher.
\newblock {Computer Science Methods for Effective and Sustainable Simulation
  Studies (Dagstuhl Seminar 22401)}.
\newblock \emph{Dagstuhl Reports}, 12\penalty0 (10):\penalty0 1--60, 2023.
\newblock ISSN 2192-5283.
\newblock \doi{10.4230/DagRep.12.10.1}.
\newblock URL \url{https://drops.dagstuhl.de/opus/volltexte/2023/17819}.

\bibitem[Zeigler et~al.(2000)Zeigler, Praehofer, and Kim]{Zeigler2000}
Bernard~P. Zeigler, Herbert Praehofer, and Tag~Gon Kim.
\newblock \emph{Theory of Modelling and Simulation: Integrating Discrete Event
  and Continuous Complex Dynamic Systems}.
\newblock Academic Press, San Diego, CA, 2000.

\bibitem[Balbo(2000)]{balbo2000introduction}
Gianfranco Balbo.
\newblock Introduction to stochastic petri nets.
\newblock In \emph{School organized by the European Educational Forum}, pages
  84--155. Springer, 2000.

\bibitem[Hillston(2005)]{hillston2005process}
Jane Hillston.
\newblock Process algebras for quantitative analysis.
\newblock In \emph{20th Annual IEEE Symposium on Logic in Computer Science
  (LICS'05)}, pages 239--248. IEEE, 2005.

\bibitem[Booch et~al.(1998)Booch, Rumbaugh, and Jacobson]{UMLuser98}
Grady Booch, James Rumbaugh, and Ivar Jacobson.
\newblock \emph{The {U}nified {M}odeling {L}anguage User Guide}.
\newblock Addison-Wesley, 1998.

\bibitem[Rumpe(2017)]{Rum17}
Bernhard Rumpe.
\newblock \emph{{Agile Modeling with UML: Code Generation, Testing,
  Refactoring}}.
\newblock Springer International, May 2017.

\bibitem[Fowler and Scott(1997)]{Fow97}
Martin Fowler and Kendall Scott.
\newblock \emph{{UML Distilled: Applying the Standard Object Modeling
  Language}}.
\newblock Addison-Wesley Longman Ltd., Essex, UK, UK, 1997.
\newblock ISBN 0-201-32563-2.

\bibitem[Weilkiens(2008)]{Wei06}
Tim Weilkiens.
\newblock \emph{{{Systems Engineering with SysML/UML: Modeling, Analysis,
  Design}}}.
\newblock Elsevier, 2008.

\bibitem[Friedenthal et~al.(2011)Friedenthal, Moore, and Steiner]{FMS11}
Sanford. Friedenthal, Alan Moore, and Rick Steiner.
\newblock \emph{A Practical Guide to SysML: The Systems Modeling Language}.
\newblock The MK/OMG Press. Elsevier Science, 2011.
\newblock ISBN 9780123852076.
\newblock URL \url{http://books.google.de/books?id=4xz6Fx50zwcC}.

\bibitem[Jansen et~al.(2022)Jansen, Pfeiffer, Rumpe, Schmalzing, and
  Wortmann]{JPR+22}
Nico Jansen, Jerome Pfeiffer, Bernhard Rumpe, David Schmalzing, and Andreas
  Wortmann.
\newblock {The Language of SysML v2 under the Magnifying Glass}.
\newblock \emph{Journal of Object Technology (JOT)}, 21:\penalty0 1--15, July
  2022.

\bibitem[Blinov et~al.(2004)Blinov, Faeder, Goldstein, and
  Hlavacek]{blinov2004bionetgen}
Michael~L Blinov, James~R Faeder, Byron Goldstein, and William~S Hlavacek.
\newblock Bionetgen: Software for rule-based modeling of signal transduction
  based on the interactions of molecular domains.
\newblock \emph{Bioinformatics}, 20\penalty0 (17):\penalty0 3289--3291, 2004.

\bibitem[Boutillier et~al.(2018)Boutillier, Maasha, Li, Medina-Abarca, Krivine,
  Feret, Cristescu, Forbes, and Fontana]{boutillier2018kappa}
Pierre Boutillier, Mutaamba Maasha, Xing Li, H{\'e}ctor~F Medina-Abarca, Jean
  Krivine, J{\'e}r{\^o}me Feret, Ioana Cristescu, Angus~G Forbes, and Walter
  Fontana.
\newblock The kappa platform for rule-based modeling.
\newblock \emph{Bioinformatics}, 34\penalty0 (13):\penalty0 i583--i592, 2018.

\bibitem[Helms et~al.(2017)Helms, Warnke, Maus, and
  Uhrmacher]{Helms2017semantics}
Tobias Helms, Tom Warnke, Carsten Maus, and Adelinde~M Uhrmacher.
\newblock Semantics and efficient simulation algorithms of an expressive
  multilevel modeling language.
\newblock \emph{ACM Transactions on Modeling and Computer Simulation (TOMACS)},
  27\penalty0 (2):\penalty0 1--25, 2017.

\bibitem[Reinhardt et~al.(2022)Reinhardt, Warnke, and
  Uhrmacher]{Reinhardt2022language}
Oliver Reinhardt, Tom Warnke, and Adelinde~M Uhrmacher.
\newblock A language for agent-based discrete-event modeling and simulation of
  linked lives.
\newblock \emph{ACM Transactions on Modeling and Computer Simulation (TOMACS)},
  32\penalty0 (1):\penalty0 1--26, 2022.

\bibitem[Pawlikowski et~al.(2002)Pawlikowski, Jeong, and
  Lee]{pawlikowski2002credibility}
Krzysztof Pawlikowski, H-DJ Jeong, and J-SR Lee.
\newblock On credibility of simulation studies of telecommunication networks.
\newblock \emph{IEEE Communications Magazine}, 40\penalty0 (1):\penalty0
  132--139, 2002.

\bibitem[Merali(2010)]{merali2010computational}
Zeeya Merali.
\newblock Computational science:... error.
\newblock \emph{Nature}, 467\penalty0 (7317):\penalty0 775--777, 2010.

\bibitem[Ewald and Uhrmacher(2014)]{ewald2014sessl}
Roland Ewald and Adelinde~M Uhrmacher.
\newblock Sessl: A domain-specific language for simulation experiments.
\newblock \emph{ACM Transactions on Modeling and Computer Simulation (TOMACS)},
  24\penalty0 (2):\penalty0 1--25, 2014.

\bibitem[Waltemath et~al.(2011{\natexlab{a}})Waltemath, Adams, Bergmann, Hucka,
  Kolpakov, Miller, Moraru, Nickerson, Sahle, Snoep,
  et~al.]{waltemath2011reproducible}
Dagmar Waltemath, Richard Adams, Frank~T Bergmann, Michael Hucka, Fedor
  Kolpakov, Andrew~K Miller, Ion~I Moraru, David Nickerson, Sven Sahle, Jacky~L
  Snoep, et~al.
\newblock Reproducible computational biology experiments with sed-ml-the
  simulation experiment description markup language.
\newblock \emph{BMC Systems Biology}, 5\penalty0 (1):\penalty0 1--10,
  2011{\natexlab{a}}.

\bibitem[Salecker et~al.(2019)Salecker, Sciaini, Meyer, and
  Wiegand]{salecker2019nlrx}
Jan Salecker, Marco Sciaini, Katrin~M Meyer, and Kerstin Wiegand.
\newblock The nlrx r package: A next-generation framework for reproducible
  netlogo model analyses.
\newblock \emph{Methods in Ecology and Evolution}, 10\penalty0 (11):\penalty0
  1854--1863, 2019.

\bibitem[G{\"o}rlach et~al.(2011)G{\"o}rlach, Sonntag, Karastoyanova, Leymann,
  and Reiter]{gorlach2011conventional}
Katharina G{\"o}rlach, Mirko Sonntag, Dimka Karastoyanova, Frank Leymann, and
  Michael Reiter.
\newblock Conventional workflow technology for scientific simulation.
\newblock \emph{Guide to e-Science: Next Generation Scientific Research and
  Discovery}, pages 323--352, 2011.

\bibitem[Waltemath et~al.(2011{\natexlab{b}})Waltemath, Adams, Beard, Bergmann,
  Bhalla, Britten, Chelliah, Cooling, Cooper, Crampin,
  et~al.]{waltemath2011minimum}
Dagmar Waltemath, Richard Adams, Daniel~A Beard, Frank~T Bergmann, Upinder~S
  Bhalla, Randall Britten, Vijayalakshmi Chelliah, Michael~T Cooling, Jonathan
  Cooper, Edmund~J Crampin, et~al.
\newblock {Minimum Information about a Simulation Experiment (MIASE)}.
\newblock \emph{PLoS Computational Biology}, 7\penalty0 (4):\penalty0 e1001122,
  2011{\natexlab{b}}.

\bibitem[Rahmandad and Sterman(2012)]{rahmandad2012reporting}
Hazhir Rahmandad and John~D Sterman.
\newblock Reporting guidelines for simulation-based research in social
  sciences.
\newblock \emph{Systems Dynamics Review}, 28\penalty0 (4):\penalty0 396--411,
  2012.

\bibitem[Grimm et~al.(2014)Grimm, Augusiak, Focks, Frank, Gabsi, Johnston, Liu,
  Martin, Meli, Radchuk, et~al.]{grimm2014towards}
Volker Grimm, Jacqueline Augusiak, Andreas Focks, B{\'e}atrice~M Frank, Faten
  Gabsi, Alice~SA Johnston, Chun Liu, Benjamin~T Martin, Mattia Meli, Viktoriia
  Radchuk, et~al.
\newblock Towards better modelling and decision support: Documenting model
  development, testing, and analysis using trace.
\newblock \emph{Ecological Modelling}, 280:\penalty0 129--139, 2014.

\bibitem[Monks et~al.(2019)Monks, Currie, Onggo, Robinson, Kunc, and
  Taylor]{monks2019strengthening}
Thomas Monks, Christine~SM Currie, Bhakti~Stephan Onggo, Stewart Robinson,
  Martin Kunc, and Simon~JE Taylor.
\newblock {Strengthening the Reporting of Empirical Simulation Studies:
  Introducing the STRESS Guidelines}.
\newblock \emph{Journal of Simulation}, 13\penalty0 (1):\penalty0 55--67, 2019.

\bibitem[Agha and Palmskog(2018)]{agha2018survey}
Gul Agha and Karl Palmskog.
\newblock A survey of statistical model checking.
\newblock \emph{ACM Transactions on Modeling and Computer Simulation (TOMACS)},
  28\penalty0 (1):\penalty0 1--39, 2018.

\bibitem[Legay et~al.(2019)Legay, Lukina, Traonouez, Yang, Smolka, and
  Grosu]{steffen_statistical_2019}
Axel Legay, Anna Lukina, Louis~Marie Traonouez, Junxing Yang, Scott~A. Smolka,
  and Radu Grosu.
\newblock Statistical {Model} {Checking}.
\newblock In Bernhard Steffen and Gerhard Woeginger, editors, \emph{Computing
  and {Software} {Science}}, volume 10000, pages 478--504. Springer
  International Publishing, Cham, 2019.
\newblock ISBN 978-3-319-91907-2 978-3-319-91908-9.
\newblock \doi{10.1007/978-3-319-91908-9_23}.
\newblock URL \url{http://link.springer.com/10.1007/978-3-319-91908-9_23}.
\newblock Series Title: Lecture Notes in Computer Science.

\bibitem[Lorig et~al.(2017)Lorig, Becker, and Timm]{lorig2017formal}
Fabian Lorig, Colja~A Becker, and Ingo~J Timm.
\newblock Formal specification of hypotheses for assisting computer simulation
  studies.
\newblock In \emph{Proceedings of the Symposium on Theory of Modeling \&
  Simulation}, pages 1--12, 2017.

\bibitem[Zeigler(1976)]{zeigler1976}
Bernhard Zeigler.
\newblock \emph{Theory of Modeling and Simulation}.
\newblock John Wiley, 1976.

\bibitem[Zeigler(1977)]{Zeigler1977}
Bernard~P. Zeigler.
\newblock Constructs for the specifications of models and experimental frames.
\newblock \emph{SIGSIM Simul. Dig.}, 9\penalty0 (1):\penalty0 12–13, sep
  1977.
\newblock ISSN 0163-6103.
\newblock \doi{10.1145/1102505.1102510}.
\newblock URL \url{https://doi.org/10.1145/1102505.1102510}.

\bibitem[Zeigler et~al.(2018)Zeigler, Muzy, and Kofman]{zeigler2018theory}
Bernard~P Zeigler, Alexandre Muzy, and Ernesto Kofman.
\newblock \emph{Theory of Modeling and Simulation: Discrete Event \& Iterative
  System Computational Foundations}.
\newblock Academic press, 2018.

\bibitem[Denil et~al.(2017)Denil, Klikovits, Mosterman, Vallecillo, and
  Vangheluwe]{denil2017experiment}
Joachim Denil, Stefan Klikovits, Pieter~J Mosterman, Antonio Vallecillo, and
  Hans Vangheluwe.
\newblock The experiment model and validity frame in m\&s.
\newblock In \emph{Proceedings of the Symposium on Theory of Modeling \&
  Simulation}, pages 1--12, 2017.

\bibitem[Deelman et~al.(2015)Deelman, Vahi, Juve, Rynge, Callaghan, Maechling,
  Mayani, Chen, {Ferreira da Silva}, Livny, and Wenger]{DEELMAN201517}
Ewa Deelman, Karan Vahi, Gideon Juve, Mats Rynge, Scott Callaghan, Philip~J.
  Maechling, Rajiv Mayani, Weiwei Chen, Rafael {Ferreira da Silva}, Miron
  Livny, and Kent Wenger.
\newblock Pegasus, a workflow management system for science automation.
\newblock \emph{Future Generation Computer Systems}, 46:\penalty0 17--35, 2015.
\newblock ISSN 0167-739X.
\newblock \doi{https://doi.org/10.1016/j.future.2014.10.008}.
\newblock URL
  \url{https://www.sciencedirect.com/science/article/pii/S0167739X14002015}.

\bibitem[Bartocci et~al.(2018)Bartocci, Deshmukh, Donzé, Fainekos, Maler,
  Ničković, and Sankaranarayanan]{bartocci_specification-based_2018survey}
Ezio Bartocci, Jyotirmoy Deshmukh, Alexandre Donzé, Georgios Fainekos, Oded
  Maler, Dejan Ničković, and Sriram Sankaranarayanan.
\newblock Specification-{Based} {Monitoring} of {Cyber}-{Physical} {Systems}:
  {A} {Survey} on {Theory}, {Tools} and {Applications}.
\newblock In \emph{Lectures on {Runtime} {Verification}}, Lecture {Notes} in
  {Computer} {Science}, pages 135--175. Springer, Cham, 2018.
\newblock ISBN 978-3-319-75631-8 978-3-319-75632-5.
\newblock \doi{10.1007/978-3-319-75632-5_5}.
\newblock URL
  \url{https://link.springer.com/chapter/10.1007/978-3-319-75632-5_5}.

\bibitem[Nenzi et~al.(2022)Nenzi, Bartocci, Bortolussi, and Loreti]{NenziBBL22}
Laura Nenzi, Ezio Bartocci, Luca Bortolussi, and Michele Loreti.
\newblock A logic for monitoring dynamic networks of spatially-distributed
  cyber-physical systems.
\newblock \emph{Logical Methods in Computer Science}, 18\penalty0 (1), 2022.
\newblock \doi{10.46298/lmcs-18(1:4)2022}.
\newblock URL \url{https://doi.org/10.46298/lmcs-18(1:4)2022}.

\bibitem[Wilsdorf et~al.(2022)Wilsdorf, Heller, Budde, Zimmermann, Warnke,
  Haubelt, Timmermann, van Rienen, and Uhrmacher]{wilsdorf2022model}
Pia Wilsdorf, Jakob Heller, Kai Budde, Julius Zimmermann, Tom Warnke, Christian
  Haubelt, Dirk Timmermann, Ursula van Rienen, and Adelinde~M Uhrmacher.
\newblock A model-driven approach for conducting simulation experiments.
\newblock \emph{Applied Sciences}, 12\penalty0 (16):\penalty0 7977, 2022.

\bibitem[Chinosi and Trombetta(2012)]{BPMN}
Michele Chinosi and Alberto Trombetta.
\newblock Bpmn: An introduction to the standard.
\newblock \emph{Computer Standards \& Interfaces}, 34\penalty0 (1):\penalty0
  124--134, 2012.

\bibitem[Mitra et~al.(2019)Mitra, Suderman, Colvin, Ionkov, Hu, Sauro, Posner,
  and Hlavacek]{mitra2019pybionetfit}
Eshan~D Mitra, Ryan Suderman, Joshua Colvin, Alexander Ionkov, Andrew Hu,
  Herbert~M Sauro, Richard~G Posner, and William~S Hlavacek.
\newblock Pybionetfit and the biological property specification language.
\newblock \emph{IScience}, 19:\penalty0 1012--1036, 2019.

\bibitem[North et~al.(2013)North, Collier, Ozik, Tatara, Macal, Bragen, and
  Sydelko]{North2013complex}
Michael~J. North, Nicholson~T. Collier, Jonathan Ozik, Eric~R. Tatara,
  Charles~M. Macal, Mark Bragen, and Pam Sydelko.
\newblock Complex adaptive systems modeling with {Repast} {Simphony}.
\newblock \emph{Complex Adaptive Systems Modeling}, 1\penalty0 (1):\penalty0 3,
  March 2013.
\newblock ISSN 2194-3206.
\newblock \doi{10.1186/2194-3206-1-3}.
\newblock URL \url{https://doi.org/10.1186/2194-3206-1-3}.

\bibitem[Johnsen et~al.(2011)Johnsen, H{\"a}hnle, Sch{\"a}fer, Schlatte, and
  Steffen]{abs}
Einar~Broch Johnsen, Reiner H{\"a}hnle, Jan Sch{\"a}fer, Rudolf Schlatte, and
  Martin Steffen.
\newblock {ABS}: A core language for abstract behavioral specification.
\newblock In Bernhard~K. Aichernig, Frank de~Boer, and Marcello~M. Bonsangue,
  editors, \emph{Proc.\ 9th Intl.\ Symp.\ on Formal Methods for Components and
  Objects ({FMCO} 2010)}, volume 6957 of \emph{LNCS}, pages 142--164. Springer,
  2011.

\bibitem[Kamburjan et~al.(2022)Kamburjan, Mitsch, and H{\"a}hnle]{kmh21}
Eduard Kamburjan, Stefan Mitsch, and Reiner H{\"a}hnle.
\newblock A hybrid programming language for formal modeling and verification of
  hybrid systems.
\newblock \emph{Leibniz Transactions on Embedded Systems}, 8\penalty0 (1),
  2022.
\newblock Special Issue on Distributed Hybrid Systems.

\bibitem[Warnke and Uhrmacher(2018)]{warnke2018complex}
Tom Warnke and Adelinde~M Uhrmacher.
\newblock Complex simulation experiments made easy.
\newblock In \emph{2018 Winter Simulation Conference (WSC)}, pages 410--424.
  IEEE, 2018.

\bibitem[Bertalanffy(1968)]{bertalanffy1968general}
Ludwig~von Bertalanffy.
\newblock \emph{General System Theory: Foundations, Development, Applications}.
\newblock G. Braziller, 1968.

\bibitem[Milner et~al.(1992)Milner, Parrow, and Walker]{milner1992calculus}
Robin Milner, Joachim Parrow, and David Walker.
\newblock A calculus of mobile processes, i.
\newblock \emph{Information and Computation}, 100\penalty0 (1):\penalty0 1--40,
  1992.

\bibitem[Priami(1995)]{priami1995stochastic}
Corrado Priami.
\newblock Stochastic $\pi$-calculus.
\newblock \emph{The Computer Journal}, 38\penalty0 (7):\penalty0 578--589,
  1995.
\newblock \doi{10.1093/comjnl/38.7.578}.

\bibitem[Phillips and Cardelli(2007)]{Cardelli2007}
Andrew Phillips and Luca Cardelli.
\newblock Efficient, correct simulation of biological processes in the
  stochastic pi-calculus.
\newblock In \emph{Proceedings of the 2007 International Conference on
  Computational Methods in Systems Biology}, CMSB'07, page 184–199, Berlin,
  Heidelberg, 2007. Springer-Verlag.
\newblock ISBN 3540751394.

\bibitem[Jensen(1996)]{jensen1996coloured}
Kurt Jensen.
\newblock \emph{Coloured Petri Nets: Basic Concepts, Analysis Methods and
  Practical Use}, volume~1.
\newblock Springer Science \& Business Media, 1996.

\bibitem[John et~al.(2010)John, Lhoussaine, Niehren, and Uhrmacher]{John2010}
Mathias John, C{\'e}dric Lhoussaine, Joachim Niehren, and Adelinde~M.
  Uhrmacher.
\newblock \emph{The Attributed Pi-Calculus with Priorities}, pages 13--76.
\newblock Springer Berlin Heidelberg, Berlin, Heidelberg, 2010.
\newblock ISBN 978-3-642-11712-1.
\newblock \doi{10.1007/978-3-642-11712-1_2}.
\newblock URL \url{https://doi.org/10.1007/978-3-642-11712-1_2}.

\bibitem[Barros(1997)]{Barros1997}
Fernando~J. Barros.
\newblock Modeling formalisms for dynamic structure systems.
\newblock \emph{ACM Transactions on Modeling and Computer Simulation},
  7\penalty0 (4):\penalty0 501–515, oct 1997.
\newblock ISSN 1049-3301.
\newblock \doi{10.1145/268403.268423}.
\newblock URL \url{https://doi.org/10.1145/268403.268423}.

\bibitem[Cellier and Greifeneder(2013)]{cellier2013continuous}
Fran{\c{c}}ois~E Cellier and Jurgen Greifeneder.
\newblock \emph{Continuous System Modeling}.
\newblock Springer Science \& Business Media, 2013.

\bibitem[Hillston et~al.(1991)Hillston, Opdahl, and Pooley]{hillston1991case}
Jane Hillston, Andreas~L Opdahl, and Rob Pooley.
\newblock A case study using the imse experimentation tool.
\newblock In \emph{Advanced Information Systems Engineering: Third
  International Conference CAiSE'91 Trondheim, Norway, May 13--15, 1991
  Proceedings 3}, pages 284--306. Springer, 1991.

\bibitem[Traor{\'e} and Muzy(2006)]{traore2006capturing}
Mamadou~K Traor{\'e} and Alexandre Muzy.
\newblock Capturing the dual relationship between simulation models and their
  context.
\newblock \emph{Simulation Modelling Practice and Theory}, 14\penalty0
  (2):\penalty0 126--142, 2006.

\bibitem[Van~Mierlo et~al.(2020)Van~Mierlo, Oakes, Van~Acker, Eslampanah,
  Denil, and Vangheluwe]{van2020exploring}
Simon Van~Mierlo, Bentley~James Oakes, Bert Van~Acker, Raheleh Eslampanah,
  Joachim Denil, and Hans Vangheluwe.
\newblock Exploring validity frames in practice.
\newblock In \emph{Systems Modelling and Management: First International
  Conference, ICSMM 2020, Bergen, Norway, June 25--26, 2020}, pages 131--148.
  Springer, 2020.

\bibitem[Ribault and Wainer(2012)]{ribault2012using}
Judica{\"e}l Ribault and Gabriel Wainer.
\newblock Using workflows and web services to manage simulation studies (wip).
\newblock In \emph{Proceedings of the 2012 Symposium on Theory of Modeling and
  Simulation-DEVS Integrative M{\&}S Symposium}, pages 1--6, 2012.

\bibitem[Page et~al.(2012)Page, Litwin, McMahon, Wickham, Shadid, and
  Chang]{page2012goal}
Ernest~H Page, Laurie Litwin, Matthew~T McMahon, Brian Wickham, Mike Shadid,
  and Elizabeth Chang.
\newblock Goal-directed grid-enabled computing for legacy simulations.
\newblock In \emph{2012 12th IEEE/ACM International Symposium on Cluster, Cloud
  and Grid Computing (ccgrid 2012)}, pages 873--879. IEEE, 2012.

\bibitem[{Cuevas-Vicentt{\'i}n} et~al.(2012){Cuevas-Vicentt{\'i}n}, Dey,
  K{\"o}hler, Riddle, and Lud{\"a}scher]{cuevas2012scientific}
V{\'i}ctor {Cuevas-Vicentt{\'i}n}, Saumen~C. Dey, Sven K{\"o}hler, Sean Riddle,
  and Bertram Lud{\"a}scher.
\newblock Scientific workflows and provenance: {{Introduction}} and research
  opportunities.
\newblock \emph{Datenbank-Spektrum}, 12\penalty0 (3):\penalty0 193--203, 2012.
\newblock \doi{10.1007/s13222-012-0100-z}.
\newblock URL \url{https://doi.org/10.1007/s13222-012-0100-z}.

\bibitem[Kahn and MacQueen(1976)]{kahn1976coroutines}
Gilles Kahn and David MacQueen.
\newblock Coroutines and networks of parallel processes.
\newblock 1976.

\bibitem[Lud{\"a}scher et~al.(2006)Lud{\"a}scher, Altintas, Berkley, Higgins,
  Jaeger, Jones, Lee, Tao, and Zhao]{ludascher2006scientific}
Bertram Lud{\"a}scher, Ilkay Altintas, Chad Berkley, Dan Higgins, Efrat Jaeger,
  Matthew Jones, Edward~A Lee, Jing Tao, and Yang Zhao.
\newblock Scientific workflow management and the kepler system.
\newblock \emph{Concurrency and Computation: Practice and Experience},
  18\penalty0 (10):\penalty0 1039--1065, 2006.

\bibitem[Oinn et~al.(2006)Oinn, Greenwood, Addis, Alpdemir, Ferris, Glover,
  Goble, Goderis, Hull, Marvin, Li, Lord, Pocock, Senger, Stevens, Wipat, and
  Wroe]{oinn2006taverna}
Tom Oinn, Mark Greenwood, Matthew Addis, M~Nedim Alpdemir, Justin Ferris, Kevin
  Glover, Carole Goble, Antoon Goderis, Duncan Hull, Darren Marvin, Peter Li,
  Phillip Lord, Matthew~R. Pocock, Martin Senger, Robert Stevens, Anil Wipat,
  and Chris Wroe.
\newblock Taverna: Lessons in creating a workflow environment for the life
  sciences.
\newblock \emph{Concurrency and Computation: Practice and Experience},
  18\penalty0 (10):\penalty0 1067--1100, 2006.

\bibitem[Leye et~al.(2009)Leye, Himmelspach, and Uhrmacher]{leye2009discussion}
Stefan Leye, Jan Himmelspach, and Adelinde~M Uhrmacher.
\newblock A discussion on experimental model validation.
\newblock In \emph{2009 11th International Conference on Computer Modelling and
  Simulation}, pages 161--167. IEEE, 2009.

\bibitem[Batt et~al.(2006)Batt, Bradley, Ewald, Fages, Hermans, Hillston,
  Kemper, Martens, Mosterman, Nielson, et~al.]{batt2006working}
Gregory Batt, JT~Bradley, Roland Ewald, Francois Fages, Holger Hermans, Jane
  Hillston, Peter Kemper, Alke Martens, Pieter Mosterman, Flemming Nielson,
  et~al.
\newblock Working groups’ report: The challenge of combining simulation and
  verification.
\newblock In \emph{Dagstuhl Seminar Proc. 06161: Simulation and Verification of
  Dynamic Systems}, 2006.

\bibitem[Clarke et~al.(1999)Clarke, Grumberg, and Peled]{Clarke1999}
Edmund~M. Clarke, Jr., Orna Grumberg, and Doron~A. Peled.
\newblock \emph{Model Checking}.
\newblock MIT Press, Cambridge, MA, USA, 1999.
\newblock ISBN 0-262-03270-8.

\bibitem[Donz{\'e} et~al.(2013)Donz{\'e}, Ferrere, and
  Maler]{donze2013efficient}
Alexandre Donz{\'e}, Thomas Ferrere, and Oded Maler.
\newblock Efficient robust monitoring for stl.
\newblock In \emph{International Conference on Computer Aided Verification},
  pages 264--279. Springer, 2013.

\bibitem[Nenzi et~al.(2018)Nenzi, Bortolussi, Ciancia, Loreti, and
  Massink]{nenzi2018qualitative}
Laura Nenzi, Luca Bortolussi, Vincenzo Ciancia, Michele Loreti, and Mieke
  Massink.
\newblock Qualitative and quantitative monitoring of spatio-temporal properties
  with {SSTL}.
\newblock \emph{Logical Methods in Computer Science}, 14\penalty0 (4), 2018.
\newblock \doi{10.23638/LMCS-14(4:2)2018}.
\newblock URL \url{https://doi.org/10.23638/LMCS-14(4:2)2018}.

\bibitem[Brim et~al.(2014)Brim, Dluhos, Safr{\'{a}}nek, and
  Vejpustek]{BrimDSV14}
Lubos Brim, Petr Dluhos, David Safr{\'{a}}nek, and Tomas Vejpustek.
\newblock Stl{$^\ast$}: Extending signal temporal logic with signal-value
  freezing operator.
\newblock \emph{Information and Computation}, 236:\penalty0 52--67, 2014.
\newblock \doi{10.1016/j.ic.2014.01.012}.
\newblock URL \url{https://doi.org/10.1016/j.ic.2014.01.012}.

\bibitem[Donz{\'{e}} et~al.(2012)Donz{\'{e}}, Maler, Bartocci, Nickovic, Grosu,
  and Smolka]{DonzeMBNGS12}
Alexandre Donz{\'{e}}, Oded Maler, Ezio Bartocci, Dejan Nickovic, Radu Grosu,
  and Scott~A. Smolka.
\newblock On temporal logic and signal processing.
\newblock In Supratik Chakraborty and Madhavan Mukund, editors, \emph{Automated
  Technology for Verification and Analysis - 10th International Symposium,
  {ATVA} 2012, Thiruvananthapuram, India, October 3-6, 2012. Proceedings},
  volume 7561 of \emph{Lecture Notes in Computer Science}, pages 92--106.
  Springer, 2012.
\newblock \doi{10.1007/978-3-642-33386-6\_9}.
\newblock URL \url{https://doi.org/10.1007/978-3-642-33386-6\_9}.

\bibitem[Vissat et~al.(2019)Vissat, Loreti, Nenzi, Hillston, and
  Marion]{Vissat2019analysis}
Ludovica~Luisa Vissat, Michele Loreti, Laura Nenzi, Jane Hillston, and Glenn
  Marion.
\newblock Analysis of spatio-temporal properties of stochastic systems using
  tstl.
\newblock \emph{ACM Transactions on Modeling and Computer Simulation},
  29\penalty0 (4), 2019.
\newblock ISSN 1049-3301.
\newblock \doi{10.1145/3326168}.
\newblock URL \url{https://doi.org/10.1145/3326168}.

\bibitem[Fowler(2010)]{Fowler2010domain}
Martin Fowler.
\newblock \emph{Domain-Specific Languages}.
\newblock Pearson Education, 2010.

\bibitem[Maoz et~al.(2011)Maoz, Ringert, and Rumpe]{MRR11b}
Shahar Maoz, Jan~Oliver Ringert, and Bernhard Rumpe.
\newblock {CDDiff: Semantic Differencing for Class Diagrams}.
\newblock In Mira Mezini, editor, \emph{ECOOP 2011 - Object-Oriented
  Programming}, pages 230--254. Springer Berlin Heidelberg, 2011.
\newblock URL
  \url{https://se-rwth.de/publications/CDDiff-Semantic-Differencing-for-Class-Diagrams.pdf}.

\bibitem[Rumpe(2016)]{Rum16}
Bernhard Rumpe.
\newblock \emph{{Modeling with UML: Language, Concepts, Methods}}.
\newblock Springer International, July 2016.
\newblock URL \url{https://mbse.se-rwth.de/}.

\bibitem[Evans et~al.(1999)Evans, France, Lano, and Rumpe]{EFLR99}
Andy Evans, Robert France, Kevin Lano, and Bernhard Rumpe.
\newblock {Meta-Modelling Semantics of UML}.
\newblock In H.~Kilov, B.~Rumpe, and I.~Simmonds, editors, \emph{Behavioral
  Specifications of Businesses and Systems}, pages 45--60. Kluver Academic
  Publisher, 1999.
\newblock ISBN 978-1-4613-7383-4.

\bibitem[Evans et~al.(1998)Evans, Bruel, France, Lano, and Rumpe]{EBF+98}
Andy Evans, Jean-Michel Bruel, Robert France, Kevin Lano, and Bernhard Rumpe.
\newblock {Making UML Precise}.
\newblock In \emph{OOPSLA'98 Workshop on ``Formalizing UML. Why and How?''},
  Vancouver, Canada, October 1998.

\bibitem[Bocciarelli et~al.(2019)Bocciarelli, D’Ambrogio, Falcone, Garro, and
  Giglio]{Bocetal19}
Paolo Bocciarelli, Andrea D’Ambrogio, Alberto Falcone, Alfredo Garro, and
  Andrea Giglio.
\newblock A model-driven approach to enable the simulation of complex systems
  on distributed architectures.
\newblock \emph{SIMULATION}, 95\penalty0 (12):\penalty0 1185--1211, 2019.
\newblock \doi{10.1177/0037549719829828}.
\newblock URL \url{https://doi.org/10.1177/0037549719829828}.

\bibitem[Casse(2017)]{CASSE20171}
Olivier Casse.
\newblock \emph{SysML in Action with Cameo Systems Modeler}.
\newblock Elsevier, 2017.
\newblock ISBN 978-1-78548-171-0.

\bibitem[Nikolaidou et~al.(2015)Nikolaidou, Kapos, Tsadimas, Dalakas, and
  Anagnostopoulos]{NKT+15}
Mara Nikolaidou, George-Dimitrios Kapos, Anargyros Tsadimas, Vassilis Dalakas,
  and Dimosthenis Anagnostopoulos.
\newblock Simulating sysml models: Overview and challenges.
\newblock In \emph{2015 10th System of Systems Engineering Conference (SoSE)},
  2015.
\newblock \doi{10.1109/SYSOSE.2015.7151961}.

\bibitem[Tolk et~al.(2013)Tolk, Diallo, Padilla, and
  Herencia-Zapana]{tolk2013reference}
Andreas Tolk, Saikou~Y Diallo, Jose~J Padilla, and Heber Herencia-Zapana.
\newblock Reference modelling in support of m\&s—foundations and
  applications.
\newblock \emph{Journal of Simulation}, 7:\penalty0 69--82, 2013.

\bibitem[Clark et~al.(2015)Clark, Brand, Combemale, and Rumpe]{CBCR15}
Tony Clark, Mark van~den Brand, Benoit Combemale, and Bernhard Rumpe.
\newblock {Conceptual Model of the Globalization for Domain-Specific
  Languages}.
\newblock In \emph{{Globalizing Domain-Specific Languages}}, LNCS 9400, pages
  7--20. Springer, 2015.
\newblock URL
  \url{http://www.se-rwth.de/publications/Conceptual-Model-of-the-Globalization-for-Domain-Specific-Languages.pdf}.

\bibitem[Faeder et~al.(2009)Faeder, Blinov, and Hlavacek]{faeder2009rule}
James~R Faeder, Michael~L Blinov, and William~S Hlavacek.
\newblock Rule-based modeling of biochemical systems with bionetgen.
\newblock In \emph{Systems Biology}, pages 113--167. Springer, 2009.

\bibitem[Hoops et~al.(2006)Hoops, Sahle, Gauges, Lee, Pahle, Simus, Singhal,
  Xu, Mendes, and Kummer]{hoops2006copasi}
Stefan Hoops, Sven Sahle, Ralph Gauges, Christine Lee, J{\"u}rgen Pahle,
  Natalia Simus, Mudita Singhal, Liang Xu, Pedro Mendes, and Ursula Kummer.
\newblock Copasi—a complex pathway simulator.
\newblock \emph{Bioinformatics}, 22\penalty0 (24):\penalty0 3067--3074, 2006.

\bibitem[Blinov et~al.(2017)Blinov, Schaff, Vasilescu, Moraru, Bloom, and
  Loew]{blinov2017compartmental}
Michael~L Blinov, James~C Schaff, Dan Vasilescu, Ion~I Moraru, Judy~E Bloom,
  and Leslie~M Loew.
\newblock Compartmental and spatial rule-based modeling with virtual cell.
\newblock \emph{Biophysical Journal}, 113\penalty0 (7):\penalty0 1365--1372,
  2017.

\bibitem[Ciocchetta and Hillston(2009)]{ciocchetta2009bio}
Federica Ciocchetta and Jane Hillston.
\newblock Bio-pepa: A framework for the modelling and analysis of biological
  systems.
\newblock \emph{Theoretical Computer Science}, 410\penalty0 (33-34):\penalty0
  3065--3084, 2009.

\bibitem[Gilmore and Hillston(1994)]{gilmore1994pepa}
Stephen Gilmore and Jane Hillston.
\newblock The pepa workbench: A tool to support a process algebra-based
  approach to performance modelling.
\newblock \emph{Computer Performance Evaluation}, 794:\penalty0 353--368, 1994.

\bibitem[Abar et~al.(2017)Abar, Theodoropoulos, Lemarinier, and
  O’Hare]{ABAR2017}
Sameera Abar, Georgios~K. Theodoropoulos, Pierre Lemarinier, and Gregory~M.P.
  O’Hare.
\newblock Agent based modelling and simulation tools: A review of the
  state-of-art software.
\newblock \emph{Computer Science Review}, 24:\penalty0 13--33, 2017.
\newblock ISSN 1574-0137.
\newblock \doi{https://doi.org/10.1016/j.cosrev.2017.03.001}.
\newblock URL
  \url{https://www.sciencedirect.com/science/article/pii/S1574013716301198}.

\bibitem[Zeigler and Sarjoughian(2005)]{devs1}
Bernard~P. Zeigler and Hessam Sarjoughian.
\newblock {Introduction to DEVS Modeling and Simulation with JAVA: Developing
  Component-based Simulation Models}.
\newblock Technical report, Arizona Center of Integrative Modeling and
  Simulation, University of Arizona, 2005.

\bibitem[Wainer et~al.(2001)Wainer, Christen, and
  Dobniewski]{wainer2001defining}
Gabriel Wainer, Gast{\'o}n Christen, and Alejandro Dobniewski.
\newblock Defining devs models with the cd++ toolkit.
\newblock In \emph{Proceedings of SCS European Simulation Symposium}, 2001.

\bibitem[North et~al.(2006)North, Collier, and Vos]{north2006experiences}
Michael~J North, Nicholson~T Collier, and Jerry~R Vos.
\newblock Experiences creating three implementations of the repast agent
  modeling toolkit.
\newblock \emph{ACM Transactions on Modeling and Computer Simulation (TOMACS)},
  16\penalty0 (1):\penalty0 1--25, 2006.

\bibitem[Luke et~al.(2005)Luke, Cioffi-Revilla, Panait, Sullivan, and
  Balan]{luke2005mason}
Sean Luke, Claudio Cioffi-Revilla, Liviu Panait, Keith Sullivan, and Gabriel
  Balan.
\newblock Mason: A multiagent simulation environment.
\newblock \emph{Simulation}, 81\penalty0 (7):\penalty0 517--527, 2005.

\bibitem[Howell and McNab(1998)]{howell1998simjava}
Fred Howell and Ross McNab.
\newblock Simjava: A discrete event simulation library for java.
\newblock \emph{Simulation Series}, 30:\penalty0 51--56, 1998.

\bibitem[McNab and Howell(1996)]{mcnab1996using}
Ross McNab and Fred Howell.
\newblock Using java for discrete event simulation.
\newblock In \emph{Proceedings of the Twelfth UK Computer and
  Telecommunications Performance Engineering Workshop}, pages 219--228.
  Citeseer, 1996.

\bibitem[Bersini(2012)]{bersini2012uml}
Hugues Bersini.
\newblock Uml for abm.
\newblock \emph{Journal of Artificial Societies and Social Simulation},
  15\penalty0 (1):\penalty0 9, 2012.

\bibitem[Wagner(2014)]{Wagner2012UML}
Gerd Wagner.
\newblock Tutorial: Information and process modeling for simulation.
\newblock In \emph{Proceedings of the 2014 Winter Simulation Conference}, WSC
  '14, page 103–117. IEEE Press, 2014.

\bibitem[Warnke et~al.(2016)Warnke, Reinhardt, and
  Uhrmacher]{warnke2016population}
Tom Warnke, Oliver Reinhardt, and Adelinde~M Uhrmacher.
\newblock Population-based ctmcs and agent-based models.
\newblock In \emph{2016 Winter Simulation Conference (WSC)}, pages 1253--1264.
  IEEE, 2016.

\bibitem[de~Boer et~al.(2017)de~Boer, Din, Fernandez-Reyes, H{\"a}hnle, Henrio,
  Johnsen, Khamespanah, Rochas, Serbanescu, Sirjani, and Yang]{ActiveObjects17}
Frank de~Boer, Crystal~Chang Din, Kiko Fernandez-Reyes, Reiner H{\"a}hnle,
  Ludovic Henrio, Einar~Broch Johnsen, Ehsan Khamespanah, Justine Rochas, Vlad
  Serbanescu, Marjan Sirjani, and Albert~Mingkun Yang.
\newblock A survey of active object languages.
\newblock \emph{ACM Computing Surveys}, 50\penalty0 (5):\penalty0 76:1--76:39,
  oct 2017.
\newblock Article 76.

\bibitem[De~Boer et~al.(2024)De~Boer, Damiani, Hähnle, Johnsen, and uard
  Kamburjan]{ABS-SoTA23}
Frank De~Boer, Ferruccio Damiani, Reiner Hähnle, Einar~Broch Johnsen, and
  Ed~uard Kamburjan, editors.
\newblock \emph{Active Object Languages: Current Research Trends}, volume 14360
  of \emph{LNCS}.
\newblock Springer, Cham, 2024.
\newblock ISBN 978-3-031-51059-5.

\bibitem[Schlatte et~al.(2022)Schlatte, Johnsen, Kamburjan, and Tarifa]{SJKT22}
Rudolf Schlatte, Einar~Broch Johnsen, Eduard Kamburjan, and Silvia Lizeth~Tapia
  Tarifa.
\newblock The {ABS} simulator toolchain.
\newblock \emph{Science of Computer Programming}, 223:\penalty0 102861, 2022.
\newblock \doi{10.1016/j.scico.2022.102861}.

\bibitem[Kamburjan et~al.(2018)Kamburjan, H{\"{a}}hnle, and Sch{\"{o}}n]{KHS18}
Eduard Kamburjan, Reiner H{\"{a}}hnle, and Sebastian Sch{\"{o}}n.
\newblock Formal modeling and analysis of railway operations with active
  objects.
\newblock \emph{Science of Computer Programming}, 166:\penalty0 167--193, 2018.
\newblock \doi{10.1016/j.scico.2018.07.001}.
\newblock URL \url{https://doi.org/10.1016/j.scico.2018.07.001}.

\bibitem[Emmanuel et~al.(2021)Emmanuel, Moy, Henrio, and Pichon]{EMHP21}
Julien Emmanuel, Matthieu Moy, Ludovic Henrio, and Gr{\'{e}}goire Pichon.
\newblock {S4BXI:} the mpi-ready portals 4 simulator.
\newblock In \emph{29th International Symposium on Modeling, Analysis, and
  Simulation of Computer and Telecommunication Systems, {MASCOTS} 2021,
  Houston, TX, USA, November 3-5, 2021}, pages 1--8. {IEEE}, 2021.
\newblock \doi{10.1109/MASCOTS53633.2021.9614285}.
\newblock URL \url{https://doi.org/10.1109/MASCOTS53633.2021.9614285}.

\bibitem[Turin et~al.(2020)Turin, Borgarelli, Donetti, Johnsen, Tarifa, and
  Damiani]{TBDJTD20}
Gianluca Turin, Andrea Borgarelli, Simone Donetti, Einar~Broch Johnsen, Silvia
  Lizeth~Tapia Tarifa, and Ferruccio Damiani.
\newblock A formal model of the kubernetes container framework.
\newblock In Tiziana Margaria and Bernhard Steffen, editors, \emph{Leveraging
  Applications of Formal Methods, Verification and Validation: Verification
  Principles - 9th International Symposium on Leveraging Applications of Formal
  Methods, ISoLA 2020, Rhodes, Greece, October 20-30, 2020, Proceedings, Part
  {I}}, volume 12476 of \emph{Lecture Notes in Computer Science}, pages
  558--577. Springer, 2020.
\newblock \doi{10.1007/978-3-030-61362-4\_32}.
\newblock URL \url{https://doi.org/10.1007/978-3-030-61362-4\_32}.

\bibitem[Johnsen et~al.(2012)Johnsen, Schlatte, and Tarifa]{JST12}
Einar~Broch Johnsen, Rudolf Schlatte, and Silvia Lizeth~Tapia Tarifa.
\newblock Modeling resource-aware virtualized applications for the cloud in
  real-time {ABS}.
\newblock In Toshiaki Aoki and Kenji Taguchi, editors, \emph{Formal Methods and
  Software Engineering - 14th International Conference on Formal Engineering
  Methods, {ICFEM} 2012, Kyoto, Japan, November 12-16, 2012. Proceedings},
  volume 7635 of \emph{Lecture Notes in Computer Science}, pages 71--86.
  Springer, 2012.
\newblock \doi{10.1007/978-3-642-34281-3\_8}.
\newblock URL \url{https://doi.org/10.1007/978-3-642-34281-3\_8}.

\bibitem[Khamespanah et~al.(2018)Khamespanah, Khosravi, and Sirjani]{KKS18}
Ehsan Khamespanah, Ramtin Khosravi, and Marjan Sirjani.
\newblock An efficient {TCTL} model checking algorithm and a reduction
  technique for verification of timed actor models.
\newblock \emph{Science of Computer Programming}, 153:\penalty0 1--29, 2018.
\newblock \doi{10.1016/j.scico.2017.11.004}.
\newblock URL \url{https://doi.org/10.1016/j.scico.2017.11.004}.

\bibitem[Sirjani(2019)]{Sirjani19}
Marjan Sirjani.
\newblock Analysing real-time distributed systems using timed actors.
\newblock In Floriano~De Rango, Carlos~T. Calafate, Miroslav Vozn{\'{a}}k,
  Alfredo Garro, and Mauro Tropea, editors, \emph{23rd {IEEE/ACM} International
  Symposium on Distributed Simulation and Real Time Applications {DS-RT} 2019,
  Cosenza, Italy, October 7-9, 2019}, page~1. {IEEE}, 2019.
\newblock \doi{10.1109/DS-RT47707.2019.8958670}.
\newblock URL \url{https://doi.org/10.1109/DS-RT47707.2019.8958670}.

\bibitem[Din et~al.(2015)Din, Bubel, and H{\"{a}}hnle]{DBH15}
Crystal~Chang Din, Richard Bubel, and Reiner H{\"{a}}hnle.
\newblock Key-abs: {A} deductive verification tool for the concurrent modelling
  language {ABS}.
\newblock In Amy~P. Felty and Aart Middeldorp, editors, \emph{Automated
  Deduction - {CADE-25} - 25th International Conference on Automated Deduction,
  Berlin, Germany, August 1-7, 2015, Proceedings}, volume 9195 of \emph{Lecture
  Notes in Computer Science}, pages 517--526. Springer, 2015.
\newblock \doi{10.1007/978-3-319-21401-6\_35}.
\newblock URL \url{https://doi.org/10.1007/978-3-319-21401-6\_35}.

\bibitem[Kamburjan et~al.(2020)Kamburjan, Din, H{\"{a}}hnle, and
  Johnsen]{KDHJ20}
Eduard Kamburjan, Crystal~Chang Din, Reiner H{\"{a}}hnle, and Einar~Broch
  Johnsen.
\newblock Behavioral contracts for cooperative scheduling.
\newblock In Wolfgang Ahrendt, Bernhard Beckert, Richard Bubel, Reiner
  H{\"a}hnle, and Mattias Ulbrich, editors, \emph{Deductive Software
  Verification: Future Perspectives}, volume 12345 of \emph{LNCS}, pages
  85--121. Springer, Cham, 2020.

\bibitem[Giachino et~al.(2016)Giachino, Laneve, and Lienhardt]{GLL16}
Elena Giachino, Cosimo Laneve, and Michael Lienhardt.
\newblock A framework for deadlock detection in core {ABS}.
\newblock \emph{Software and Systems Modeling}, 15\penalty0 (4):\penalty0
  1013--1048, 2016.
\newblock \doi{10.1007/s10270-014-0444-y}.

\bibitem[Albert et~al.(2015)Albert, Fern{\'{a}}ndez, Puebla, and
  Rom{\'{a}}n{-}D{\'{\i}}ez]{AFPR15}
Elvira Albert, Jes{\'{u}}s~Correas Fern{\'{a}}ndez, Germ{\'{a}}n Puebla, and
  Guillermo Rom{\'{a}}n{-}D{\'{\i}}ez.
\newblock Quantified abstract configurations of distributed systems.
\newblock \emph{Formal Aspects of Computing}, 27\penalty0 (4):\penalty0
  665--699, 2015.
\newblock \doi{10.1007/s00165-014-0321-z}.

\bibitem[Bubel et~al.(2014)Bubel, Flores{-}Montoya, and H{\"{a}}hnle]{BMH14}
Richard Bubel, Antonio Flores{-}Montoya, and Reiner H{\"{a}}hnle.
\newblock Analysis of executable software models.
\newblock In Marco Bernardo, Ferruccio Damiani, Reiner H{\"{a}}hnle,
  Einar~Broch Johnsen, and Ina Schaefer, editors, \emph{Formal Methods for
  Executable Software Models: 14th Intl.\ School on Formal Methods for the
  Design of Computer, Communication, and Software Systems, {SFM}, Bertinoro,
  Italy}, volume 8483 of \emph{LNCS}, pages 1--25. Springer, 2014.
\newblock \doi{10.1007/978-3-319-07317-0\_1}.

\bibitem[Henrio and Rochas(2017)]{HenrioRochas17}
Ludovic Henrio and Justine Rochas.
\newblock Multiactive objects and their applications.
\newblock \emph{Logical Methods in Computer Science}, 13\penalty0 (4), 2017.
\newblock \doi{10.23638/LMCS-13(4:12)2017}.
\newblock URL \url{https://doi.org/10.23638/LMCS-13(4:12)2017}.

\bibitem[Matloff(2008)]{matloff2008introduction}
Norm Matloff.
\newblock Introduction to discrete-event simulation and the simpy language.
\newblock \emph{Davis, CA. Dept of Computer Science. University of California
  at Davis.}, 2\penalty0 (2009):\penalty0 1--33, 2008.

\bibitem[Vasudevan et~al.(2021)Vasudevan, Zafar, Xingran, Singh, and van~der
  Aalst]{pourbafrani2021python}
Sandhya Vasudevan, Faizan Zafar, Yuan Xingran, Ravikumar Singh, and Wil~MP
  van~der Aalst.
\newblock A python extension to simulate petri nets in process mining.
\newblock \emph{arXiv preprint arXiv:2102.08774}, 2021.

\bibitem[Van~Tendeloo and Vangheluwe(2014)]{van2014modular}
Yentl Van~Tendeloo and Hans Vangheluwe.
\newblock The modular architecture of the python (p) devs simulation kernel.
\newblock In \emph{Proceedings of the 2014 Symposium on Theory of Modeling and
  Simulation-DEVS}, pages 387--392, 2014.

\bibitem[Douglas-Smith et~al.(2020)Douglas-Smith, Iwanaga, Croke, and
  Jakeman]{douglas2020certain}
Dominique Douglas-Smith, Takuya Iwanaga, Barry~FW Croke, and Anthony~J Jakeman.
\newblock Certain trends in uncertainty and sensitivity analysis: An overview
  of software tools and techniques.
\newblock \emph{Environmental Modelling \& Software}, 124:\penalty0 104588,
  2020.

\bibitem[Peyman et~al.(2021)Peyman, Copado, Panadero, Juan, and
  Dehghanimohammadabadi]{Pythonwsc2021}
Mohammad Peyman, Pedro Copado, Javier Panadero, Angel~A. Juan, and Mohammad
  Dehghanimohammadabadi.
\newblock A tutorial on how to connect python with different simulation
  software to develop rich simheuristics.
\newblock In \emph{2021 Winter Simulation Conference (WSC)}, pages 1--12, 2021.
\newblock \doi{10.1109/WSC52266.2021.9715511}.

\bibitem[Peng et~al.(2016)Peng, Warnke, Haack, and Uhrmacher]{PengWHU16}
Danhua Peng, Tom Warnke, Fiete Haack, and Adelinde~M. Uhrmacher.
\newblock Reusing simulation experiment specifications to support developing
  models by successive extension.
\newblock \emph{Simulation Modelling Practice and Theory}, 68:\penalty0 33--53,
  2016.
\newblock \doi{10.1016/j.simpat.2016.07.006}.
\newblock URL \url{https://doi.org/10.1016/j.simpat.2016.07.006}.

\bibitem[Wilsdorf et~al.(2023{\natexlab{a}})Wilsdorf, Wolpers, Hilton, Haack,
  and Uhrmacher]{wilsdorf2022automatic}
Pia Wilsdorf, Anja Wolpers, Jason Hilton, Fiete Haack, and Adelinde Uhrmacher.
\newblock Automatic reuse, adaption, and execution of simulation experiments
  via provenance patterns.
\newblock \emph{ACM Transactions on Modeling and Computer Simulation},
  33\penalty0 (1–2), feb 2023{\natexlab{a}}.
\newblock ISSN 1049-3301.
\newblock \doi{10.1145/3564928}.
\newblock URL \url{https://doi.org/10.1145/3564928}.

\bibitem[Ludäscher et~al.(2009)Ludäscher, Weske, McPhillips, and
  Bowers]{LudascherWMB09}
Bertram Ludäscher, Mathias Weske, Timothy~M. McPhillips, and Shawn Bowers.
\newblock Scientific workflows: {Business} as usual?
\newblock In Umeshwar Dayal, Johann Eder, Jana Koehler, and Hajo~A. Reijers,
  editors, \emph{Business Process Management (BPM), 7th Intl. Conference},
  volume 5701 of \emph{LNCS}, pages 31--47, Ulm, Germany, 2009. Springer.
\newblock \doi{10.1007/978-3-642-03848-8\_4}.
\newblock URL \url{https://doi.org/10.1007/978-3-642-03848-8_4}.

\bibitem[McPhillips et~al.(2009)McPhillips, Bowers, Zinn, and
  Lud{\"a}scher]{McPhillips2009workflows}
Timothy~M. McPhillips, Shawn Bowers, Daniel Zinn, and Bertram Lud{\"a}scher.
\newblock Scientific workflow design for mere mortals.
\newblock \emph{Future Generation Computer Systems}, 25\penalty0 (5):\penalty0
  541--551, 2009.
\newblock \doi{10.1016/j.future.2008.06.013}.
\newblock URL \url{https://doi.org/10.1016/j.future.2008.06.013}.

\bibitem[Zinn et~al.(2009)Zinn, Bowers, McPhillips, and Ludäscher]{ZinnBML09}
Daniel Zinn, Shawn Bowers, Timothy~M. McPhillips, and Bertram Ludäscher.
\newblock Scientific workflow design with data assembly lines.
\newblock In Ewa Deelman and Ian~J. Taylor, editors, \emph{4th Workshop on
  Workflows in Support of Large-Scale Scienc (WORKS)}. ACM, 2009.
\newblock \doi{10.1145/1645164.1645178}.
\newblock URL \url{https://doi.org/10.1145/1645164.1645178}.

\bibitem[Crusoe et~al.(2022)Crusoe, Abeln, Iosup, Amstutz, Chilton, Tijanić,
  Ménager, Soiland-Reyes, Gavrilović, Goble, and
  Community]{crusoe_methods_2022}
Michael~R. Crusoe, Sanne Abeln, Alexandru Iosup, Peter Amstutz, John Chilton,
  Nebojša Tijanić, Hervé Ménager, Stian Soiland-Reyes, Bogdan Gavrilović,
  Carole Goble, and The~CWL Community.
\newblock Methods included: Standardizing computational reuse and portability
  with the {Common} {Workflow} {Language}.
\newblock \emph{Communications of the ACM}, 65\penalty0 (6):\penalty0 54--63,
  May 2022.
\newblock ISSN 0001-0782.
\newblock \doi{10.1145/3486897}.
\newblock URL \url{https://dl.acm.org/doi/10.1145/3486897}.

\bibitem[Otter et~al.(2015)Otter, Thuy, Bouskela, Buffoni, Elmqvist, Fritzson,
  Garro, Jardin, Olsson, Payelleville, Schamai, Thomas, and Tundis]{dlr99941}
Martin Otter, Nguyen Thuy, Daniel Bouskela, Lena Buffoni, Hilding Elmqvist,
  Peter Fritzson, Alfredo Garro, Audrey Jardin, Hans Olsson, Maxime
  Payelleville, Wladimir Schamai, Eric Thomas, and Andrea Tundis.
\newblock Formal requirements modeling for simulation-based verification.
\newblock In \emph{11th International Modelica Conference}, 2015.

\bibitem[Wilsdorf et~al.(2023{\natexlab{b}})Wilsdorf, Zuska, Andelfinger, and
  Uhrmacher]{Wilsdorf2023}
Pia Wilsdorf, Marian Zuska, Philipp Andelfinger, and Adelinde~M. Uhrmacher.
\newblock Validation without data - formalizing stylized facts of time series.
\newblock In \emph{Winter Simulation Conference (WSC 2023)}. IEEE Press,
  2023{\natexlab{b}}.
\newblock accepted.

\bibitem[Kwiatkowska et~al.(2005)Kwiatkowska, Norman, and
  Parker]{KwiatkowskaNP05}
Marta~Z. Kwiatkowska, Gethin Norman, and David Parker.
\newblock Probabilistic model checking in practice: Case studies with {PRISM}.
\newblock \emph{{SIGMETRICS} Performance Evaluation Review}, 32\penalty0
  (4):\penalty0 16--21, 2005.
\newblock \doi{10.1145/1059816.1059820}.
\newblock URL \url{https://doi.org/10.1145/1059816.1059820}.

\bibitem[Bortolussi et~al.(2016)Bortolussi, Milios, and
  Sanguinetti]{BortolussiMS16}
Luca Bortolussi, Dimitrios Milios, and Guido Sanguinetti.
\newblock Smoothed model checking for uncertain continuous-time markov chains.
\newblock \emph{Information and Computation}, 247:\penalty0 235--253, 2016.
\newblock \doi{10.1016/j.ic.2016.01.004}.
\newblock URL \url{https://doi.org/10.1016/j.ic.2016.01.004}.

\bibitem[Bartocci et~al.(2022)Bartocci, Mateis, Nesterini, and
  Nickovic]{BartocciMNN22}
Ezio Bartocci, Cristinel Mateis, Eleonora Nesterini, and Dejan Nickovic.
\newblock Survey on mining signal temporal logic specifications.
\newblock \emph{Information and Computation}, 289\penalty0 (Part):\penalty0
  104957, 2022.
\newblock \doi{https://doi.org/10.1016/j.ic.2022.104957}.

\bibitem[Paech and Rumpe(1994)]{PR94}
Barbara Paech and Bernhard Rumpe.
\newblock {A new Concept of Refinement used for Behaviour Modelling with
  Automata}.
\newblock In \emph{Proceedings of the Industrial Benefit of Formal Methods
  (FME'94)}, LNCS 873, pages 154--174. Springer, 1994.

\bibitem[Philipps and Rumpe(1997)]{PR97}
Jan Philipps and Bernhard Rumpe.
\newblock {Refinement of Information Flow Architectures}.
\newblock In M.~Hinchey, editor, \emph{ICFEM'97 Proceedings}, Hiroshima, Japan,
  1997. IEEE CS Press.

\bibitem[Sepúlveda et~al.(2016)Sepúlveda, Cravero, and
  Cachero]{SEPULVEDA201616}
Samuel Sepúlveda, Ania Cravero, and Cristina Cachero.
\newblock Requirements modeling languages for software product lines: A
  systematic literature review.
\newblock \emph{Information and Software Technology}, 69:\penalty0 16--36,
  2016.
\newblock ISSN 0950-5849.
\newblock \doi{https://doi.org/10.1016/j.infsof.2015.08.007}.
\newblock URL
  \url{https://www.sciencedirect.com/science/article/pii/S0950584915001494}.

\bibitem[Combemale et~al.(2016)Combemale, France, J{\'e}z{\'e}quel, Rumpe,
  Steel, and Vojtisek]{CFJ+16}
Benoit Combemale, Robert France, Jean-Marc J{\'e}z{\'e}quel, Bernhard Rumpe,
  James Steel, and Didier Vojtisek.
\newblock \emph{{Engineering Modeling Languages: Turning Domain Knowledge into
  Tools}}.
\newblock Chapman \& Hall/CRC Innovations in Software Engineering and Software
  Development Series, November 2016.

\bibitem[Cengarle et~al.(2009)Cengarle, Gr{\"o}nniger, and Rumpe]{CGR09}
Mar{\'i}a~Victoria Cengarle, Hans Gr{\"o}nniger, and Bernhard Rumpe.
\newblock {Variability within Modeling Language Definitions}.
\newblock In \emph{Conference on Model Driven Engineering Languages and Systems
  (MODELS'09)}, LNCS 5795, pages 670--684. Springer, 2009.
\newblock URL
  \url{http://www.se-rwth.de/publications/Variability-within-Modeling-Language-Definitions.pdf}.

\bibitem[H{\"o}lldobler et~al.(2021)H{\"o}lldobler, Kautz, and Rumpe]{HKR21}
Katrin H{\"o}lldobler, Oliver Kautz, and Bernhard Rumpe.
\newblock \emph{{MontiCore Language Workbench and Library Handbook: Edition
  2021}}.
\newblock {Aachener Informatik-Berichte, Software Engineering, Band 48}. Shaker
  Verlag, May 2021.
\newblock ISBN 978-3-8440-8010-0.
\newblock URL \url{http://www.monticore.de/handbook.pdf}.

\bibitem[Clark et~al.(2004)Clark, Evans, Sammut, and Willans]{CESW04}
Tony Clark, Andy Evans, Paul Sammut, and James Willans.
\newblock An executable metamodelling facility for domain specific language
  design.
\newblock In \emph{Proc. 4th OOPSLA Workshop on Domain-Specific Modeling},
  2004.

\bibitem[Atkinson and Kuhne(2003)]{AK03}
Colin Atkinson and Thomas Kuhne.
\newblock Model-driven development: a metamodeling foundation.
\newblock \emph{IEEE Software}, 20\penalty0 (5), 2003.
\newblock ISSN 0740-7459.

\bibitem[Kleppe(2008)]{Kle08}
Anneke Kleppe.
\newblock \emph{Software Language Engineering: Creating Domain-Specific
  Languages using Metamodels}.
\newblock Pearson Education, 2008.

\bibitem[Zschaler et~al.(2009)Zschaler, Kolovos, Drivalos, Paige, and
  Rashid]{ZKD+09}
Steffen Zschaler, Dimitrios~S Kolovos, Nikolaos Drivalos, Richard~F Paige, and
  Awais Rashid.
\newblock Domain-specific metamodelling languages for software language
  engineering.
\newblock In \emph{International Conference on Software Language Engineering},
  pages 334--353. Springer, 2009.

\bibitem[{O}bject~{M}anagement {G}roup(2008)]{OMG08b}
{O}bject~{M}anagement {G}roup.
\newblock {Meta Object Facility (MOF) 2.0 Query/View/Transformation
  Specification}, April 2008.
\newblock URL \url{https://www.omg.org/spec/QVT/1.3/PDF}.
\newblock Accessed: 2023-09-30.

\bibitem[Kapos et~al.(2021)Kapos, Tsadimas, Kotronis, Dalakas, Nikolaidou, and
  Anagnostopoulos]{kapos2019declarative}
George-Dimitrios Kapos, Anargyros Tsadimas, Christos Kotronis, Vassilis
  Dalakas, Mara Nikolaidou, and Dimosthenis Anagnostopoulos.
\newblock A declarative approach for transforming sysml models to executable
  simulation models.
\newblock \emph{IEEE Transactions on Systems, Man, and Cybernetics: Systems},
  51\penalty0 (6):\penalty0 3330--3345, 2021.
\newblock \doi{10.1109/TSMC.2019.2922153}.

\bibitem[Cetinkaya et~al.(2012)Cetinkaya, Verbraeck, and
  Seck]{Cetinkaya2012model}
Deniz Cetinkaya, Alexander Verbraeck, and Mamadou~D. Seck.
\newblock Model transformation from bpmn to devs in the mdd4ms framework.
\newblock In \emph{Proceedings of the 2012 Symposium on Theory of Modeling and
  Simulation - DEVS Integrative M{\&}S Symposium}, TMS/DEVS '12, San Diego, CA,
  USA, 2012. Society for Computer Simulation International.
\newblock ISBN 9781618397867.

\bibitem[Sanders et~al.(2003)Sanders, Courtney, Deavours, Daly, Derisavi, and
  Lam]{Sanders2003}
William~H Sanders, Tod Courtney, Daniel Deavours, David Daly, Salem Derisavi,
  and Vinh Lam.
\newblock Multi-formalism and multi-solution-method modeling frameworks: The
  m{\"o}bius approach.
\newblock In \emph{Proc. Symp. Performance Evaluation - Stories and
  Perspectives}, pages 241--256, 2003.

\bibitem[de~Lara et~al.(2004)de~Lara, Vangheluwe, and
  Alfonseca]{DeLara2004meta}
Juan de~Lara, Hans Vangheluwe, and Manuel Alfonseca.
\newblock Meta-modelling and graph grammars for multi-paradigm modelling in
  {AToM3}.
\newblock \emph{Software \& Systems Modeling}, 3\penalty0 (3):\penalty0
  194--209, August 2004.
\newblock ISSN 1619-1374.
\newblock \doi{10.1007/s10270-003-0047-5}.
\newblock URL \url{https://doi.org/10.1007/s10270-003-0047-5}.

\bibitem[Mellor et~al.(2002)Mellor, Scott, Uhl, and Weise]{mellor2002model}
Stephen~J Mellor, Kendall Scott, Axel Uhl, and Dirk Weise.
\newblock Model-driven architecture.
\newblock In \emph{International Conference on Object-Oriented Information
  Systems}, pages 290--297. Springer, 2002.

\bibitem[Teran-Somohano et~al.(2015)Teran-Somohano, Smith, Ledet, Yilmaz, and
  Oğuztüzün]{Teran-Somohano2015model}
Alejandro Teran-Somohano, Alice~E. Smith, Joseph Ledet, Levent Yilmaz, and
  Halit Oğuztüzün.
\newblock A model-driven engineering approach to simulation experiment design
  and execution.
\newblock In \emph{2015 Winter Simulation Conference (WSC)}, pages 2632--2643,
  2015.
\newblock \doi{10.1109/WSC.2015.7408371}.

\bibitem[Zschaler and Polack(2020)]{zschaler2020family}
Steffen Zschaler and Fiona A.~C. Polack.
\newblock A family of languages for trustworthy agent-based simulation.
\newblock In \emph{Proceedings of the 13th {ACM} {SIGPLAN} {International}
  {Conference} on {Software} {Language} {Engineering}}, {SLE} 2020, pages
  16--21, New York, NY, USA, November 2020. Association for Computing
  Machinery.
\newblock \doi{10.1145/3426425.3426929}.

\bibitem[Kelly and Weaver(2004)]{kelly2004goal}
Tim Kelly and Rob Weaver.
\newblock The goal structuring notation–a safety argument notation.
\newblock In \emph{Proceedings of the dependable systems and networks 2004
  workshop on assurance cases}, volume~6. Citeseer, 2004.

\bibitem[Robinson(2008)]{robinson2008conceptual}
Stewart Robinson.
\newblock Conceptual modelling for simulation part i: Definition and
  requirements.
\newblock \emph{Journal of the Operational Research Society}, 59\penalty0
  (3):\penalty0 278--290, 2008.

\bibitem[Robinson et~al.(2015)Robinson, Arbez, Birta, Tolk, and
  Wagner]{robinson2015}
Stewart Robinson, Gilbert Arbez, Louis~G. Birta, Andreas Tolk, and Gerd Wagner.
\newblock Conceptual modeling: Definition, purpose and benefits.
\newblock In \emph{2015 Winter Simulation Conference (WSC)}, pages 2812--2826,
  2015.
\newblock \doi{10.1109/WSC.2015.7408386}.

\bibitem[Fujimoto et~al.(2017)Fujimoto, Bock, Chen, Page, and
  Panchal]{fujimoto2017research}
Richard Fujimoto, Conrad Bock, Wei Chen, Ernest Page, and Jitesh~H Panchal.
\newblock \emph{Research Challenges in Modeling and Simulation for Engineering
  Complex Systems}.
\newblock Springer, 2017.

\bibitem[Wilsdorf et~al.(2020)Wilsdorf, Haack, and
  Uhrmacher]{wilsdorf2020conceptual}
Pia Wilsdorf, Fiete Haack, and Adelinde~M Uhrmacher.
\newblock Conceptual models in simulation studies: Making it explicit.
\newblock In \emph{2020 Winter Simulation Conference (WSC)}, pages 2353--2364.
  IEEE, 2020.

\bibitem[Haack et~al.(2020)Haack, Budde, and Uhrmacher]{Haack2020}
Fiete Haack, Kai Budde, and Adelinde~M. Uhrmacher.
\newblock Exploring mechanistic and temporal regulation of lrp6 endocytosis in
  canonical wnt signaling.
\newblock \emph{Journal of Cell Science}, 133\penalty0 (15), August 2020.

\bibitem[Grimm et~al.(2017)Grimm, Polhill, and Touza]{grimm2017documenting}
Volker Grimm, Gary Polhill, and Julia Touza.
\newblock Documenting social simulation models: the odd protocol as a standard.
\newblock In \emph{Simulating Social Complexity}, pages 349--365. Springer,
  2017.

\bibitem[Grimm et~al.(2020)Grimm, Railsback, Vincenot, Berger, Gallagher,
  DeAngelis, Edmonds, Ge, Giske, Groeneveld, et~al.]{grimm2020odd}
Volker Grimm, Steven~F Railsback, Christian~E Vincenot, Uta Berger, Cara
  Gallagher, Donald~L DeAngelis, Bruce Edmonds, Jiaqi Ge, Jarl Giske, Juergen
  Groeneveld, et~al.
\newblock The odd protocol for describing agent-based and other simulation
  models: A second update to improve clarity, replication, and structural
  realism.
\newblock \emph{Journal of Artificial Societies and Social Simulation},
  23\penalty0 (2), 2020.

\bibitem[Erdemir et~al.(2012)Erdemir, Guess, Halloran, Tadepalli, and
  Morrison]{erdemir2012considerations}
Ahmet Erdemir, Trent~M Guess, Jason Halloran, Srinivas~C Tadepalli, and Tina~M
  Morrison.
\newblock Considerations for reporting finite element analysis studies in
  biomechanics.
\newblock \emph{Journal of Biomechanics}, 45\penalty0 (4):\penalty0 625--633,
  2012.

\bibitem[Krause et~al.(2010)Krause, Uhlendorf, Lubitz, Schulz, Klipp, and
  Liebermeister]{krause2010annotation}
Falko Krause, Jannis Uhlendorf, Timo Lubitz, Marvin Schulz, Edda Klipp, and
  Wolfram Liebermeister.
\newblock Annotation and merging of sbml models with semantic sbml.
\newblock \emph{Bioinformatics}, 26\penalty0 (3):\penalty0 421--422, 2010.

\bibitem[Moreau and Groth(2013)]{moreau2013provenance}
Luc Moreau and Paul Groth.
\newblock Provenance: an introduction to prov.
\newblock \emph{Synthesis Lectures on the Semantic Web: Theory and Technology},
  3\penalty0 (4):\penalty0 1--129, 2013.

\bibitem[Ruscheinski and Uhrmacher(2017)]{ruscheinski2017provenance}
Andreas Ruscheinski and Adelinde Uhrmacher.
\newblock Provenance in modeling and simulation studies—bridging gaps.
\newblock In \emph{2017 Winter Simulation Conference (WSC)}, pages 872--883.
  IEEE, 2017.

\bibitem[Budde et~al.(2021)Budde, Smith, Wilsdorf, Haack, and
  Uhrmacher]{budde2021relating}
Kai Budde, Jacob Smith, Pia Wilsdorf, Fiete Haack, and Adelinde~M Uhrmacher.
\newblock Relating simulation studies by provenance—developing a family of
  wnt signaling models.
\newblock \emph{PLOS Computational Biology}, 17\penalty0 (8):\penalty0
  e1009227, 2021.

\bibitem[Lud{\"a}scher et~al.(2009)Lud{\"a}scher, Bowers, and
  McPhillips]{ludascher2009scientifica}
Bertram Lud{\"a}scher, Shawn Bowers, and Timothy McPhillips.
\newblock Scientific {{Workflows}}.
\newblock In Ling Liu and M.~Tamer {\"O}zsu, editors, \emph{Encyclopedia of
  {{Database Systems}}}, pages 2507--2511. {Springer US}, {Boston, MA}, 2009.
\newblock ISBN 978-0-387-39940-9.
\newblock \doi{10.1007/978-0-387-39940-9_1471}.
\newblock URL \url{https://doi.org/10.1007/978-0-387-39940-9_1471}.

\bibitem[Liew et~al.(2016)Liew, Atkinson, Galea, Ang, Martin, and
  Hemert]{liew2016scientific}
Chee~Sun Liew, Malcolm~P. Atkinson, Michelle Galea, Tan~Fong Ang, Paul Martin,
  and Jano I.~Van Hemert.
\newblock Scientific {{Workflows}}: {{Moving Across Paradigms}}.
\newblock \emph{ACM Computing Surveys}, 49\penalty0 (4):\penalty0 66:1--66:39,
  December 2016.
\newblock ISSN 0360-0300.
\newblock \doi{10.1145/3012429}.
\newblock URL \url{https://dl.acm.org/doi/10.1145/3012429}.

\bibitem[Herschel et~al.(2017)Herschel, Diestelk{\"a}mper, and
  Lahmar]{herschel2017survey}
Melanie Herschel, Ralf Diestelk{\"a}mper, and Houssem~Ben Lahmar.
\newblock A survey on provenance: {{What}} for? {{What}} form? {{What}} from?
\newblock \emph{The VLDB Journal}, 26\penalty0 (6):\penalty0 881--906, December
  2017.
\newblock ISSN 1066-8888, 0949-877X.
\newblock \doi{10.1007/s00778-017-0486-1}.
\newblock URL
  \url{https://link-springer-com.proxy2.library.illinois.edu/article/10.1007/s00778-017-0486-1}.

\bibitem[Ruscheinski et~al.(2019{\natexlab{a}})Ruscheinski, Wilsdorf,
  Dombrowsky, and Uhrmacher]{ruscheinski2019capturing}
Andreas Ruscheinski, Pia Wilsdorf, Marcus Dombrowsky, and Adelinde~M.
  Uhrmacher.
\newblock Capturing and {Reporting} {Provenance} {Information} of {Simulation}
  {Studies} {Based} on an {Artifact}-{Based} {Workflow} {Approach}.
\newblock In \emph{Proceedings of the 2019 {ACM} {SIGSIM} {Conference} on
  {Principles} of {Advanced} {Discrete} {Simulation}}, {SIGSIM}-{PADS} '19,
  pages 185--196, New York, NY, USA, 2019{\natexlab{a}}. ACM.
\newblock ISBN 978-1-4503-6723-3.
\newblock event-place: Chicago, IL, USA.

\bibitem[Vacul{\'\i}n et~al.(2011)Vacul{\'\i}n, Hull, Heath, Cochran, Nigam,
  and Sukaviriya]{vaculin2011declarative}
Roman Vacul{\'\i}n, Richard Hull, Terry Heath, Craig Cochran, Anil Nigam, and
  Piyawadee Sukaviriya.
\newblock Declarative business artifact centric modeling of decision and
  knowledge intensive business processes.
\newblock In \emph{2011 IEEE 15th International Enterprise Distributed Object
  Computing Conference}, pages 151--160. IEEE, 2011.

\bibitem[Ruscheinski et~al.(2019{\natexlab{b}})Ruscheinski, Warnke, and
  Uhrmacher]{ruscheinski2019artifact}
Andreas Ruscheinski, Tom Warnke, and Adelinde~M Uhrmacher.
\newblock Artifact-based workflows for supporting simulation studies.
\newblock \emph{IEEE Transactions on Knowledge and Data Engineering},
  32\penalty0 (6):\penalty0 1064--1078, 2019{\natexlab{b}}.

\bibitem[Klabunde et~al.(2015)Klabunde, Willekens, Zinn, Leuchter,
  et~al.]{klabunde2015agent}
Anna Klabunde, F~Willekens, Sabine Zinn, Matthias Leuchter, et~al.
\newblock An agent-based decision model of migration, embedded in the life
  course-model description in odd+ d format.
\newblock \emph{Max Planck Institute for Demographic Research}, 2015.

\bibitem[Sch{\"u}tzel et~al.(2014)Sch{\"u}tzel, Peng, Uhrmacher, and
  Perrone]{Schuetzel2014}
Johannes Sch{\"u}tzel, Danhua Peng, Adelinde~M. Uhrmacher, and L.~Felipe
  Perrone.
\newblock Perspectives on languages for specifying simulation experiments.
\newblock In \emph{Winter Simulation Conference (WSC 2014)}, pages 2836--2847,
  Electronic ISSN: 1558-4305 Print ISSN: 0891-7736, 2014. IEEE.
\newblock URL \url{http://eprints.mosi.informatik.uni-rostock.de/32/}.
\newblock doi:10.1109/WSC.2014.7020125.

\bibitem[Al-Aswadi et~al.(2020)Al-Aswadi, Chan, and Gan]{al2020automatic}
Fatima~N Al-Aswadi, Huah~Yong Chan, and Keng~Hoon Gan.
\newblock Automatic ontology construction from text: a review from shallow to
  deep learning trend.
\newblock \emph{Artificial Intelligence Review}, 53\penalty0 (6):\penalty0
  3901--3928, 2020.

\bibitem[Bergmann et~al.(2014)Bergmann, Adams, Moodie, Cooper, Glont,
  Golebiewski, Hucka, Laibe, Miller, Nickerson, et~al.]{bergmann2014combine}
Frank~T Bergmann, Richard Adams, Stuart Moodie, Jonathan Cooper, Mihai Glont,
  Martin Golebiewski, Michael Hucka, Camille Laibe, Andrew~K Miller, David~P
  Nickerson, et~al.
\newblock Combine archive and omex format: One file to share all information to
  reproduce a modeling project.
\newblock \emph{BMC Bioinformatics}, 15\penalty0 (1):\penalty0 1--9, 2014.

\bibitem[Ayll{\'o}n et~al.(2021)Ayll{\'o}n, Railsback, Gallagher, Augusiak,
  Baveco, Berger, Charles, Martin, Focks, Galic, et~al.]{ayllon2021keeping}
Daniel Ayll{\'o}n, Steven~F Railsback, Cara Gallagher, Jacqueline Augusiak,
  Hans Baveco, Uta Berger, Sandrine Charles, Romina Martin, Andreas Focks, Nika
  Galic, et~al.
\newblock Keeping modelling notebooks with trace: Good for you and good for
  environmental research and management support.
\newblock \emph{Environmental Modelling \& Software}, 136:\penalty0 104932,
  2021.

\bibitem[Pimentel et~al.(2017)Pimentel, Murta, Braganholo, and
  Freire]{pimentel2017noworkflow}
Joao~Felipe Pimentel, Leonardo Murta, Vanessa Braganholo, and Juliana Freire.
\newblock noworkflow: a tool for collecting, analyzing, and managing provenance
  from python scripts.
\newblock \emph{Proceedings of the VLDB Endowment}, 10\penalty0 (12), 2017.

\bibitem[Apel et~al.(2013)Apel, Batory, K{\"a}stner, and Saake]{ABKS13}
Sven Apel, Don~S. Batory, Christian K{\"a}stner, and Gunter Saake.
\newblock \emph{Feature-Oriented Software Product Lines: Concepts and
  Implementation}.
\newblock Springer, 2013.
\newblock ISBN 978-3-642-37520-0.

\bibitem[Schobbens et~al.(2006)Schobbens, Heymans, and Trigaux]{SHT06}
Pierre{-}Yves Schobbens, Patrick Heymans, and Jean{-}Christophe Trigaux.
\newblock Feature diagrams: {A} survey and a formal semantics.
\newblock In \emph{14th {IEEE} International Conference on Requirements
  Engineering {(RE} 2006), 11-15 September 2006, Minneapolis/St.Paul,
  Minnesota, {USA}}, pages 136--145. {IEEE} Computer Society, 2006.
\newblock \doi{10.1109/RE.2006.23}.
\newblock URL \url{https://doi.org/10.1109/RE.2006.23}.

\bibitem[Schaefer et~al.(2010)Schaefer, Bettini, Bono, Damiani, and
  Tanzarella]{dop}
Ina Schaefer, Lorenzo Bettini, Viviana Bono, Ferruccio Damiani, and Nico
  Tanzarella.
\newblock {Delta-Oriented Programming of Software Product Lines}.
\newblock In Jan Bosch and Jaejoon Lee, editors, \emph{Software Product Lines:
  Going Beyond (SPLC 2010)}, volume 6287 of \emph{LNCS}, pages 77--91, 2010.
\newblock ISBN 978-3-642-15578-9.
\newblock \doi{10.1007/978-3-642-15579-6\_6}.

\bibitem[Th{\"{u}}m et~al.(2012)Th{\"{u}}m, Schaefer, Hentschel, and
  Apel]{TSHA12}
Thomas Th{\"{u}}m, Ina Schaefer, Martin Hentschel, and Sven Apel.
\newblock Family-based deductive verification of software product lines.
\newblock In Klaus Ostermann and Walter Binder, editors, \emph{Generative
  Programming and Component Engineering, GPCE'12, Dresden, Germany}, pages
  11--20. {ACM}, 2012.

\bibitem[Davis et~al.(2000)Davis, Bigelow, and McEver]{davis2000exploratory}
Paul~K Davis, James Bigelow, and Jimmie McEver.
\newblock \emph{Exploratory Analysis and a Case History of Multiresolution,
  Multiperspective Modeling}.
\newblock Number RP-925. Rand Corporation, 2000.

\bibitem[Pohl et~al.(2005)Pohl, B\"{o}ckle, and Linden]{SWPL05}
Klaus Pohl, G\"{u}nter B\"{o}ckle, and Frank J. van~der Linden.
\newblock \emph{Software Product Line Engineering: Foundations, Principles and
  Techniques}.
\newblock Springer-Verlag, 2005.

\bibitem[Slaats(2020)]{slaats2020declarative}
Tijs Slaats.
\newblock Declarative and hybrid process discovery: Recent advances and open
  challenges.
\newblock \emph{Journal on Data Semantics}, 9\penalty0 (1):\penalty0 3--20,
  2020.

\bibitem[Ruscheinski et~al.(2022)Ruscheinski, Wilsdorf, Zimmermann, van Rienen,
  and Uhrmacher]{ruscheinski2022artefact}
Andreas Ruscheinski, Pia Wilsdorf, Julius Zimmermann, Ursula van Rienen, and
  Adelinde~M Uhrmacher.
\newblock An artefact-based workflow for finite element simulation studies.
\newblock \emph{Simulation Modelling Practice and Theory}, 116:\penalty0
  102464, 2022.

\bibitem[Pidd(2002)]{pidd2002simulation}
Michael Pidd.
\newblock Simulation software and model reuse: A polemic.
\newblock In \emph{Proceedings of the Winter Simulation Conference}, volume~1,
  pages 772--775. IEEE, 2002.

\bibitem[Robinson et~al.(2004)Robinson, Nance, Paul, Pidd, and
  Taylor]{robinson2004simulation}
Stewart Robinson, Richard~E Nance, Ray~J Paul, Michael Pidd, and Simon~JE
  Taylor.
\newblock Simulation model reuse: Definitions, benefits and obstacles.
\newblock \emph{Simulation Modelling Practice and Theory}, 12\penalty0
  (7-8):\penalty0 479--494, 2004.

\bibitem[Petty and Weisel(2019)]{petty2019model}
Mikel~D Petty and Eric~W Weisel.
\newblock Model composition and reuse.
\newblock In \emph{Model Engineering for Simulation}, pages 57--85. Elsevier,
  2019.

\bibitem[Paul and Taylor(2002)]{paul2002use}
Ray~J Paul and Simon~JE Taylor.
\newblock What use is model reuse: Is there a crook at the end of the rainbow?
\newblock In \emph{Proceedings of the Winter Simulation Conference}, volume~1,
  pages 648--652. IEEE, 2002.

\bibitem[Szabo and Teo(2007)]{szabo2007syntactic}
Claudia Szabo and Yong~Meng Teo.
\newblock On syntactic composability and model reuse.
\newblock In \emph{First Asia International Conference on Modelling \&
  Simulation (AMS'07)}, pages 230--237. IEEE, 2007.

\bibitem[Janssen et~al.(2020)Janssen, Pritchard, and Lee]{janssen2020code}
Marco~A Janssen, Calvin Pritchard, and Allen Lee.
\newblock On code sharing and model documentation of published individual and
  agent-based models.
\newblock \emph{Environmental Modelling \& Software}, 134:\penalty0 104873,
  2020.

\bibitem[Popper et~al.(2020)Popper, Zechmeister, Brunmeir, Rippinger,
  Weibrecht, Urach, Bicher, Schneckenreither, and Rauber]{popper2020synthetic}
Nikolas Popper, Melanie Zechmeister, Dominik Brunmeir, Claire Rippinger, Nadine
  Weibrecht, Christoph Urach, Martin Bicher, G{\"u}nter Schneckenreither, and
  Andreas Rauber.
\newblock Synthetic reproduction and augmentation of covid-19 case reporting
  data by agent-based simulation.
\newblock \emph{medRxiv}, pages 2020--11, 2020.

\bibitem[Bicher et~al.(2018)Bicher, Urach, and Popper]{bicher2018gepoc}
Martin Bicher, Christoph Urach, and Niki Popper.
\newblock Gepoc abm: A generic agent-based population model for austria.
\newblock In \emph{2018 Winter Simulation Conference (WSC)}, pages 2656--2667.
  IEEE, 2018.

\bibitem[Zhu et~al.(2019)Zhu, Yao, Li, and Tang]{zhu2019reusability}
Feng Zhu, Yiping Yao, Jin Li, and Wenjie Tang.
\newblock Reusability and composability analysis for an agent-based
  hierarchical modelling and simulation framework.
\newblock \emph{Simulation Modelling Practice and Theory}, 90:\penalty0 81--97,
  2019.

\bibitem[Bartholet et~al.(2004)Bartholet, Brogan, Reynolds, and Carnahan]{phil}
Robert~G. Bartholet, David~C. Brogan, Paul~F. Reynolds, and Joseph~C. Carnahan.
\newblock {In Search of the Philosopher's Stone: Simulation Composability
  Versus Component-Based Software Design}.
\newblock In \emph{Proceedings of the Fall Simulation Interoperability
  Workshop}, Orlando, {USA}, 2004.

\bibitem[Dalle(2006)]{osa}
Olivier Dalle.
\newblock {OSA: An Open Component-based Architecture for Discrete-event
  Simulation}.
\newblock In \emph{Proceedings of the $20^{th}$ European Conference on Modeling
  and Simulation}, Prague, Czech Republic, 2006.

\bibitem[Himmelspach and Uhrmacher(2007)]{Himmelspach2007}
Jan Himmelspach and Adelinde~M. Uhrmacher.
\newblock Plug'n simulate.
\newblock In \emph{Proceedings of the 40th Annual Simulation Symposium}, ANSS
  '07, page 137–143, USA, 2007. IEEE Computer Society.
\newblock ISBN 0769528147.
\newblock \doi{10.1109/ANSS.2007.34}.
\newblock URL \url{https://doi.org/10.1109/ANSS.2007.34}.

\bibitem[Tolk and Muguira(2003)]{tolk2003levels}
Andreas Tolk and James~A Muguira.
\newblock The levels of conceptual interoperability model.
\newblock In \emph{Proceedings of the 2003 Fall Simulation Interoperability
  Workshop}, volume~7, pages 1--11. Citeseer, 2003.

\bibitem[Rohl and Uhrmacher(2008)]{Roehl2008}
Mathias Rohl and Adelinde~M. Uhrmacher.
\newblock Definition and analysis of composition structures for discrete-event
  models.
\newblock In \emph{2008 Winter Simulation Conference}, pages 942--950, 2008.
\newblock \doi{10.1109/WSC.2008.4736160}.

\bibitem[Wang et~al.(2009)Wang, Tolk, and Wang]{wang2009interoperability}
Wenguang Wang, Andreas Tolk, and Weiping Wang.
\newblock The levels of conceptual interoperability model: Applying systems
  engineering principles to m\&s.
\newblock In \emph{Proceedings of the 2009 Spring Simulation Multiconference},
  SpringSim '09, San Diego, CA, USA, 2009. Society for Computer Simulation
  International.

\bibitem[Lara and Vangheluwe(2002)]{lara2002atom}
Juan~de Lara and Hans Vangheluwe.
\newblock Atom 3: A tool for multi-formalism and meta-modelling.
\newblock In \emph{International Conference on Fundamental Approaches to
  Software Engineering}, pages 174--188. Springer, 2002.

\bibitem[Liu and Lee(2002)]{Liu2002component}
Jie Liu and Edward~A. Lee.
\newblock A component-based approach to modeling and simulating mixed-signal
  and hybrid systems.
\newblock \emph{ACM Transactions on Modeling and Computer Simulation},
  12\penalty0 (4):\penalty0 343–368, 2002.
\newblock ISSN 1049-3301.
\newblock \doi{10.1145/643120.643125}.
\newblock URL \url{https://doi.org/10.1145/643120.643125}.

\bibitem[Nov{\`e}re et~al.(2005)Nov{\`e}re, Finney, Hucka, Bhalla, Campagne,
  Collado-Vides, Crampin, Halstead, Klipp, Mendes, et~al.]{novere2005minimum}
Nicolas~Le Nov{\`e}re, Andrew Finney, Michael Hucka, Upinder~S Bhalla, Fabien
  Campagne, Julio Collado-Vides, Edmund~J Crampin, Matt Halstead, Edda Klipp,
  Pedro Mendes, et~al.
\newblock Minimum information requested in the annotation of biochemical models
  (miriam).
\newblock \emph{Nature Biotechnology}, 23\penalty0 (12):\penalty0 1509--1515,
  2005.

\bibitem[Lloyd et~al.(2004)Lloyd, Halstead, and Nielsen]{lloyd2004cellml}
Catherine~M Lloyd, Matt~DB Halstead, and Poul~F Nielsen.
\newblock Cellml: its future, present and past.
\newblock \emph{Progress in Biophysics and Molecular Biology}, 85\penalty0
  (2-3):\penalty0 433--450, 2004.

\bibitem[Hucka et~al.(2003)Hucka, Finney, Sauro, Bolouri, Doyle, Kitano, Arkin,
  Bornstein, Bray, Cornish-Bowden, et~al.]{hucka2003systems}
Michael Hucka, Andrew Finney, Herbert~M Sauro, Hamid Bolouri, John~C Doyle,
  Hiroaki Kitano, Adam~P Arkin, Benjamin~J Bornstein, Dennis Bray, Athel
  Cornish-Bowden, et~al.
\newblock The systems biology markup language (sbml): a medium for
  representation and exchange of biochemical network models.
\newblock \emph{Bioinformatics}, 19\penalty0 (4):\penalty0 524--531, 2003.

\bibitem[Smith et~al.(2014)Smith, Butterworth, Bassingthwaighte, and
  Sauro]{smith2014sbml}
Lucian~P Smith, Erik Butterworth, James~B Bassingthwaighte, and Herbert~M
  Sauro.
\newblock Sbml and cellml translation in antimony and jsim.
\newblock \emph{Bioinformatics}, 30\penalty0 (7):\penalty0 903--907, 2014.

\bibitem[Page and Opper(1999)]{page1999observations}
Ernest~H Page and Jeffrey~M Opper.
\newblock Observations on the complexity of composable simulation.
\newblock In \emph{Proceedings of the 31st Winter Simulation Conference}, pages
  553--560. IEEE, 1999.

\bibitem[Schaff et~al.(2023)Schaff, Lakshminarayana, Murphy, Bergmann,
  Funahashi, Sullivan, and Smith]{schaff2023sbml}
James~C Schaff, Anuradha Lakshminarayana, Robert~F Murphy, Frank~T Bergmann,
  Akira Funahashi, Devin~P Sullivan, and Lucian~P Smith.
\newblock Sbml level 3 package: spatial processes, version 1, release 1.
\newblock \emph{Journal of Integrative Bioinformatics}, 20\penalty0
  (1):\penalty0 20220054, 2023.

\bibitem[Smith et~al.(2021)Smith, Bergmann, Garny, Helikar, Karr, Nickerson,
  Sauro, Waltemath, and K{\"o}nig]{smith2021simulation}
Lucian~P Smith, Frank~T Bergmann, Alan Garny, Tom{\'a}{\v{s}} Helikar, Jonathan
  Karr, David Nickerson, Herbert Sauro, Dagmar Waltemath, and Matthias
  K{\"o}nig.
\newblock The simulation experiment description markup language (sed-ml):
  language specification for level 1 version 4.
\newblock \emph{Journal of integrative bioinformatics}, 18\penalty0
  (3):\penalty0 20210021, 2021.

\bibitem[Silver et~al.(2011)Silver, Miller, Hybinette, Baramidze, and
  York]{silver2011ontology}
Gregory~A Silver, John~A Miller, Maria Hybinette, Gregory Baramidze, and
  William~S York.
\newblock An ontology for discrete-event modeling and simulation.
\newblock \emph{Simulation}, 87\penalty0 (9):\penalty0 747--773, 2011.

\bibitem[Li et~al.(2010)Li, Donizelli, Rodriguez, Dharuri, Endler, Chelliah,
  Li, He, Henry, Stefan, et~al.]{li2010biomodels}
Chen Li, Marco Donizelli, Nicolas Rodriguez, Harish Dharuri, Lukas Endler,
  Vijayalakshmi Chelliah, Lu~Li, Enuo He, Arnaud Henry, Melanie~I Stefan,
  et~al.
\newblock Biomodels database: An enhanced, curated and annotated resource for
  published quantitative kinetic models.
\newblock \emph{BMC Systems Biology}, 4\penalty0 (1):\penalty0 1--14, 2010.

\bibitem[ACM(accessed 2024)]{acm}
ACM.
\newblock Artifact reviewing and badging, accessed 2024.
\newblock URL
  \url{https://www.acm.org/publications/policies/artifact-review-and-badging-current}.

\bibitem[Mili et~al.(1995)Mili, Mili, and Mili]{mili1995reusing}
Hafedh Mili, Fatma Mili, and Ali Mili.
\newblock Reusing software: Issues and research directions.
\newblock \emph{IEEE Transactions on Software Engineering}, 21\penalty0
  (6):\penalty0 528--562, 1995.

\bibitem[Mohagheghi and Conradi(2007)]{mohagheghi2007quality}
Parastoo Mohagheghi and Reidar Conradi.
\newblock Quality, productivity and economic benefits of software reuse: A
  review of industrial studies.
\newblock \emph{Empirical Software Engineering}, 12:\penalty0 471--516, 2007.

\bibitem[M{\"u}ller et~al.(2013)M{\"u}ller, Bohn, Dre{\ss}ler, Groeneveld,
  Klassert, Martin, Schl{\"u}ter, Schulze, Weise, and
  Schwarz]{muller2013describing}
Birgit M{\"u}ller, Friedrich Bohn, Gunnar Dre{\ss}ler, J{\"u}rgen Groeneveld,
  Christian Klassert, Romina Martin, Maja Schl{\"u}ter, Jule Schulze, Hanna
  Weise, and Nina Schwarz.
\newblock Describing human decisions in agent-based models--odd+ d, an
  extension of the odd protocol.
\newblock \emph{Environmental Modelling \& Software}, 48:\penalty0 37--48,
  2013.

\bibitem[Beikirch et~al.(2018)Beikirch, Cramer, Frank, Otte, Pabich, and
  Trimborn]{RePEc:arx:papers:1812.02726}
Maximilian Beikirch, Simon Cramer, Martin Frank, Philipp Otte, Emma Pabich, and
  Torsten Trimborn.
\newblock {Simulation of Stylized Facts in Agent-Based Computational Economic
  Market Models}.
\newblock Papers 1812.02726, arXiv.org, November 2018.
\newblock URL \url{https://ideas.repec.org/p/arx/papers/1812.02726.html}.

\bibitem[Steiniger and Uhrmacher(2016)]{Steiniger2016}
Alexander Steiniger and Adelinde~M. Uhrmacher.
\newblock Intensional couplings in variable-structure models: An exploration
  based on multilevel-devs.
\newblock \emph{ACM Transactions on Modeling and Computer Simulation},
  26\penalty0 (2):\penalty0 9:1--9:27, 2016.
\newblock \doi{10.1145/2818641}.
\newblock URL \url{https://doi.org/10.1145/2818641}.

\bibitem[Elmqvist and Mattsson(1997)]{elmqvist1997modelica}
Hilding Elmqvist and Sven-Erik Mattsson.
\newblock Modelica-the next generation modeling language: An international
  design effort.
\newblock In \emph{Proceedings of First World Congress of System Simulation},
  pages 1--3. Citeseer, 1997.

\bibitem[Randhawa et~al.(2010)Randhawa, Shaffer, and Tyson]{Randhawa2010}
Ranjit Randhawa, Cliff Shaffer, and John Tyson.
\newblock Model composition for macromolecular regulatory networks.
\newblock \emph{IEEE/ACM Transactions on Computational Biology and
  Bioinformatics}, 7\penalty0 (2):\penalty0 278--287, 2010.
\newblock \doi{10.1109/TCBB.2008.64}.

\bibitem[Schulz et~al.(2006)Schulz, Klipp, Uhlendorf, and
  Liebermeister]{schulz2006sbmlmerge}
Marvin Schulz, Edda Klipp, Jannis Uhlendorf, and Wolfram Liebermeister.
\newblock Sbmlmerge, a system for combining biochemical network models.
\newblock \emph{Genome Informatics}, 17\penalty0 (1):\penalty0 62--71, 2006.

\bibitem[Hussain et~al.(2022)Hussain, Masoudi, Mocko, and
  Paredis]{hussain2022approaches}
Mohammad Hussain, Nafiseh Masoudi, Gregory Mocko, and Chris Paredis.
\newblock Approaches for simulation model reuse in systems design—a review.
\newblock \emph{SAE Technical Paper}, \penalty0 (2022-01-0355), 2022.

\bibitem[Clements and Northrop(2001)]{ClementsNorthrop01}
Paul Clements and Linda Northrop.
\newblock \emph{{Software Product Lines: Practices \& Patterns}}.
\newblock Addison Wesley Longman, 2001.

\bibitem[Setyautami and H{\"{a}}hnle(2021)]{SetyautamiHaehnle21}
Maya Retno~Ayu Setyautami and Reiner H{\"{a}}hnle.
\newblock An architectural pattern to realize multi software product lines in
  {Java}.
\newblock In Paul Gr{\"{u}}nbacher, Christoph Seidl, Deepak Dhungana, and
  Helena Lovasz{-}Bukvova, editors, \emph{VaMoS'21: 15th Intl.\ Working Conf.\
  on Variability Modelling of Software-Intensive Systems, Virtual Event}, pages
  9:1--9:9. {ACM}, 2021.
\newblock \doi{10.1145/3442391}.

\bibitem[Henkel et~al.(2018)Henkel, Hoehndorf, Kacprowski, Kn{\"u}pfer,
  Liebermeister, and Waltemath]{henkel2018notions}
Ron Henkel, Robert Hoehndorf, Tim Kacprowski, Christian Kn{\"u}pfer, Wolfram
  Liebermeister, and Dagmar Waltemath.
\newblock Notions of similarity for systems biology models.
\newblock \emph{Briefings in Bioinformatics}, 19\penalty0 (1):\penalty0 77--88,
  2018.

\bibitem[Waltz and Buchanan(2009)]{waltz2009automating}
David Waltz and Bruce~G Buchanan.
\newblock Automating science.
\newblock \emph{Science}, 324\penalty0 (5923):\penalty0 43--44, 2009.

\bibitem[Maass and Storey(2021)]{maass2021pairing}
Wolfgang Maass and Veda~C. Storey.
\newblock Pairing conceptual modeling with machine learning.
\newblock \emph{Data \& Knowledge Engineering}, 134:\penalty0 101909, July
  2021.
\newblock ISSN 0169-023X.
\newblock \doi{10.1016/j.datak.2021.101909}.
\newblock URL
  \url{https://www.sciencedirect.com/science/article/pii/S0169023X21000367}.

\bibitem[Shuttleworth and Padilla(2022)]{Shuttleworth2022narratives}
David Shuttleworth and Jose Padilla.
\newblock From narratives to conceptual models via natural language processing.
\newblock In \emph{2022 Winter Simulation Conference (WSC)}, pages 2222--2233,
  2022.
\newblock \doi{10.1109/WSC57314.2022.10015274}.

\bibitem[Auer et~al.(2007)Auer, Bizer, Kobilarov, Lehmann, Cyganiak, and
  Ives]{DBpedia}
S{\"o}ren Auer, Christian Bizer, Georgi Kobilarov, Jens Lehmann, Richard
  Cyganiak, and Zachary Ives.
\newblock Dbpedia: A nucleus for a web of open data.
\newblock In Karl Aberer, Key-Sun Choi, Natasha Noy, Dean Allemang, Kyung-Il
  Lee, Lyndon Nixon, Jennifer Golbeck, Peter Mika, Diana Maynard, Riichiro
  Mizoguchi, Guus Schreiber, and Philippe Cudr{\'e}-Mauroux, editors, \emph{The
  Semantic Web}, pages 722--735, Berlin, Heidelberg, 2007. Springer Berlin
  Heidelberg.
\newblock ISBN 978-3-540-76298-0.

\bibitem[Gütebier et~al.(2022)Gütebier, Bleimehl, Henkel, Munro, Müller,
  Morgner, Laenge, Pachauer, Erdl, Weimar, Walther~Langendorf, Vialard, Liebig,
  Preusse, Waltemath, and Jarasch]{Guetebier2022covid}
Lea Gütebier, Tim Bleimehl, Ron Henkel, Jamie Munro, Sebastian Müller, Axel
  Morgner, Jakob Laenge, Anke Pachauer, Alexander Erdl, Jens Weimar, Kirsten
  Walther~Langendorf, Vincent Vialard, Thorsten Liebig, Martin Preusse, Dagmar
  Waltemath, and Alexander Jarasch.
\newblock {CovidGraph: A Graph to Fight COVID-19}.
\newblock \emph{Bioinformatics}, 38\penalty0 (20):\penalty0 4843--4845, 08
  2022.
\newblock ISSN 1367-4803.
\newblock \doi{10.1093/bioinformatics/btac592}.
\newblock URL \url{https://doi.org/10.1093/bioinformatics/btac592}.

\bibitem[Maneschijn et~al.(2022)Maneschijn, Bemthuis, Bukhsh, and
  Iacob]{maneschijn2022methodology}
Dennis~GJC Maneschijn, Rob~H Bemthuis, Faiza~Allah Bukhsh, and Maria-Eugenia
  Iacob.
\newblock A {Methodology} for {Aligning} {Process} {Model} {Abstraction}
  {Levels} and {Stakeholder} {Needs}.
\newblock In \emph{{ICEIS} (1)}, pages 137--147, 2022.

\bibitem[Cairoli et~al.(2023)Cairoli, Anselmi, d'Onofrio, and
  Bortolussi]{CAIROLI2023114169}
Francesca Cairoli, Fabio Anselmi, Alberto d'Onofrio, and Luca Bortolussi.
\newblock Generative abstraction of markov population processes.
\newblock \emph{Theoretical Computer Science}, 977:\penalty0 114169, 2023.
\newblock ISSN 0304-3975.
\newblock \doi{https://doi.org/10.1016/j.tcs.2023.114169}.

\bibitem[Jackson and Saenz(2023)]{jackson2023natural}
Ilya Jackson and Maria~Jesus Saenz.
\newblock From natural language to simulations: Applying gpt-3 codex to
  automate simulation modeling of logistics systems, 2023.

\bibitem[Brown et~al.(2020)Brown, Mann, Ryder, Subbiah, Kaplan, Dhariwal,
  Neelakantan, Shyam, Sastry, Askell, Agarwal, Herbert-Voss, Krueger, Henighan,
  Child, Ramesh, Ziegler, Wu, Winter, Hesse, Chen, Sigler, Litwin, Gray, Chess,
  Clark, Berner, McCandlish, Radford, Sutskever, and Amodei]{brown2020language}
Tom Brown, Benjamin Mann, Nick Ryder, Melanie Subbiah, Jared~D Kaplan, Prafulla
  Dhariwal, Arvind Neelakantan, Pranav Shyam, Girish Sastry, Amanda Askell,
  Sandhini Agarwal, Ariel Herbert-Voss, Gretchen Krueger, Tom Henighan, Rewon
  Child, Aditya Ramesh, Daniel Ziegler, Jeffrey Wu, Clemens Winter, Chris
  Hesse, Mark Chen, Eric Sigler, Mateusz Litwin, Scott Gray, Benjamin Chess,
  Jack Clark, Christopher Berner, Sam McCandlish, Alec Radford, Ilya Sutskever,
  and Dario Amodei.
\newblock Language models are few-shot learners.
\newblock In H.~Larochelle, M.~Ranzato, R.~Hadsell, M.F. Balcan, and H.~Lin,
  editors, \emph{Advances in Neural Information Processing Systems}, volume~33,
  pages 1877--1901. Curran Associates, Inc., 2020.
\newblock URL
  \url{https://proceedings.neurips.cc/paper/2020/file/1457c0d6bfcb4967418bfb8ac142f64a-Paper.pdf}.

\bibitem[Hansen and Hebart(2022)]{Hansen2022computation}
Hannes Hansen and Martin~N. Hebart.
\newblock Semantic features of object concepts generated with gpt-3, 2022.
\newblock URL \url{https://arxiv.org/abs/2202.03753}.

\bibitem[Džeroski and Todorovski(2008)]{DZEROSKI2008360}
Sašo Džeroski and Ljupčo Todorovski.
\newblock Equation discovery for systems biology: Finding the structure and
  dynamics of biological networks from time course data.
\newblock \emph{Current Opinion in Biotechnology}, 19\penalty0 (4):\penalty0
  360--368, 2008.
\newblock \doi{https://doi.org/10.1016/j.copbio.2008.07.002}.
\newblock Protein technologies / Systems biology.

\bibitem[North et~al.(2022)North, Wikle, and Schliep]{north2022review}
Joshua~S. North, Christopher~K. Wikle, and Erin~M. Schliep.
\newblock A review of data-driven discovery for dynamic systems, 2022.

\bibitem[Sun et~al.(2019)Sun, Ouyang, Zhang, and Zhang]{sun2019data}
Sheng Sun, Runhai Ouyang, Bochao Zhang, and Tong-Yi Zhang.
\newblock Data-driven discovery of formulas by symbolic regression.
\newblock \emph{MRS Bulletin}, 44\penalty0 (7):\penalty0 559--564, 2019.

\bibitem[Brunton et~al.(2016)Brunton, Proctor, and Kutz]{brunton}
Steven~L. Brunton, Joshua~L. Proctor, and J.~Nathan Kutz.
\newblock Discovering governing equations from data by sparse identification of
  nonlinear dynamical systems.
\newblock \emph{Proceedings of the National Academy of Sciences}, 113\penalty0
  (15):\penalty0 3932--3937, 2016.
\newblock \doi{10.1073/pnas.1517384113}.

\bibitem[Burrage et~al.(2024)Burrage, Weerasinghe, and
  Burrage]{burrage_using_2024}
Pamela~M. Burrage, Hasitha~N. Weerasinghe, and Kevin Burrage.
\newblock Using a library of chemical reactions to fit systems of ordinary
  differential equations to agent-based models: a machine learning approach.
\newblock \emph{Numerical Algorithms}, 2024.
\newblock ISSN 1572-9265.
\newblock \doi{10.1007/s11075-023-01737-0}.

\bibitem[Jiang et~al.(2022)Jiang, Singh, Wrede, Hellander, and Petzold]{Jiang}
Richard Jiang, Prashant Singh, Fredrik Wrede, Andreas Hellander, and Linda
  Petzold.
\newblock Identification of dynamic mass-action biochemical reaction networks
  using sparse bayesian methods.
\newblock \emph{PLOS Computational Biology}, 18\penalty0 (1):\penalty0 1--21,
  01 2022.
\newblock \doi{10.1371/journal.pcbi.1009830}.

\bibitem[Ahmed et~al.(2023)Ahmed, Telmer, Zhou, and
  Miskov-Zivanov]{Ahmed2022context}
Yasmine Ahmed, Cheryl~A. Telmer, Gaoxiang Zhou, and Natasa Miskov-Zivanov.
\newblock Context-aware knowledge selection and reliable model recommendation
  with accordion.
\newblock \emph{bioRxiv}, 2023.
\newblock \doi{10.1101/2022.01.22.477231}.
\newblock URL
  \url{https://www.biorxiv.org/content/early/2023/01/15/2022.01.22.477231}.

\bibitem[Lorig(2019)]{lorig2019hypothesis}
Fabian Lorig.
\newblock \emph{Hypothesis-Driven Simulation Studies: Assistance for the
  Systematic Design and Conducting of Computer Simulation Experiments}.
\newblock Springer Vieweg, Wiesbaden, 2019.

\bibitem[Perrone et~al.(2012)Perrone, Main, and Ward]{perrone2012safe}
L~Felipe Perrone, Christopher~S Main, and Bryan~C Ward.
\newblock Safe: Simulation automation framework for experiments.
\newblock In \emph{Proceedings of the 2012 Winter Simulation Conference (WSC)},
  pages 1--12. IEEE, 2012.

\bibitem[Wilsdorf et~al.(2021)Wilsdorf, Fischer, Haack, and
  Uhrmacher]{wilsdorf2021exploiting}
Pia Wilsdorf, Nadine Fischer, Fiete Haack, and Adelinde~M. Uhrmacher.
\newblock Exploiting provenance and ontologies in supporting best practices for
  simulation experiments: A case study on sensitivity analysis.
\newblock In \emph{2021 Winter Simulation Conference (WSC)}, pages 1--12, 2021.
\newblock \doi{10.1109/WSC52266.2021.9715362}.

\bibitem[Leye et~al.(2014)Leye, Ewald, and Uhrmacher]{leye2014composing}
Stefan Leye, Roland Ewald, and Adelinde~M. Uhrmacher.
\newblock Composing problem solvers for simulation experimentation: A case
  study on steady state estimation.
\newblock \emph{PLOS ONE}, 9\penalty0 (4):\penalty0 1--13, 04 2014.
\newblock \doi{10.1371/journal.pone.0091948}.
\newblock URL \url{https://doi.org/10.1371/journal.pone.0091948}.

\bibitem[Suleimenova et~al.(2021)Suleimenova, Arabnejad, Edeling, and
  Groen]{suleimenova2012sensitivity}
Diana Suleimenova, Hamid Arabnejad, Wouter Edeling, and Derek Groen.
\newblock Sensitivity-driven simulation development: A case study in forced
  migration.
\newblock \emph{Philosophical Transactions of the Royal Society A:
  Mathematical, Physical and Engineering Sciences}, 379\penalty0
  (2197):\penalty0 20200077, may 2021.
\newblock \doi{10.1098/rsta.2020.0077}.
\newblock Publisher: Royal Society.

\bibitem[Villaverde et~al.(2018)Villaverde, Fröhlich, Weindl, Hasenauer, and
  Banga]{Villaverde2018benchmarking}
Alejandro~F Villaverde, Fabian Fröhlich, Daniel Weindl, Jan Hasenauer, and
  Julio~R Banga.
\newblock {Benchmarking Optimization Methods for Parameter Estimation in Large
  Kinetic Models}.
\newblock \emph{Bioinformatics}, 35\penalty0 (5):\penalty0 830--838, 08 2018.
\newblock ISSN 1367-4803.
\newblock \doi{10.1093/bioinformatics/bty736}.
\newblock URL \url{https://doi.org/10.1093/bioinformatics/bty736}.

\bibitem[Kreikemeyer and Andelfinger(2023)]{kreikemeyer2023}
Justin~N. Kreikemeyer and Philipp Andelfinger.
\newblock Smoothing methods for automatic differentiation across conditional
  branches.
\newblock \emph{IEEE Access}, 11:\penalty0 143190--143211, 2023.
\newblock \doi{10.1109/ACCESS.2023.3342136}.

\bibitem[K{\"o}nig et~al.(2022)K{\"o}nig, Hoos, and Rijn]{Koenig2022speeding}
Matthias K{\"o}nig, Holger~H Hoos, and Jan N~van Rijn.
\newblock Speeding up neural network robustness verification via algorithm
  configuration and an optimised mixed integer linear programming solver
  portfolio.
\newblock \emph{Machine Learning}, pages 1--20, 2022.

\bibitem[Alemi et~al.(2016)Alemi, Chollet, Een, Irving, Szegedy, and
  Urban]{Irving2016deepmath}
Alexander~A. Alemi, Fran\c{c}ois Chollet, Niklas Een, Geoffrey Irving,
  Christian Szegedy, and Josef Urban.
\newblock Deepmath - deep sequence models for premise selection.
\newblock In \emph{Proceedings of the 30th International Conference on Neural
  Information Processing Systems}, NIPS'16, page 2243–2251, Red Hook, NY,
  USA, 2016. Curran Associates Inc.
\newblock ISBN 9781510838819.

\bibitem[Zela et~al.(2018)Zela, Klein, Falkner, and Hutter]{Zela2018towards}
Arber Zela, Aaron Klein, Stefan Falkner, and Frank Hutter.
\newblock Towards automated deep learning: Efficient joint neural architecture
  and hyperparameter search.
\newblock 2018.
\newblock \doi{10.48550/ARXIV.1807.06906}.
\newblock URL \url{https://arxiv.org/abs/1807.06906}.

\bibitem[Helms et~al.(2015)Helms, Ewald, Rybacki, and Uhrmacher]{Helms2015}
Tobias Helms, Roland Ewald, Stefan Rybacki, and Adelinde~M Uhrmacher.
\newblock Automatic runtime adaptation for component-based simulation
  algorithms.
\newblock \emph{ACM Transactions on Modeling and Computer Simulation (TOMACS)},
  26\penalty0 (1):\penalty0 1--24, 2015.

\bibitem[Hunter et~al.(2019)Hunter, Applegate, Arora, Chong, Cooper,
  Rinc\'{o}n-Guevara, and Vivas-Valencia]{Hunter2019introduction}
Susan~R. Hunter, Eric~A. Applegate, Viplove Arora, Bryan Chong, Kyle Cooper,
  Oscar Rinc\'{o}n-Guevara, and Carolina Vivas-Valencia.
\newblock An introduction to multiobjective simulation optimization.
\newblock \emph{ACM Transactions on Modeling and Computer Simulation},
  29\penalty0 (1), 2019.
\newblock ISSN 1049-3301.
\newblock \doi{10.1145/3299872}.
\newblock URL \url{https://doi.org/10.1145/3299872}.

\bibitem[Xiao et~al.(2020)Xiao, Andelfinger, Cai, Richmond, Knoll, and
  Eckhoff]{xiao2020openablext}
Jiajian Xiao, Philipp Andelfinger, Wentong Cai, Paul Richmond, Alois Knoll, and
  David Eckhoff.
\newblock Openablext: An automatic code generation framework for agent-based
  simulations on cpu-gpu-fpga heterogeneous platforms.
\newblock \emph{Concurrency and Computation: Practice and Experience},
  32\penalty0 (21):\penalty0 e5807, 2020.
\newblock \doi{https://doi.org/10.1002/cpe.5807}.

\bibitem[Noor and Hemmati(2015)]{Noor2015similarity}
Tanzeem~Bin Noor and Hadi Hemmati.
\newblock A similarity-based approach for test case prioritization using
  historical failure data.
\newblock In \emph{2015 IEEE 26th International Symposium on Software
  Reliability Engineering (ISSRE)}, pages 58--68, 2015.
\newblock \doi{10.1109/ISSRE.2015.7381799}.

\bibitem[Bocciarelli et~al.(2013)Bocciarelli, D'Ambrogio, Giglio, and
  Gianni]{Bocciarelli2013saas}
Paolo Bocciarelli, Andrea D'Ambrogio, Andrea Giglio, and Daniele Gianni.
\newblock A saas-based automated framework to build and execute distributed
  simulations from sysml models.
\newblock In \emph{2013 Winter Simulations Conference (WSC)}, pages 1371--1382,
  2013.
\newblock \doi{10.1109/WSC.2013.6721523}.

\bibitem[Feldkamp et~al.(2020{\natexlab{a}})Feldkamp, Bergmann, and
  Strassburger]{feldkamp2020knowledge}
Niclas Feldkamp, Soeren Bergmann, and Steffen Strassburger.
\newblock Knowledge {Discovery} in {Simulation} {Data}.
\newblock \emph{ACM Transactions on Modeling and Computer Simulation},
  30\penalty0 (4):\penalty0 24:1--24:25, November 2020{\natexlab{a}}.
\newblock ISSN 1049-3301.
\newblock \doi{10.1145/3391299}.
\newblock URL \url{https://doi.org/10.1145/3391299}.

\bibitem[Ali et~al.(2019)Ali, Alqahtani, Jones, and Xie]{Ali2019clustering}
Mohammed Ali, Ali Alqahtani, Mark~W. Jones, and Xianghua Xie.
\newblock Clustering and classification for time series data in visual
  analytics: A survey.
\newblock \emph{IEEE Access}, 7:\penalty0 181314--181338, 2019.
\newblock \doi{10.1109/ACCESS.2019.2958551}.

\bibitem[Dambros et~al.(2019)Dambros, Trierweiler, and
  Farenzena]{Dambros2019oscillation}
Jônathan~W.V. Dambros, Jorge~O. Trierweiler, and Marcelo Farenzena.
\newblock Oscillation detection in process industries – part i: Review of the
  detection methods.
\newblock \emph{Journal of Process Control}, 78:\penalty0 108--123, 2019.
\newblock ISSN 0959-1524.
\newblock \doi{https://doi.org/10.1016/j.jprocont.2019.04.002}.
\newblock URL
  \url{https://www.sciencedirect.com/science/article/pii/S0959152419302239}.

\bibitem[Bl\'{a}zquez-Garc\'{\i}a et~al.(2021)Bl\'{a}zquez-Garc\'{\i}a, Conde,
  Mori, and Lozano]{Blazquez-Garcia2022}
Ane Bl\'{a}zquez-Garc\'{\i}a, Angel Conde, Usue Mori, and Jose~A. Lozano.
\newblock A review on outlier/anomaly detection in time series data.
\newblock \emph{ACM Computing Surveys}, 54\penalty0 (3), apr 2021.
\newblock ISSN 0360-0300.
\newblock \doi{10.1145/3444690}.
\newblock URL \url{https://doi.org/10.1145/3444690}.

\bibitem[Feng and Staum(2017)]{feng2017green}
Mingbin Feng and Jeremy Staum.
\newblock Green {Simulation}: {Reusing} the {Output} of {Repeated}
  {Experiments}.
\newblock \emph{ACM Transactions on Modeling and Computer Simulation},
  27\penalty0 (4):\penalty0 23:1--23:28, October 2017.
\newblock ISSN 1049-3301.
\newblock \doi{10.1145/3129130}.
\newblock URL \url{https://doi.org/10.1145/3129130}.

\bibitem[Blount et~al.(2021)Blount, Chapman, Johnson, and
  Ludascher]{blount2021observed}
Tom Blount, Adriane Chapman, Michael Johnson, and Bertram Ludascher.
\newblock Observed vs. {Possible} {Provenance} ({Research} {Track}).
\newblock 2021.
\newblock URL
  \url{https://www.usenix.org/conference/tapp2021/presentation/blount}.

\bibitem[Allen et~al.(2010)Allen, Seligman, Blaustein, and
  Chapman]{Allen2010provenance}
M~David Allen, Len Seligman, Barbara Blaustein, and Adriane Chapman.
\newblock Provenance capture and use: A practical guide.
\newblock Technical report, Mitre CorpP McLean VA, 2010.

\bibitem[Gebhardt et~al.(2022)Gebhardt, Touré, Waltemath, Wolkenhauer, and
  Scharm]{gebhardt2022exploring}
Tom Gebhardt, Vasundra Touré, Dagmar Waltemath, Olaf Wolkenhauer, and Martin
  Scharm.
\newblock Exploring the evolution of biochemical models at the network level.
\newblock \emph{PLOS ONE}, 17\penalty0 (3):\penalty0 e0265735, March 2022.
\newblock ISSN 1932-6203.
\newblock \doi{10.1371/journal.pone.0265735}.
\newblock URL
  \url{https://journals.plos.org/plosone/article?id=10.1371/journal.pone.0265735}.
\newblock Publisher: Public Library of Science.

\bibitem[Berti et~al.(2019)Berti, Van~Zelst, and van~der
  Aalst]{berti2019process}
Alessandro Berti, Sebastiaan~J Van~Zelst, and Wil van~der Aalst.
\newblock Process mining for python (pm4py): Bridging the gap between process
  and data science.
\newblock \emph{arXiv preprint arXiv:1905.06169}, 2019.

\bibitem[Lin et~al.(2022{\natexlab{a}})Lin, Wang, Liu, and Qiu]{LIN2022111}
Tianyang Lin, Yuxin Wang, Xiangyang Liu, and Xipeng Qiu.
\newblock A survey of transformers.
\newblock \emph{AI Open}, 3:\penalty0 111--132, 2022{\natexlab{a}}.
\newblock ISSN 2666-6510.
\newblock \doi{https://doi.org/10.1016/j.aiopen.2022.10.001}.
\newblock URL
  \url{https://www.sciencedirect.com/science/article/pii/S2666651022000146}.

\bibitem[Giabbanelli(2023)]{giabbanelli2023gptbased}
Philippe~J. Giabbanelli.
\newblock Gpt-based models meet simulation: How to efficiently use large-scale
  pre-trained language models across simulation tasks, 2023.

\bibitem[Vaithilingam et~al.(2022)Vaithilingam, Zhang, and
  Glassman]{Vaithilingam2022}
Priyan Vaithilingam, Tianyi Zhang, and Elena~L. Glassman.
\newblock Expectation vs.~experience: Evaluating the usability of code
  generation tools powered by large language models.
\newblock In \emph{Extended Abstracts of the 2022 CHI Conference on Human
  Factors in Computing Systems}, CHI EA '22, New York, NY, USA, 2022.
  Association for Computing Machinery.
\newblock ISBN 9781450391566.
\newblock \doi{10.1145/3491101.3519665}.
\newblock URL \url{https://doi.org/10.1145/3491101.3519665}.

\bibitem[Woo(2020)]{woo2020rise}
Marcus Woo.
\newblock The rise of no/low code software development—no experience needed?
\newblock \emph{Engineering (Beijing, China)}, 6\penalty0 (9):\penalty0 960,
  2020.

\bibitem[Jeschke et~al.(2011)Jeschke, Ewald, and Uhrmacher]{JESCHKE20112562}
Matthias Jeschke, Roland Ewald, and Adelinde~M. Uhrmacher.
\newblock Exploring the performance of spatial stochastic simulation
  algorithms.
\newblock \emph{Journal of Computational Physics}, 230\penalty0 (7):\penalty0
  2562--2574, 2011.
\newblock ISSN 0021-9991.
\newblock \doi{https://doi.org/10.1016/j.jcp.2010.12.030}.
\newblock URL
  \url{https://www.sciencedirect.com/science/article/pii/S0021999110007047}.

\bibitem[Kleijnen(1995)]{Kleijnen1995sensitivity}
Jack P.~C. Kleijnen.
\newblock Sensitivity analysis and optimization in simulation: Design of
  experiments and case studies.
\newblock In \emph{Proceedings of the 27th Conference on Winter Simulation},
  WSC '95, page 133–140, USA, 1995. IEEE Computer Society.
\newblock ISBN 0780330188.
\newblock \doi{10.1145/224401.224454}.
\newblock URL \url{https://doi.org/10.1145/224401.224454}.

\bibitem[161(2019)]{SFB-TRR161}
Collaborative Research Center SFB-TRR 161.
\newblock Quantitative methods for visual computing, 2019.
\newblock URL \url{https://www.sfbtrr161.de/}.
\newblock Accessed: 2023-01-27.

\bibitem[Bylinskii et~al.(2022)Bylinskii, Herman, Hertzmann, Hutka, and
  Zhang]{Bylinskii2022towards}
Zoya Bylinskii, Laura Herman, Aaron Hertzmann, Stefanie Hutka, and Yile Zhang.
\newblock Towards better user studies in computer graphics and vision, 2022.
\newblock URL \url{https://arxiv.org/abs/2206.11461}.

\bibitem[Von~Elm et~al.(2007)Von~Elm, Altman, Egger, Pocock, G{\o}tzsche, and
  Vandenbroucke]{vonElm2007strengthening}
Erik Von~Elm, Douglas~G Altman, Matthias Egger, Stuart~J Pocock, Peter~C
  G{\o}tzsche, and Jan~P Vandenbroucke.
\newblock The strengthening the reporting of observational studies in
  epidemiology (strobe) statement: Guidelines for reporting observational
  studies.
\newblock \emph{The Lancet}, 370\penalty0 (9596):\penalty0 1453--1457, 2007.

\bibitem[Sadowski and Grabau(1999)]{sadowski1999tips}
Deborah~A Sadowski and Mark~R Grabau.
\newblock Tips for successful practice of simulation.
\newblock In \emph{WSC'99. 1999 Winter Simulation Conference}, volume~1, pages
  60--66. IEEE, 1999.

\bibitem[Sturrock(2020)]{sturrock2020tested}
David~T Sturrock.
\newblock Tested success tips for simulation project excellence.
\newblock In \emph{2020 Winter Simulation Conference (WSC)}, pages 1143--1151.
  IEEE, 2020.

\bibitem[Sturrock(2015)]{sturrock2015tutorial}
David~T Sturrock.
\newblock Tutorial: Tips for successful practice of simulation.
\newblock In \emph{2015 Winter Simulation Conference (WSC)}, pages 1756--1764.
  IEEE, 2015.

\bibitem[Baldwin et~al.(2004)Baldwin, Eldabi, and Paul]{baldwin2004simulation}
Lynne~P Baldwin, Tillal Eldabi, and Ray~J Paul.
\newblock Simulation in healthcare management: A soft approach (mapiu).
\newblock \emph{Simulation, Modelling, Practice and Theory}, 12\penalty0
  (7-8):\penalty0 541--557, 2004.

\bibitem[Blomkamp(2018)]{blomkamp2018promise}
Emma Blomkamp.
\newblock The promise of co-design for public policy.
\newblock \emph{Australian journal of public administration}, 77\penalty0
  (4):\penalty0 729--743, 2018.

\bibitem[Orton et~al.(2011)Orton, Lloyd-Williams, Taylor-Robinson, O'Flaherty,
  and Capewell]{orton2011use}
Lois Orton, Ffion Lloyd-Williams, David Taylor-Robinson, Martin O'Flaherty, and
  Simon Capewell.
\newblock The use of research evidence in public health decision making
  processes: Systematic review.
\newblock \emph{PloS One}, 6\penalty0 (7):\penalty0 e21704, 2011.

\bibitem[Smith and Stewart(2015)]{smith2015black}
Katherine~E Smith and Ellen Stewart.
\newblock ‘black magic’ and ‘gold dust’: the epistemic and political
  uses of evidence tools in public health policy making.
\newblock \emph{Evidence \& Policy}, 11\penalty0 (3):\penalty0 415--437, 2015.

\bibitem[Oliver et~al.(2014)Oliver, Innvar, Lorenc, Woodman, and
  Thomas]{oliver2014systematic}
Kathryn Oliver, Simon Innvar, Theo Lorenc, Jenny Woodman, and James Thomas.
\newblock A systematic review of barriers to and facilitators of the use of
  evidence by policymakers.
\newblock \emph{BMC Health Services Research}, 14:\penalty0 1--12, 2014.

\bibitem[Taylor-Robinson et~al.(2008)Taylor-Robinson, Milton, Lloyd-Williams,
  O'Flaherty, and Capewell]{TaylorRobinson2008}
David Taylor-Robinson, Beth Milton, Efion Lloyd-Williams, Martin O'Flaherty,
  and Simon Capewell.
\newblock Policy-makers' attitudes to decision support models for coronary
  heart disease: a qualitative study.
\newblock \emph{Journal of Health Services Research \& Policy}, 13\penalty0
  (4):\penalty0 209--2014, 2008.

\bibitem[Burns et~al.(2003)Burns, O'Connor, and Stocklmayer]{burns2003science}
Terry~W Burns, D~John O'Connor, and Susan~M Stocklmayer.
\newblock Science communication: A contemporary definition.
\newblock \emph{Public Understanding of Science}, 12\penalty0 (2):\penalty0
  183--202, 2003.

\bibitem[Kappel and Holmen(2019)]{kappel2019science}
Klemens Kappel and Sebastian~Jon Holmen.
\newblock Why science communication, and does it work? a taxonomy of science
  communication aims and a survey of the empirical evidence.
\newblock \emph{Frontiers in Communication}, 4:\penalty0 55, 2019.

\bibitem[Guenther and Joubert(2017)]{Guenther&Jouberg2017}
Lars Guenther and Marina Joubert.
\newblock Science communication as a field of research: Identifying trends,
  challenges and gaps by analysing research papers.
\newblock \emph{Journal of Science Communication}, 16:\penalty0 A02, 2017.
\newblock \doi{https://doi.org/10.22323/2.16020202}.

\bibitem[Guo et~al.(2021)Guo, Qiu, Wang, and Cohen]{Guo+2021}
Yue Guo, Wei Qiu, Yizhong Wang, and Trevor Cohen.
\newblock Automated lay language summarization of biomedical scientific
  reviews.
\newblock In \emph{Proceedings of the AAAI Conference on Artificial
  Intelligence}, pages 160--168, 2021.
\newblock \doi{https://doi.org/10.1609/aaai.v35i1.16089}.

\bibitem[Davis(2016)]{Davis2016}
Paul~K. Davis.
\newblock \emph{Capabilities for Joint Analysis in the Department of Defense:
  Rethinking Support for Strategic Analysis}.
\newblock RAND Corporation, Santa Monica, CA, 2016.
\newblock \doi{10.7249/RR1469}.

\bibitem[Board(2020)]{GEMS2020}
Defense~Science Board.
\newblock \emph{AD1155605 - Task Force on Gaming, Exercising, Modeling, and
  Simulation (GEMS)}.
\newblock US DoD, Defense Science Board, OUSD(R\&E), 1400 Defense Pentagon,
  Washington, DC, 2020.

\bibitem[Chen and Pu(2004)]{chen2004survey}
Li~Chen and Pearl Pu.
\newblock Survey of preference elicitation methods.
\newblock Technical Report IC/2004/67, Ecole Politechnique Federale de Lausanne
  (EPFL), 2004.

\bibitem[Best et~al.(2020)Best, Dallow, and Montague]{best2020prior}
Nicky Best, Nigel Dallow, and Timothy Montague.
\newblock Prior elicitation.
\newblock \emph{Bayesian Methods in Pharmaceutical Research}, pages 87--109,
  2020.

\bibitem[Wallenius et~al.(2008)Wallenius, Dyer, Fishburn, Steuer, Zionts, and
  Deb]{wallenius2008multiple}
Jyrki Wallenius, James~S Dyer, Peter~C Fishburn, Ralph~E Steuer, Stanley
  Zionts, and Kalyanmoy Deb.
\newblock Multiple criteria decision making, multiattribute utility theory:
  Recent accomplishments and what lies ahead.
\newblock \emph{Management Science}, 54\penalty0 (7):\penalty0 1336--1349,
  2008.

\bibitem[Knowles(2006)]{knowles2006parego}
Joshua Knowles.
\newblock Parego: A hybrid algorithm with on-line landscape approximation for
  expensive multiobjective optimization problems.
\newblock \emph{IEEE Transactions on Evolutionary Computation}, 10\penalty0
  (1):\penalty0 50--66, 2006.

\bibitem[Daulton et~al.(2021)Daulton, Balandat, and
  Bakshy]{daulton2021parallel}
Samuel Daulton, Maximilian Balandat, and Eytan Bakshy.
\newblock Parallel bayesian optimization of multiple noisy objectives with
  expected hypervolume improvement.
\newblock \emph{Advances in Neural Information Processing Systems},
  34:\penalty0 2187--2200, 2021.

\bibitem[Astudillo and Frazier(2020)]{astudillo2020multi}
Raul Astudillo and Peter Frazier.
\newblock Multi-attribute bayesian optimization with interactive preference
  learning.
\newblock In \emph{International Conference on Artificial Intelligence and
  Statistics}, pages 4496--4507. PMLR, 2020.

\bibitem[Lin et~al.(2022{\natexlab{b}})Lin, Astudillo, Frazier, and
  Bakshy]{lin2022preference}
Zhiyuan~Jerry Lin, Raul Astudillo, Peter Frazier, and Eytan Bakshy.
\newblock Preference exploration for efficient bayesian optimization with
  multiple outcomes.
\newblock In \emph{International Conference on Artificial Intelligence and
  Statistics}, pages 4235--4258. PMLR, 2022{\natexlab{b}}.

\bibitem[Gonz{\'a}lez et~al.(2017)Gonz{\'a}lez, Dai, Damianou, and
  Lawrence]{gonzalez2017preferential}
Javier Gonz{\'a}lez, Zhenwen Dai, Andreas Damianou, and Neil~D Lawrence.
\newblock Preferential bayesian optimization.
\newblock In \emph{International Conference on Machine Learning}, pages
  1282--1291. PMLR, 2017.

\bibitem[Harper et~al.(2021)Harper, Mustafee, and Yearworth]{Harper+2021trust}
Alison Harper, Navonil Mustafee, and Mike Yearworth.
\newblock Facets of trust in simulation studies.
\newblock \emph{European Journal of Operational Research}, 289\penalty0
  (1):\penalty0 197--213, 2021.
\newblock ISSN 0377-2217.
\newblock \doi{https://doi.org/10.1016/j.ejor.2020.06.043}.

\bibitem[Vernon-Bido et~al.(2015)Vernon-Bido, Collins, and
  Sokolowski]{Vernon-Bido+2015}
Daniele Vernon-Bido, Andrew Collins, and John Sokolowski.
\newblock Effective visualization in modeling \& simulation.
\newblock In \emph{Proceedings of the 48th Annual Simulation Symposium}, ANSS
  '15, page 33–40, San Diego, CA, USA, 2015. Society for Computer Simulation
  International.

\bibitem[Lane(2008)]{Lane2008}
David~C. Lane.
\newblock The emergence and use of diagramming in system dynamics: a critical
  account.
\newblock \emph{Systems Research and Behavioral Science}, 25\penalty0
  (1):\penalty0 3--23, 2008.
\newblock \doi{https://doi.org/10.1002/sres.826}.

\bibitem[Siebers and Kl{\"u}gl(2017)]{Siebers+Klugl2017}
Peer-Olaf Siebers and Franziska Kl{\"u}gl.
\newblock \emph{What Software Engineering Has to Offer to Agent-Based Social
  Simulation}, pages 81--117.
\newblock Springer International Publishing, Cham, 2017.
\newblock \doi{10.1007/978-3-319-66948-9_6}.

\bibitem[Tominski et~al.(2009)Tominski, Abello, and Schumann]{tominski2009cgv}
Christian Tominski, James Abello, and Heidrun Schumann.
\newblock Cgv—an interactive graph visualization system.
\newblock \emph{Computers \& Graphics}, 33\penalty0 (6):\penalty0 660--678,
  2009.

\bibitem[Schulz(2011)]{schulz2011treevis}
Hans-Jorg Schulz.
\newblock Treevis.net: A tree visualization reference.
\newblock \emph{IEEE Computer Graphics and Applications}, 31\penalty0
  (6):\penalty0 11--15, 2011.

\bibitem[St-Aubin et~al.(2023)St-Aubin, Wainer, and Loor]{St-Aubin+2023}
Bruno St-Aubin, Gabriel Wainer, and Fernando Loor.
\newblock A survey of visualization capabilities for simulation environments.
\newblock In \emph{2023 Annual Modeling and Simulation Conference (ANNSIM)},
  pages 13--24, 2023.

\bibitem[Unger and Schumann(2009)]{Unger&Schumann2009}
Andrea Unger and Heidrun Schumann.
\newblock Visual support for the understanding of simulation processes.
\newblock In \emph{2009 IEEE Pacific Visualization Symposium}, pages 57--64,
  2009.
\newblock \doi{10.1109/PACIFICVIS.2009.4906838}.

\bibitem[Eichner et~al.(2014)Eichner, Bittig, Schumann, and
  Tominski]{Eichner+2014}
Christian Eichner, Arne Bittig, Heidrun Schumann, and Christian Tominski.
\newblock Analyzing simulations of biochemical systems with feature-based
  visual analytics.
\newblock \emph{Computers \& Graphics}, 38:\penalty0 18--26, 2014.
\newblock ISSN 0097-8493.
\newblock \doi{https://doi.org/10.1016/j.cag.2013.09.001}.

\bibitem[Feldkamp et~al.(2020{\natexlab{b}})Feldkamp, Bergmann, and
  Strassburger]{Feldkamp+2020}
Niclas Feldkamp, Soeren Bergmann, and Steffen Strassburger.
\newblock Knowledge discovery in simulation data.
\newblock \emph{ACM Transactions on Modeling and Computer Simulation},
  30\penalty0 (4), nov 2020{\natexlab{b}}.
\newblock ISSN 1049-3301.
\newblock \doi{10.1145/3391299}.

\bibitem[Matković et~al.(2018)Matković, Gračanin, and Hauser]{Kresimir2018}
Krešimir Matković, Denis Gračanin, and Helwig Hauser.
\newblock Visual analytics for simulation ensembles.
\newblock In \emph{2018 Winter Simulation Conference (WSC)}, pages 321--335,
  2018.
\newblock \doi{10.1109/WSC.2018.8632312}.

\bibitem[Hakanen et~al.(2022)Hakanen, Radoš, Misitano, Saini, Miettinen, and
  Matković]{Kresimir2022a}
Jussi Hakanen, Sanjin Radoš, Giovanni Misitano, Bhupinder~S. Saini, Kaisa
  Miettinen, and Krešimir Matković.
\newblock Interactivized: Visual interaction for better decisions with
  interactive multiobjective optimization.
\newblock \emph{IEEE Access}, 10:\penalty0 33661--33678, 2022.
\newblock \doi{10.1109/ACCESS.2022.3161465}.

\bibitem[Andrienko et~al.(2018)Andrienko, Lammarsch, Andrienko, Fuchs, Keim,
  Miksch, and Rind]{andrienko2018viewing}
Natalia Andrienko, Tim Lammarsch, Gennady Andrienko, Georg Fuchs, Daniel Keim,
  Silvia Miksch, and Andrea Rind.
\newblock Viewing visual analytics as model building.
\newblock In \emph{Computer Graphics Forum}, volume~37, pages 275--299. Wiley
  Online Library, 2018.

\bibitem[Sondag et~al.(2022)Sondag, Turkay, Xu, Matthews, Mohr, and
  Archambault]{sondag2022visual}
Max Sondag, Cagatay Turkay, Kai Xu, Louise Matthews, Sibylle Mohr, and Daniel
  Archambault.
\newblock Visual analytics of contact tracing policy simulations during an
  emergency response.
\newblock In \emph{Computer Graphics Forum}, volume~41, pages 29--41. Wiley
  Online Library, 2022.

\bibitem[Afzal et~al.(2020)Afzal, Ghani, Jenkins-Smith, Ebert, Hadwiger, and
  Hoteit]{afzal2020visual}
Shehzad Afzal, Sohaib Ghani, Hank~C Jenkins-Smith, David~S Ebert, Markus
  Hadwiger, and Ibrahim Hoteit.
\newblock A visual analytics based decision making environment for covid-19
  modeling and visualization.
\newblock In \emph{2020 IEEE Visualization Conference (VIS)}, pages 86--90.
  IEEE, 2020.

\bibitem[Vennix(1999)]{Vennix+1999}
Jac A.~M. Vennix.
\newblock Group model-building: Tackling messy problems.
\newblock \emph{System Dynamics Review}, 15\penalty0 (4):\penalty0 379--401,
  1999.
\newblock
  \doi{https://doi.org/10.1002/(SICI)1099-1727(199924)15:4<379::AID-SDR179>3.0.CO;2-E}.

\bibitem[Ramanath and Gilbert(2004)]{Ramanath&Gilbert2004}
Ana~Maria Ramanath and Nigel Gilbert.
\newblock The design of participatory agent-based social simulations.
\newblock \emph{Journal of Artificial Societies and Social Simulation},
  7\penalty0 (4):\penalty0 1, 2004.
\newblock URL \url{https://www.jasss.org/7/4/1.html}.

\bibitem[Voinov et~al.(2018)Voinov, Jenni, Gray, Kolagani, Glynn, Bommel,
  Prell, Zellner, Paolisso, Jordan, Sterling, {Schmitt Olabisi}, Giabbanelli,
  Sun, {Le Page}, Elsawah, BenDor, Hubacek, Laursen, Jetter, Basco-Carrera,
  Singer, Young, Brunacini, and Smajgl]{Voinov+2018}
Alexey Voinov, Karen Jenni, Steven Gray, Nagesh Kolagani, Pierre~D. Glynn,
  Pierre Bommel, Christina Prell, Moira Zellner, Michael Paolisso, Rebecca
  Jordan, Eleanor Sterling, Laura {Schmitt Olabisi}, Philippe~J. Giabbanelli,
  Zhanli Sun, Christophe {Le Page}, Sondoss Elsawah, Todd~K. BenDor, Klaus
  Hubacek, Bethany~K. Laursen, Antonie Jetter, Laura Basco-Carrera, Alison
  Singer, Laura Young, Jessica Brunacini, and Alex Smajgl.
\newblock Tools and methods in participatory modeling: Selecting the right tool
  for the job.
\newblock \emph{Environmental Modelling \& Software}, 109:\penalty0 232--255,
  2018.
\newblock ISSN 1364-8152.
\newblock \doi{https://doi.org/10.1016/j.envsoft.2018.08.028}.

\bibitem[Barreteau et~al.(2017)Barreteau, Bots, Daniell, Etienne, Perez,
  Barnaud, Bazile, Becu, Castella, Dar{\'e}, and Trebuil]{Barreteau+2017}
Olivier Barreteau, Pieter Bots, Katherine Daniell, Michel Etienne, Pascal
  Perez, C{\'e}cile Barnaud, Didier Bazile, Nicolas Becu, Jean-Christophe
  Castella, William's Dar{\'e}, and Guy Trebuil.
\newblock \emph{Participatory Approaches, in Simulating Social Complexity: A
  Handbook}, pages 253--292.
\newblock Springer International Publishing, 2017.
\newblock \doi{10.1007/978-3-319-66948-9_12}.
\newblock URL \url{https://doi.org/10.1007/978-3-319-66948-9_12}.

\bibitem[Barnaud et~al.(2008)Barnaud, Bousquet, and Trebuil]{Barnaud+2008}
Cécile Barnaud, François Bousquet, and Guy Trebuil.
\newblock Multi-agent simulations to explore rules for rural credit in a
  highland farming community of northern thailand.
\newblock \emph{Ecological Economics}, 66\penalty0 (4):\penalty0 615--627,
  2008.
\newblock \doi{https://doi.org/10.1016/j.ecolecon.2007.10.022}.

\bibitem[Sahraoui et~al.(2021)Sahraoui, {De Godoy Leski}, Benot, Revers,
  Salles, {van Halder}, Barneix, and Carassou]{Sahraoui+2021}
Yohan Sahraoui, Charles {De Godoy Leski}, Marie-Lise Benot, Frédéric Revers,
  Denis Salles, Inge {van Halder}, Marie Barneix, and Laure Carassou.
\newblock Integrating ecological networks modelling in a participatory approach
  for assessing impacts of planning scenarios on landscape connectivity.
\newblock \emph{Landscape and Urban Planning}, 209:\penalty0 104039, 2021.
\newblock \doi{https://doi.org/10.1016/j.landurbplan.2021.104039}.

\bibitem[{van Bruggen} et~al.(2019){van Bruggen}, Nikolic, and
  Kwakkel]{Bruggen+2019}
Anna {van Bruggen}, Ior Nikolic, and Jan Kwakkel.
\newblock Modeling with stakeholders for transformative change.
\newblock \emph{Sustainability}, 11\penalty0 (3):\penalty0 825, 2019.
\newblock \doi{https://doi.org/10.3390/su11030825}.

\bibitem[Barreteau and Others(2003)]{Barreteau+2003}
Olivier Barreteau and Others.
\newblock Our companion modelling approach.
\newblock \emph{Journal of Artificial Societies and Social Simulation},
  6\penalty0 (2):\penalty0 1, 2003.
\newblock URL \url{https://jasss.soc.surrey.ac.uk/6/2/1.html}.

\bibitem[Will et~al.(2021)Will, Dressler, Kreuer, Thulke, Grêt-Regamey, and
  Müller]{Will+2021}
Meike Will, Gunnar Dressler, David Kreuer, Hans-Hermann Thulke, Adrienne
  Grêt-Regamey, and Birgit Müller.
\newblock How to make socio-environmental modelling more useful to support
  policy and management?
\newblock \emph{People and Nature}, 3\penalty0 (3):\penalty0 560--572, 2021.
\newblock \doi{https://doi.org/10.1002/pan3.10207}.

\bibitem[Bouchet et~al.(2022)Bouchet, Thoms, and Parsons]{bouchet2022using}
Louis Bouchet, Martin~C Thoms, and Melissa Parsons.
\newblock Using causal loop diagrams to conceptualize groundwater as a
  social-ecological system.
\newblock \emph{Frontiers in Environmental Science}, 10:\penalty0 836206, 2022.

\bibitem[Parry and Evans(2008)]{parry2008comparative}
Hazel~R Parry and Andrew~J Evans.
\newblock A comparative analysis of parallel processing and super-individual
  methods for improving the computational performance of a large
  individual-based model.
\newblock \emph{Ecological Modelling}, 214\penalty0 (2-4):\penalty0 141--152,
  2008.

\bibitem[Keim et~al.(2008)Keim, Andrienko, Fekete, G{\"{o}}rg, Kohlhammer, and
  Melan{\c{c}}on]{Keim2008}
Daniel~A. Keim, Gennady~L. Andrienko, Jean{-}Daniel Fekete, Carsten G{\"{o}}rg,
  J{\"{o}}rn Kohlhammer, and Guy Melan{\c{c}}on.
\newblock Visual analytics: Definition, process, and challenges.
\newblock In Andreas Kerren, John~T. Stasko, Jean{-}Daniel Fekete, and Chris
  North, editors, \emph{Information Visualization - Human-Centered Issues and
  Perspectives}, volume 4950 of \emph{Lecture Notes in Computer Science}, pages
  154--175. Springer, 2008.
\newblock \doi{10.1007/978-3-540-70956-5\_7}.
\newblock URL \url{https://doi.org/10.1007/978-3-540-70956-5\_7}.

\end{thebibliography}

\end{document}